\documentclass[smallcondensed,final]{svjour3}
\pdfoutput=1

\smartqed  

\usepackage{array}
\usepackage{amsmath}
\usepackage{amssymb}
\usepackage{bm}
\usepackage{graphics}
\usepackage{mathrsfs}
\usepackage{verbatim}
\usepackage{subfigure}
\usepackage{placeins}
\usepackage{overpic}

\usepackage[square,numbers]{natbib}
\usepackage{marvosym}
\usepackage{xcolor}
\usepackage[normalem]{ulem}

\begin{document}

\title{Influence of Different Subgrid Scale Models in Low-Order LES of Supersonic Jet Flows}

\titlerunning{Influence of Subgrid Scale Models in Low-Order LES} 


\author{Carlos Junqueira-Junior \and Sami Yamouni \and Jo\~{a}o Luiz F. 
Azevedo \and William R. Wolf}

\institute{Carlos Junqueira-Junior (\Letter) \and Jo\~{a}o Luiz F. Azevedo 
	       \at Instituto de Aeron\'{a}utica e Espa\c{c}o, DCTA/IAE/ALA, S\~{a}o 
		   Jos\'{e} dos Campos,
		   SP, 12228-904, Brazil \\ \email{junior.hmg@gmail.com}
	  \and Sami Yamouni 
           \at Instituto Tecnol\'{o}gico de Aeron\'{a}utica, DCTA/ITA, S\~{a}o 
		   Jos\'{e} dos Campos, SP, 
           12228-900, Brazil 
	  \and  William R. Wolf
		   \at Faculdade de Engenharia {  Mec\^{a}nica}, 
		   Universidade Estadual de Campinas, Rua Mendeleyev, 200,
		   Campinas, SP, 13083-860, Brazil
		   }

\date{2016}

\maketitle

\begin{abstract}

The present work is concerned with a study of large eddy simulations (LES) 
of unsteady turbulent jet flows. In particular, the present analysis is 
focused on the effects of the subgrid scale modeling used when a second-order 
spatial discretization methodology is employed for the numerical simulations. 
The present effort addresses perfectly expanded supersonic jets, because the 
authors want to emphasize the effects of the jet mixing phenomena. The LES 
formulation is discretized using the finite difference approach, after the 
equations are rewritten in a generalized coordinate system. Both space and 
time discretizations are second order accurate, and an explicit time march 
is adopted. Special care is dedicated to the discretization of the energy 
equation in order to appropriately model the energy equation of the filtered 
Navier-Stokes formulation. The classical Smagorinsky, the dynamic Smagorinsky 
and the Vreman models are the subgrid scale closures selected for the present 
work. 
The computational results are compared to data 
in the literature in order to validate the present simulation tool. 
{  Results indicate that the characteristics of numerical 
discretization can overcome effects of the subgrid scale models. A detailed 
analysis is presented for the performance of each subgrid closure in the numerical 
context here considered.}


\end{abstract}

\section{Introduction}

A novel compressible large eddy simulation (LES) tool has recently been developed at 
Instituto de Aeron\'{a}utica e Espa\c{c}o (IAE) \cite{JuniorAIAA2015}. This numerical 
tool was developed, to a large extent, to generate unsteady flow data on large launch 
vehicle propulsion exhaust jets, for which the noise generated on certain flight 
conditions can be extremely relevant for dimensioning of payload structures being 
carried by the launcher. The approach intended was to study the aeroacoustics of 
such jets using a hybrid approach, based on the Ffowcs Williams and Hawkings analogy 
\cite{Wolf2012}. As such, and considering previous capabilities available in the research 
group \cite{BIGA02,BIGA07}, the decision was to use a low order spatial discretization 
approach for such development. The LES tool developed was denoted JAZzY, and further 
details of its formulation and computational performance can be found in 
Refs.\ \cite{Junior16,JuniorAIAA2015}.

It should be emphasized that the use of low order spatial discretization for LES calculations 
is not a new idea, and it is actually suggested by several recognized research groups in the 
world, as indicated, for instance, in Ref.\ \cite{moin2012}. The main aspect involved in 
such proposition is computational efficiency, which can typically be achieved with the low 
order spatial discretization. However, the critical question that 
arises in such context is the effectiveness of the subgrid scale modeling \cite{Li15,WangAIAA2015}. 
Therefore, the present paper is precisely directed towards the study of the effects of the 
subgrid scale modeling used when a second-order spatial discretization methodology is employed 
for the numerical simulations with an LES formulation. The subgrid scale closures included in 
the present study are the classical Smagorinsky model \cite{Lilly65,Lilly67,Smagorinsky63}, 
the dynamic Smagorinsky model \cite{Germano91,moin91} and the Vreman model \cite{vreman2004}.

In the present effort, the LES formulation is discretized using the finite difference 
approach, after the governing equations are rewritten in general curvilinear coordinates 
\cite{Junior16,JuniorAIAA2015}. Inviscid and viscous numerical fluxes are calculated using a 
second-order accurate centered scheme with the explicit addition of artificial dissipation 
terms. Time march uses a five-stage, second-order accurate, explicit Runge-Kutta scheme. 
The present formulation for the energy equation is based on the System I set of equations 
\cite{Vreman1995}, in order to appropriately model the filtered terms of the energy equation. 
The test cases address numerical simulations of perfectly expanded jet flows, and the current 
results are compared to both numerical \cite{Mendez10} and experimental \cite{bridges2008turbulence} 
independent data.
%


\section{Large Eddy Simulation Filtering}

The large eddy simulation is based on the principle 
of scale separation, which is addressed as a 
filtering procedure in a mathematical formalism. A 
modified version of the the System I filtering 
approach \cite{Vreman1995} is used in present work, 
which is given by
\begin{equation}
\begin{array}{c}
\displaystyle \frac{\partial \overline{\rho} }{\partial t} + \frac{\partial}{\partial x_{j}} 
\left( \overline{\rho} \widetilde{  u_{j} } \right) = 0 \, \mbox{,}\\
\displaystyle \frac{\partial}{\partial t} \left( \overline{ \rho } \widetilde{ u_{i} } \right) 
+ \frac{\partial}{\partial x_{j}} 
\left( \overline{ \rho } \widetilde{ u_{i} } \widetilde{ u_{j} } \right)
+ \frac{\partial \overline{p}}{\partial x_{i}} 
- \frac{\partial {\tau}_{ij}}{\partial x_{j}}  
+ \frac{1}{3} \frac{\partial}{\partial x_{j}}\left( {\delta}_{ij} \sigma_{kk}\right)
= 0 \, \mbox{,} \\ 
\displaystyle \frac{\partial \overline{e}}{\partial t} 
+ \frac{\partial}{\partial x_{j}} 
\left[ \left( \overline{e} + \overline{p} \right)\widetilde{u_{j}} \right]
- \frac{\partial}{\partial x_{j}}\left({\tau}_{ij} \widetilde{u_{i}} \right)
+ \frac{1}{3} \frac{\partial}{\partial x_{j}}
\left[ \left( \delta_{ij}{\sigma}_{kk} \right) \widetilde{u_{i}} \right]
+ \frac{\partial {q}_{j}}{\partial x_{j}} = 0 \, \mbox{,}
\end{array}
\label{eq:modified_system_I}
\end{equation}
in which $t$ and $x_{i}$ are independent variables 
representing time and spatial coordinates of a 
Cartesian coordinate system, $\textbf{x}$, respectively. 
The components of the velocity vector, $\textbf{u}$, are 
written as $u_{i}$ and $i=1,2,3$. Density, pressure and 
total energy {  per unit volume} are denoted by $\rho$, $p$ and 
$e$, respectively. The $\left( \overline{\cdot} \right)$ and
$\left( \tilde{\cdot} \right)$ operators are used in order 
to represent filtered and Favre averaged properties, 
respectively. The System I formulation neglects the double 
correlation term and the total energy {  per unit volume}
is written as 
\begin{equation}
	\overline{e} = \frac{\overline{p}}{\gamma - 1} 
	+ \frac{1}{2} \rho \widetilde{u}_{i} \widetilde{u}_{i} \, \mbox{.} 
\end{equation}
The heat flux, $q_{j}$, is given by
\begin{equation}
	{q}_{j} = \left(\kappa+{\kappa}_{sgs}\right) 
	\frac{\partial \widetilde{T}}{\partial x_{j}} 
	\, \mbox{.}
	\label{eq:q_mod}
\end{equation}
where $T$ is the static temperature and $\kappa$ is the thermal 
conductivity {  coefficient}, which can by expressed {  as}
\begin{equation}
\kappa = \frac{\mu C_{p}}{Pr} \, \mbox{,}
\end{equation}
The thermal conductivity {  coefficient} is a function of the specific heat at 
constant pressure, $Cp$, of the Prandtl number, $Pr$, which is 
equal to $0.72$ for air, and of the dynamic viscosity {  coefficient}, $\mu$.
The SGS thermal conductivity {  coefficient}, $\kappa_{sgs}$, is written as
\begin{equation}
	\kappa_{sgs} = \frac{\mu_{sgs} C_{p}}{ {Pr}_{sgs} } 
	\, \mbox{,}
	\label{eq:kappa_sgs}
\end{equation}
where ${Pr}_{sgs}$ is the SGS Prandtl number, which is
equal to $0.9$ for static SGS models and $\mu_{sgs}$
is the eddy viscosity {  coefficient} which is calculated by the SGS
closure. The dynamic viscosity {  coefficient}, $\mu$, can be calculated 
using the Sutherland Law,
\begin{eqnarray}
	\mu \left( \widetilde{T} \right) = \mu_{\infty} 
	\left( \frac{\widetilde{T}}{\widetilde{T}_{\infty}}
	\right)^{\frac{3}{2}} 
	\frac{\widetilde{T}_{0}+S_{1}}{\widetilde{T}+S_{1}} \mbox{ , }
	& \mbox{with} \: S_{1} = 110.4K \, \mbox{.}
\label{eq:sutherland}
\end{eqnarray}
Density, static pressure and static temperature are correlated 
by the equation of state, given by
\begin{equation}
	\overline{p} = {  \overline{\rho}} R \widetilde{T} \, \mbox{,}
\end{equation}
where $R$ is the gas constant, written as
\begin{equation}
R = C_{p} - C_{v} \, \mbox{,}
\end{equation}
and $C_{v}$ is the specif heat at constant volume.
The shear-stress tensor, $\tau_{ij}$, is written 
according to the Stokes hypothesis and includes
the eddy viscosity {  coefficient}, $\mu_{sgs}$,
\begin{equation}
	{\tau}_{ij} = 2 \left(\mu+{\mu}_{sgs}\right) 
	\left( \tilde{S}_{ij} - \frac{1}{3} \delta_{ij} \tilde{S}_{kk} \right) \,
	\label{eq:tau_mod}
\end{equation}
in which the components of the rate-of-strain tensor, $\tilde{S}_{ij}$, are 
given by
\begin{equation}
	\tilde{S}_{ij} = \frac{1}{2} 
	\left( \frac{\partial \tilde{u}_{i}}{\partial x_{j}} 
	+ \frac{\partial \tilde{u}_{j}}{\partial x_{i}} 
\right) \, \mbox{.}
\end{equation}
The SGS stress tensor components are written using the eddy 
viscosity {  coefficient} \cite{Sagaut05},
\begin{equation}
    \sigma_{ij} = - 2 \mu_{sgs} \left( \tilde{S}_{ij} 
	- \frac{1}{3} \tilde{S}_{kk} \right)
    + \frac{1}{3} \delta_{ij} \sigma_{kk}
    \, \mbox{.}
    \label{eq:sgs_visc}
\end{equation}
The eddy viscosity {  coefficient}, $\mu_{sgs}$, and the components of the 
isotropic part of the SGS stress tensor, $\sigma_{kk}$, are
modeled by the SGS closure.

\section{Subgrid Scale Modeling}

The present section is directed towards the description of the turbulence 
modeling and the theoretical formulation of subgrid {  scale} closures 
included in the present work. The {  closure} models presented here are 
{  based} on the homogeneous turbulence theory, which is usually developed 
in the spectral space as an {  attempt} to quantify the interaction between 
the different scales of turbulence.

\subsection{Smagorinsky Model}

The Smagorinsky model \cite{Smagorinsky63} is one of the simplest 
{  algebraic} models for the deviatoric part of the SGS 
tensor used in large eddy simulations. The isotropic part of the 
SGS tensor is neglected for {  the} Smagorinsky model in the current work. 
This SGS closure is a classical model {  based on} the 
large scale properties and {  it} is written as 
\begin{equation}
{  \mu_{sgs} = \overline{\rho} \, \left( C_{s} \Delta \right)^{2} 
	| \widetilde{S} | \, \mbox{,}}
\end{equation}
where
\begin{equation}
| \tilde{S} | = 
\left( 2 \tilde{S}_{ij} \tilde{S}_{ij} \right)^{\frac{1}{2}} \, \mbox{,}
\end{equation}
$\Delta$ is the filter size and $C_{s}$ is the Smagorinsky constant. 
Several attempts can be found in the literature regarding the 
evaluation of the Smagorinsky constant. The value of this constant 
is adjusted to improve the results for different flow 
configurations. In {  practical} terms, the Smagorinsky subgrid 
model has a flow dependency on the constant, which takes values 
ranging from $0.1$ to $0.2$ depending on the flow. The {  value suggested 
by} Lilly \cite{Lilly67}, $C_{s}=0.148$, is used in the current 
work.

This model is generally over dissipative in regions of large mean strain. 
This is {  particularly} true in the transitional region between laminar and 
turbulent flows. Moreover, the limiting behavior near the wall is not
correct, and the model predictions correlate poorly with the exact subgrid 
scale tensor \cite{Garnier09}. However, it is a very simple model and, 
with the use of damping functions and good calibration, it can be successfully 
applied in large eddy simulations.

\newpage
\FloatBarrier

\subsection{Vreman Model} 

Vreman \cite{vreman2004} proposed a turbulence model that can correctly predict
inhomogeneous turbulent flows. For such flows, the eddy viscosity should become 
small in laminar and transitional regions. This requirement is unfortunately not 
satisfied by existing simple eddy viscosity closures such as the classic 
Smagorinsky model \cite{Deardorff70,Lilly65,Smagorinsky63}. The Vreman SGS model
is also very simple and it is given by
\begin{equation}
	{
	  \mu_{sgs} = \overline{\rho} \, \bm{c} \, 
	\sqrt{\frac{B_{\beta}}{\alpha_{ij} \alpha_{ij}}
	} 
	\,\mbox{ ,}}
\end{equation}
with
\begin{equation}
	\alpha_{ij} = \frac{\partial \tilde{u}_{j}}{\partial x_{i}} 
	\, \mbox{ ,}
\end{equation}
\begin{equation}
	{  B_{\beta} = \beta_{11}\beta_{22} - \beta_{12}^{2} 
	          + \beta_{11}\beta_{33} - \beta_{13}^{2}
			  + \beta_{22}\beta_{33} - \beta_{23}^{2}
			  \, \mbox{ ,}}
\end{equation}
and 
\begin{equation}
	{ \beta_{ij} = \Delta^{2}_{m}\alpha_{mi}\alpha_{mj}} \mbox{ .}
\end{equation}
The $\bm{c}$ constant is related to the Smagorinsky constant, $C_{s}$, 
and it is given by
\begin{equation}
	\bm{c} = 2.5 \, C_{s}^{2} 
	\, \mbox{,}
\end{equation}
and $\Delta_{m}$ is the filter width in each direction. In the present work,
the isotropic part of the SGS tensor is neglected for the Vreman model.
The $\alpha$ symbol represents the matrix of {  first order}
derivatives of the filtered components of velocity, $\tilde{u}_{i}$. The SGS 
eddy viscosity {  coefficient} is defined as zero when $\alpha_{ij}\alpha_{ij}$ equals zero. 
Vreman \cite{vreman2004} states that the $\beta_{ij}$ tensor is proportional to 
the Clark model \cite{Clark79,Leonard74} in its general anisotropic form 
\cite{vreman1996}.

The Vreman model can be classified as a very simple model because it is {  expressed}
in first-order derivatives and it {  does} not involves explicit filtering, 
averaging and clipping procedures, and it is rotationally invariant for isotropic 
filter widths. The model was originally created for incompressible flows and it 
has presented good results for two incompressible flows configurations: the 
transitional and turbulent mixing layer at high Reynolds number and the 
turbulent channel flow \cite{vreman1996}. In both cases, the Vreman model is 
found to be more accurate than the classical Smagorinsky model and as good as 
the dynamic Smagorinsky model.

\subsection{Dynamic Smagorinsky Model}

Germano {\it et al.} \cite{germano90} developed a dynamic SGS model in order to 
overcome the issues of the classical Smagorinsky closure. The model uses the strain 
rate fields at two different scales and, thus, extracts spectral information in 
the large scale field to extrapolate the small stresses \cite{moin91}. 
The coefficients of the model are computed instantaneously in the dynamic model. 
They are a function of the positioning in space and time rather than being specified 
{\em a priori}. Moin {\it et al.} \cite{moin91} extended the work of Germano for compressible 
flows. The dynamic Smagorinsky model for compressible flow configurations is detailed in 
the present section.

The dynamic model introduces the test filter, $\widehat{\left( \cdot \right)}$, which
has a larger filter width, $\widehat{\Delta}$, than the one of the resolved grid filter, 
$\overline{\left( \cdot \right)}$. The use of test filters generates a second field 
with larger scales than the resolved field. The Yoshizawa model \cite{Yoshizawa86} is 
used for the isotropic portion of the SGS tensor and it is written as
\begin{equation}
	\sigma_{ll} = 2 C_{I} \overline{\rho} {\Delta}^{2}|\tilde{S}|^{2}
	\, \mbox{,}
	\label{eq:yoshizawa}
\end{equation}
where $C_{I}$ is defined by
\begin{equation}
	C_{I} = \frac{\biggl\langle \widehat{\overline{\rho} \tilde{u}_{l} \tilde{u}_{l}} -
	        \left( \widehat{\overline{\rho}\tilde{u}_{l}}
	               \widehat{\overline{\rho}\tilde{u}_{l}}/
				   \widehat{\overline{\rho}} \right)\biggr\rangle }
		   {\biggl\langle 2 \widehat{\Delta}^{2} \widehat{\overline{\rho}}
		   |\widehat{\overline{S}}|^{2} - 
	        2 {\Delta}^{2} \widehat{ \overline{\rho}
		   |\overline{S}|^{2}}\biggr\rangle}
		   \, \mbox{.}
		   \label{eq:av_ci}
\end{equation}
A volume averaging, here indicated by $\langle \,\, \rangle$, is {  
suggested by} Moin {\it et al} \cite{moin91} and by Garnier {\it et al} 
\cite{Garnier09} in order to avoid numerical issues. The eddy viscosity, 
$\mu_{sgs}$, is calculated using the same approach used by static Smagorinsky 
model,
\begin{equation}
\mu_{sgs} = \left( \rho C_{ds} \Delta \right)^{2} | \tilde{S} | \, \mbox{,}
\end{equation}
where
\begin{equation}
| \tilde{S} | = \left( 2 \tilde{S}_{ij} \tilde{S}_{ij} \right)^{\frac{1}{2}} 
\, \mbox{,}
\end{equation}
and $C_{ds}$ is the dynamic constant of the model, which is given by
\begin{equation}
	C_{ds} = \frac{\biggl\langle
	        \left[ \widehat{\overline{\rho} \tilde{u}_{i} \tilde{u}_{j}} -
	        \left( \widehat{\overline{\rho}\tilde{u}_{i}}
	        \widehat{\overline{\rho}\tilde{u}_{j}}/
			\widehat{\overline{\rho}} \right) \right]\tilde{S}_{ij} - 
            \frac{1}{3}\tilde{S}_{mm}
	        \left(\mathscr{T}_{ll} - \widehat{\sigma}_{ll}\right)
			\biggr\rangle}{\biggl\langle
			2 {\Delta}^{2}\left[
			\widehat{\overline{\rho}|\tilde{S}|\tilde{S}_{ij}}\tilde{S}_{ij}
			- \frac{1}{3}\left(\overline{\rho}|\tilde{S}|\tilde{S}_{mm}\right)^{\widehat{ }}
			\tilde{S}_{ll}\right] -
            2 \widehat{\Delta}^{2}\left(
			\widehat{\overline{\rho}}|\widehat{\tilde{S}}|
			\widehat{\tilde{S}}_{ij}\tilde{S}_{ij} -
            \frac{1}{3}\widehat{\overline{\rho}}|\widehat{\tilde{S}}|
            \widehat{\tilde{S}}_{mm} \tilde{S}_{ll}\right)
			\biggr\rangle}
	\, \mbox{.}
    \label{eq:av_cd}
\end{equation}
The SGS Prandtl number is computed using the dynamic constant, $C_{ds}$, and 
written as
\begin{equation}
	{Pr}_{sgs} = C_{ds} \frac{\biggl\langle \Delta^{2} 
	\biggl(\overline{\rho}|\tilde{S}|\frac{\partial \overset{\sim}{T}}{\partial x_{j}}
	\biggr)^{\widehat{ }} \,
	\frac{\partial\overset{\sim}{T}}{\partial x_{j}} -
	\widehat{\Delta}^{2}\widehat{\overline{\rho}}|\widehat{\tilde{S}}|
	\frac{\partial \overset{\sim}{T}}{\partial x_{j}}
	\frac{\partial \overset{\sim}{T}}{\partial x_{j}} \biggr\rangle}
	{\biggl\langle\left[ \widehat{\overline{\rho} \tilde{u}_{j} \overset{\sim}{T}} - 
	\left( \widehat{
		\overline{\rho} \tilde{u_{j}}
	               } 
	\widehat{\overline{\rho} \overset{\sim}{T}} \right)/
		   \widehat{\overline{\rho}}\right] 
		   \frac{\partial \overset{\sim}{T}}{\partial x_{j}}\biggr\rangle}
	\, \mbox{.}
	\label{eq:av_Pr_sgs}
\end{equation}
%



\section{Transformation of Coordinates}

In the present work, the filtered Navier-Stokes equations are written in strong 
conservation law form for a 3-D general curvilinear coordinate system as
\begin{equation}
	\frac{\partial \hat{Q}}{\partial t} 
	+ \frac{\partial }{\partial \xi}\left(\hat{\mathbf{E}}_{e}-\hat{\mathbf{E}}_{v}\right) 
	+ \frac{\partial}{\partial \eta}\left(\hat{\mathbf{F}}_{e}-\hat{\mathbf{F}}_{v}\right)
	+ \frac{\partial}{\partial \zeta}\left(\hat{\mathbf{G}}_{e}-\hat{\mathbf{G}}_{v}\right) 
	= 0 \, \mbox{.}
	\label{eq:vec-LES}
\end{equation}
The general coordinate transformation adopted in the present case can be written as
\begin{eqnarray}
	\xi & = & \xi {  \left(x,y,z \right)}  \, \mbox{,} \nonumber\\
	\eta & = & \eta {  \left(x,y,z \right)}  \, \mbox{,} \\
	\zeta & = & \zeta {  \left(x,y,z \right)}\ \, \mbox{.} \nonumber
\end{eqnarray}
For the simulations performed in the present paper, $\xi$ is the axial jet flow direction, 
$\eta$ is the radial direction and $\zeta$ is the azimuthal direction. The new vector of
conserved properties in general curvilinear coordinates can be written as
\begin{equation}
	\hat{Q} = J^{-1} \left[ \overline{\rho} \quad \overline{\rho}\tilde{u} \quad 
	\overline{\rho}\tilde{v} \quad \overline{\rho}\tilde{w} \quad \overline{e} \right]^{T} 
	\quad \mbox{ .}
	\label{eq:hat_Q_vec}
\end{equation}
Here, $J$ is the Jacobian of the transformation, which could be expressed as
\begin{equation}
	J = \left( x_{\xi} y_{\eta} z_{\zeta} + x_{\eta}y_{\zeta}z_{\xi} +
	           x_{\zeta} y_{\xi} z_{\eta} - x_{\xi}y_{\zeta}z_{\eta} -
			   x_{\eta} y_{\xi} z_{\zeta} - x_{\zeta}y_{\eta}z_{\xi} 
	    \right)^{-1} \, \mbox{ .}
\end{equation}
The inverse metric terms of the transformation, which are used to compute 
the transformation Jacobian, can be directly computed by central finite 
differences from the mesh information. {  Such computation is given by}
\begin{eqnarray}
	\displaystyle {  x_{\xi}   = \frac{\partial x}{\partial \xi}  \, \mbox{,}} & 
	\displaystyle {  x_{\eta}  = \frac{\partial x}{\partial \eta} \, \mbox{,}} & 
	\displaystyle {  x_{\zeta} = \frac{\partial x}{\partial \zeta}\, \mbox{,}} \nonumber \\
	\displaystyle {  y_{\xi}   = \frac{\partial y}{\partial \xi}  \, \mbox{,}} & 
	\displaystyle {  y_{\eta}  = \frac{\partial y}{\partial \eta} \, \mbox{,}} & 
	\displaystyle {  y_{\zeta} = \frac{\partial y}{\partial \zeta}\, \mbox{,}} \\
	\displaystyle {  z_{\xi}   = \frac{\partial z}{\partial \xi}  \, \mbox{,}} & 
	\displaystyle {  z_{\eta}  = \frac{\partial z}{\partial \eta} \, \mbox{,}} & 
	\displaystyle {  z_{\zeta} = \frac{\partial z}{\partial \zeta}\, \mbox{.}} \nonumber
\end{eqnarray}

The inviscid flux vectors in general curvilinear coordinates, $\hat{\mathbf{E}}_{e}$, 
$\hat{\mathbf{F}}_{e}$ and $\hat{\mathbf{G}}_{e}$, can be written as
\begin{eqnarray}
	\hat{\mathbf{E}}_{e} = J^{-1} \left\{\begin{array}{c}
		\overline{\rho} U \\
		\overline{\rho}\tilde{u} U + \overline{p} \xi_{x} \\
		\overline{\rho}\tilde{v} U + \overline{p} \xi_{y} \\
		\overline{\rho}\tilde{w} U + \overline{p} \xi_{z} \\
		\left( \overline{e} + \overline{p} \right) U 
\end{array}\right\} \, \mbox{ ,} &
	\hat{\mathbf{F}}_{e} = J^{-1} \left\{\begin{array}{c}
		\overline{\rho} V \\
		\overline{\rho}\tilde{u} V + \overline{p} \eta_{x} \\
		\overline{\rho}\tilde{v} V + \overline{p} \eta_{y} \\
		\overline{\rho}\tilde{w} V + \overline{p} \eta_{z} \\
		\left( \overline{e} + \overline{p} \right) V 
\end{array}\right\} \, \mbox{ ,} & \nonumber \\
    &
	\hat{\mathbf{G}}_{e} = J^{-1} \left\{\begin{array}{c}
		\overline{\rho} W \\
		\overline{\rho}\tilde{u} W + \overline{p} \zeta_{x} \\
		\overline{\rho}\tilde{v} W + \overline{p} \zeta_{y} \\
		\overline{\rho}\tilde{w} W + \overline{p} \zeta_{z} \\
		\left( \overline{e} + \overline{p} \right) W 
	\end{array}\right\} \, \mbox{.} & 
	\label{eq:hat-flux-G}
\end{eqnarray}
{  The contravariant velocity components, $U$, $V$ and $W$, are 
calculated as}
\begin{eqnarray}
	{  U = \xi_{x}\overline{u} + \xi_{y}\overline{v} + \xi_{z}\overline{w} 
	\, \mbox{,}} \nonumber \\
	{  V = \eta_{x}\overline{u} + \eta_{y}\overline{v} + \eta_{z}\overline{w} 
	\, \mbox{,}} \\
	{  W = \zeta_{x}\overline{u} + \zeta_{y}\overline{v} + \zeta_{z}\overline{w} 
	\, \mbox{.}} \nonumber
  \label{eq:vel_contrv}
\end{eqnarray}
{  The metric terms are given by}
\begin{eqnarray}
	{  \xi_{x} = J \left( y_{\eta}z_{\zeta} - y_{\zeta}z_{\eta} \right) \, \mbox{,}} & 
	{  \xi_{y} = J \left( z_{\eta}x_{\zeta} - z_{\zeta}x_{\eta} \right) \, \mbox{,}} & 
	{  \xi_{z} = J \left( x_{\eta}y_{\zeta} - x_{\zeta}y_{\eta} \right) \, \mbox{,}} \nonumber \\
	{  \eta_{x} = J \left( y_{\eta}z_{\xi} - y_{\xi}z_{\eta} \right) \, \mbox{,}} & 
	{  \eta_{y} = J \left( z_{\eta}x_{\xi} - z_{\xi}x_{\eta} \right) \, \mbox{,}} & 
	{  \eta_{z} = J \left( x_{\eta}y_{\xi} - x_{\xi}y_{\eta} \right) \, \mbox{,}} \\
	{  \zeta_{x} = J \left( y_{\xi}z_{\eta} - y_{\eta}z_{\xi} \right) \, \mbox{,}} & 
	{  \zeta_{y} = J \left( z_{\xi}x_{\eta} - z_{\eta}x_{\xi} \right) \, \mbox{,}} & 
	{  \zeta_{z} = J \left( x_{\xi}y_{\eta} - x_{\eta}y_{\xi} \right) \, \mbox{.}} \nonumber \\
\end{eqnarray}
The viscous flux vectors in general curvilinear coordinates, $\hat{\mathbf{E}}_{v}$, $\hat{\mathbf{F}}_{v}$ and 
$\hat{\mathbf{G}}_{v}$, are written as 
\begin{equation}
	\hat{\mathbf{E}}_{v} = J^{-1} \left\{\begin{array}{c}
		0 \\
		\xi_{x}{\tau}_{xx} +  \xi_{y}{\tau}_{xy} + \xi_{z}{\tau}_{xz} \\
		\xi_{x}{\tau}_{xy} +  \xi_{y}{\tau}_{yy} + \xi_{z}{\tau}_{yz} \\
		\xi_{x}{\tau}_{xz} +  \xi_{y}{\tau}_{yz} + \xi_{z}{\tau}_{zz} \\
		\xi_{x}{\beta}_{x} +  \xi_{y}{\beta}_{y} + \xi_{z}{\beta}_{z} 
	\end{array}\right\} \, \mbox{,}
	\label{eq:hat-flux-Ev}
\end{equation}
\begin{equation}
	\hat{\mathbf{F}}_{v} = J^{-1} \left\{\begin{array}{c}
		0 \\
		\eta_{x}{\tau}_{xx} +  \eta_{y}{\tau}_{xy} + \eta_{z}{\tau}_{xz} \\
		\eta_{x}{\tau}_{xy} +  \eta_{y}{\tau}_{yy} + \eta_{z}{\tau}_{yz} \\
		\eta_{x}{\tau}_{xz} +  \eta_{y}{\tau}_{yz} + \eta_{z}{\tau}_{zz} \\
		\eta_{x}{\beta}_{x} +  \eta_{y}{\beta}_{y} + \eta_{z}{\beta}_{z} 
	\end{array}\right\} \, \mbox{,}
	\label{eq:hat-flux-Fv}
\end{equation}
\begin{equation}
	\hat{\mathbf{G}}_{v} = J^{-1} \left\{\begin{array}{c}
		0 \\
		\zeta_{x}{\tau}_{xx} +  \zeta_{y}{\tau}_{xy} + \zeta_{z}{\tau}_{xz} \\
		\zeta_{x}{\tau}_{xy} +  \zeta_{y}{\tau}_{yy} + \zeta_{z}{\tau}_{yz} \\
		\zeta_{x}{\tau}_{xz} +  \zeta_{y}{\tau}_{yz} + \zeta_{z}{\tau}_{zz} \\
		\zeta_{x}{\beta}_{x} +  \zeta_{y}{\beta}_{y} + \zeta_{z}{\beta}_{z} 
	\end{array}\right\} \, \mbox{ .}
	\label{eq:hat-flux-Gv}
\end{equation}
The $\beta_{x}$, $\beta_{y}$ and $\beta_{z}$ terms, which appear in the energy equation in the viscous 
flux vectors, can be calculated as
\begin{eqnarray}
	\beta_{x} = {\tau}_{xx}\tilde{u} + {\tau}_{xy}\tilde{v} +
	{\tau}_{xz}\tilde{w} - \overline{q}_{x} \, \mbox{,} \nonumber \\
	\beta_{y} = {\tau}_{xy}\tilde{u} + {\tau}_{yy}\tilde{v} +
	{\tau}_{yz}\tilde{w} - \overline{q}_{y} \, \mbox{,} \\
	\beta_{z} = {\tau}_{xz}\tilde{u} + {\tau}_{yz}\tilde{v} +
	{\tau}_{zz}\tilde{w} - \overline{q}_{z} \mbox{.} \nonumber
\end{eqnarray}
%


\section{Dimensionless Formulation}

In the present effort, the governing equations, given by Eq.\ 
\eqref{eq:vec-LES}, are made dimensionless by an appropriate selection of 
reference variables. From the perspective of the authors, the main advantage 
of the nondimensionalization process is that all flow properties are scaled 
to the same order of magnitude, which has important computational advantages 
\cite{BIGA02}.
In the present work, the dimensionless time, $\underline{t}$, is obtained 
as a function of the speed of sound of the jet at the inlet, $a_{j}$, and 
the jet entrance diameter, $D$. Hence, it can be written as 
\begin{equation}
	\underline{t} = t \frac{a_{j}}{D} \, \mbox{ .}
	\label{eq:non-dim-time}
\end{equation}
Dimensionless velocity components are referred to the speed of sound of the 
jet at the inlet as
\begin{equation}
	\underline{\textbf{u}} = \frac{\textbf{u}}{a_{j}} \, \mbox{ .}
	\label{eq:non-dim-vel}
\end{equation}
Density, pressure and total energy per unit of volume are made dimensionless 
with regard to the density and speed of the sound of the jet at the inlet. 
Hence, they can be written as
\begin{equation}
    \underline{\rho} = \frac{\rho}{\rho_{j}} \hspace*{0.5 cm} \mbox{,} \hspace*{0.5 cm} 
	\underline{p} = \frac{p}{\rho_{j}a_{j}^{2}} \hspace*{0.5 cm} \mbox{,} \hspace*{0.5 cm}
	\underline{e} = \frac{e}{\rho_{j}a_{j}^{2}} \, \mbox{ .}
\end{equation}
Similarly, the viscosity coefficients, both bulk viscosity and subgrid scale 
viscosity coefficients, are nondimensionalized by the laminar viscosity 
coefficient at the jet exit temperature, $\mu_{j}$. The governing equations 
can, then, be rewritten, in terms of dimensionless variables, as 
\begin{equation}
	\frac{\partial \underline{Q}}{\partial t} + 
	\frac{\partial \underline{\mathbf{E}}_{e}}{\partial \xi} +
	\frac{\partial \underline{\mathbf{F}}_{e}}{\partial \eta} + 
	\frac{\partial \underline{\mathbf{G}}_{e}}{\partial \zeta} =
	\frac{M_{j}}{Re} \left( \frac{\partial \underline{\mathbf{E}}_{v}}{\partial \xi} 
	+ \frac{\partial \underline{\mathbf{F}}_{v}}{\partial \eta} 
	+ \frac{\partial \underline{\mathbf{G}}_{v}}{\partial \zeta} \right)	\, \mbox{ .}
	\label{eq:vec-underline-split-RANS}
\end{equation}
The jet exit Mach number and jet exit Reynolds number are given, respectively, 
by
\begin{eqnarray}
	M_{j}=\frac{U_j}{a_{j}} & \hspace*{0.5 cm} \mbox{and} 
	\hspace*{0.5 cm} & 	Re = \frac{\rho_{j}U_{j}D}{\mu_{j}} \, \mbox{.}
\end{eqnarray}
%


\section{Numerical Formulation}

{  The governing equations previously described are discretized in a 
structured finite difference context for a general curvilinear 
coordinate system \cite{BIGA02}. The numerical flux is calculated 
through a central difference scheme with the explicit addition 
of the anisotropic scalar artificial dissipation model of Turkel and Vatsa
\cite{Turkel_Vatsa_1994}. The time integration is performed by an 
explicit, 2nd-order, 5-stage Runge-Kutta scheme 
\cite{jameson_mavriplis_86, Jameson81}.  Conserved properties
and artificial dissipation terms are properly treated near boundaries in order
to assure the physical correctness of the numerical formulation.}
 
\subsection{Spatial Discretization}

For the remainder of the paper, the authors will drop all underbars and tildes 
in the formulation for the sake of simplicity. Nevertheless, the reader should 
be advised that all equations are referring to filtered dimensionless 
quantities. Furthermore, the $M_{j} / Re$ factor is assumed to be incorporated 
into the definition of the viscous flux vectors, again with the objective of 
simplifying the notation for the forthcoming discussion. The work uses a 
finite difference framework in order to discretize the governing equations, 
Eq.\ \eqref{eq:vec-underline-split-RANS}. Hence, the result of the 
discretization of the spatial derivatives in the governing equations can be 
written as
\begin{equation}
	\left(\frac{\partial Q}{\partial t}\right)_{i,j,k} \, = \,  - \, RHS_{i,j,k} \, \mbox{ .}	
	\label{eq:spatial_discret}
\end{equation}
Here, $RHS_{i,j,k}$ represents the residue for the $(i,j,k)$ grid point. It is 
very convenient to write $RHS_{i,j,k}$ as a function of the numerical flux 
vectors at the interfaces between grid points, following a nomenclature 
similar to the one used in Ref.\ \cite{Turkel_Vatsa_1994}. Therefore, the 
residue can be written as
\begin{eqnarray}
	{RHS}_{i,j,k} & = & 
	\frac{1}{\Delta \xi} \left( 
	{\mathbf{E}_{e}}_{(i+\frac{1}{2},j,k)} - {\mathbf{E}_{e}}_{(i-\frac{1}{2},j,k)} - 
	{\mathbf{E}_{v}}_{(i+\frac{1}{2},j,k)} + {\mathbf{E}_{v}}_{(i-\frac{1}{2},j,k)} 
	\right) \nonumber \\
	& & \frac{1}{\Delta \eta} \left( 
	{\mathbf{F}_{e}}_{(i,j+\frac{1}{2},k)} - {\mathbf{F}_{e}}_{(i,j-\frac{1}{2},k)} - 
	{\mathbf{F}_{v}}_{(i,j+\frac{1}{2},k)} + {\mathbf{F}_{v}}_{(i,j-\frac{1}{2},k)} 
	\right) \\
	& & \frac{1}{\Delta \zeta} \left( 
	{\mathbf{G}_{e}}_{(i,j,k+\frac{1}{2})} - {\mathbf{G}_{e}}_{(i,j,k-\frac{1}{2})} - 
	{\mathbf{G}_{v}}_{(i,j,k+\frac{1}{2})} + {\mathbf{G}_{v}}_{(i,j,k-\frac{1}{2})} 
	\right) \, \mbox{.} \nonumber
\end{eqnarray}

Since a centered spatial discretization is being considered, the interface numerical flux 
vectors are defined as the arithmetic average of the corresponding physical flux vectors at 
the two grid points that share that interface. However, still due to the use of a centered 
scheme, the inviscid numerical fluxes must be augmented by artificial dissipation terms, in 
order to maintain numerical stability. In the present case, the scalar, non-isotropic, artificial 
dissipation model proposed by Turkel and Vatsa \cite{Turkel_Vatsa_1994} is used. Hence, 
the numerical inviscid interface fluxes are written as
\begin{eqnarray}
	{\mathbf{E}_{e}}_{(i \pm \frac{1}{2},j,k)} 
	= \frac{1}{2} \left( {\mathbf{E}_{e}}_{(i,j,k)} + {\mathbf{E}_{e}}_{(i \pm 1,j,k)} \right)
	- J^{-1} \mathbf{d}_{(i \pm \frac{1}{2},j,k)} \, \mbox{,} \nonumber \\
	{\mathbf{F}_{e}}_{(i,j\pm \frac{1}{2},k)} 
	= \frac{1}{2} \left( {\mathbf{F}_{e}}_{(i,j,k)} + {\mathbf{F}_{e}}_{(i,j \pm 1,k)} \right)
	- J^{-1} \mathbf{d}_{(i,j \pm \frac{1}{2},k)} \, \mbox{,} \label{eq:inv_flux_vec}\\
	{\mathbf{G}_{e}}_{(i,j,k\pm \frac{1}{2})} 
	= \frac{1}{2} \left( {\mathbf{G}_{e}}_{(i,j,k)} + {\mathbf{G}_{e}}_{(i,j,k \pm 1)} \right)
	- J^{-1} \mathbf{d}_{(i,j,k \pm \frac{1}{2})} \, \mbox{,} \nonumber
\end{eqnarray}
where the $\mathbf{d}_{(i\pm \frac{1}{2},j,k)}$, 
$\mathbf{d}_{(i,j\pm \frac{1}{2},k)}$ and 
$\mathbf{d}_{(i,j,k\pm \frac{1}{2})}$ terms are the artificial dissipation 
operators. For instance, the operator in the $\xi$ direction, at the 
$(i + 1/2)$ interface, can be expressed as 
\begin{eqnarray}
	\mathbf{d}_{(i + \frac{1}{2},j,k)} & = & 
	\lambda_{(i + \frac{1}{2},j,k)} \left[ \epsilon_{(i + \frac{1}{2},j,k)}^{(2)}
	\left( \mathcal{W}_{(i+1,j,k)} - \mathcal{W}_{(i,j,k)} \right) \right. \label{eq:dissip_term}\\
	& & \epsilon_{(i + \frac{1}{2},j,k)}^{(4)} \left( \mathcal{W}_{(i+2,j,k)} 
	- 3 \mathcal{W}_{(i+1,j,k)} + 3 \mathcal{W}_{(i,j,k)} 
	- \mathcal{W}_{(i-1,j,k)} \right) \left. \right] \, \mbox{ .} \nonumber
\end{eqnarray}
In this equation, the $\epsilon^{(2)}$ and $\epsilon^{(4)}$ {  terms} are written as
\begin{eqnarray}
	\epsilon_{(i + \frac{1}{2},j,k)}^{(2)} & = &
	k^{(2)} \mbox{max} \left( \nu_{(i+1,j,k)}^{d}, 
	\nu_{(i,j,k)}^{d} \right) \, \mbox{,} \label{eq:eps_2_dissip} \\
	\epsilon_{(i + \frac{1}{2},j,k)}^{(4)} & = &
	\mbox{max} \left[ 0, k^{(4)} - \epsilon_{(i + \frac{1}{2},j,k)}^{(2)} \right] 
	\, \mbox{ .} \label{eq:eps_4_dissip}
\end{eqnarray}
The pressure gradient sensor operator, $\nu_{(i,j,k)}^{d}$, for the $\xi$ 
direction, as indicated in Ref.\ \cite{Turkel_Vatsa_1994}, is 
defined as
\begin{equation}
	\nu_{(i,j,k)}^{d} = \frac{|p_{(i+1,j,k)} - 2 p_{(i,j,k)} + p_{(i-1,j,k)}|}
	                          {p_{(i+1,j,k)} - 2 p_{(i,j,k)} + p_{(i-1,j,k)}} 
	\, \mbox{ .}
\label{eq:p_grad_sensor}
\end{equation}
The $\mathcal{W}$ vector in Eq.\ \eqref{eq:dissip_term} is calculated as a function of the
conserved variable vector, $\hat{Q}$.
{  The formulation intends to keep the total enthalpy constant in the final converged 
solution, for steady state cases, which is the correct result for the Euler equations, and hence reduce 
the effect of the artificial dissipation terms. This approach is also valid for the viscous 
formulation because the artificial dissipation terms are added to the inviscid flux terms, in which they 
are really necessary to avoid nonlinear instabilities of the numerical formulation. 
The $\mathcal{W}$ vector is given by}
\begin{equation}
	{ 
	\mathcal{W} = \hat{Q} + \left[0 \,\, 0 \,\, 0 \,\, 0 \,\, p \right]^{T} \, \mbox{.}
	}
	\label{eq:W_dissip}
\end{equation}
{  The spectral radius-based scaling factor, $\lambda$, for the i-th 
direction is written as}
\begin{equation}
	{ 
	\lambda_{(i+\frac{1}{2},j,k)} = \frac{1}{2} \left[ 
	\left( \overline{\lambda_{\xi}}\right)_{(i,j,k)} + 
	\left( \overline{\lambda_{\xi}}\right)_{(i+1,j,k)}
	\right] \, \mbox{,} 
	}
\end{equation}
{  where}
\begin{equation}
	{  
	\overline{\lambda_{\xi}}_{(i,j,k)} = \lambda_{\xi} \left[ 1 + 
	\left(\frac{\lambda_{\eta}}{\lambda_{\xi}} \right)^{0.5} + 
	\left(\frac{\lambda_{\zeta}}{\lambda_{\xi}} \right)^{0.5} \right] 
	\, \mbox{.}
	}
\end{equation}
{ The spectral radii, $\lambda_{\xi}$, $\lambda_{\eta}$ and $\lambda_{\zeta}$ are given
by}
\begin{eqnarray}
	{  \lambda_{\xi}} &{  =} & 
	{  |U| + a \sqrt{\xi_{x}^{2} + \eta_{y}^{2} + \zeta_{z}^{2}} 
	\, \mbox{,}} \nonumber \\
	{  \lambda_{\xi}} &{  =} & 
	{  |V| + a \sqrt{\xi_{x}^{2} + \eta_{y}^{2} + \zeta_{z}^{2}} 
	\, \mbox{,}} \\
	{  \lambda_{\xi}} &{  =} & 
	{  |W| + a \sqrt{\xi_{x}^{2} + \eta_{y}^{2} + \zeta_{z}^{2}} 
	\, \mbox{,}} \nonumber
\end{eqnarray}
{  in which, $U$, $V$ and $W$ are the contravariant velocity components in the $\xi$, $\eta$
and $\zeta$ directions, previously given in Eq.\ \eqref{eq:vel_contrv}, and $a$ is the local 
speed of sound, which can be written as}
\begin{equation}
	{  a = \sqrt{\frac{\gamma p}{\rho}} \, \mbox{.}}
\end{equation}
{  The calculation of artificial dissipation terms for the other coordinate directions
is completely similar and, therefore, it is not discussed here.

It should be emphasized that the present artificial dissipation model is 
nonlinear and, hence, it allows for the selection between second and fourth 
difference artificial dissipation terms. Therefore, for the problem of interest here, this approach 
ensures that, throughout most of the computational domain, only the 3rd-order fourth difference 
artificial dissipation terms are active, thus reducing the amount of artificial dissipation 
introduced in the solution. Furthermore, the scaling of the 
artificial dissipation operator in each coordinate direction, for instance, 
$\lambda_{(i + \frac{1}{2},j,k)}$ in Eq.\ \eqref{eq:dissip_term}, is primarily 
weighted by its own spectral radius of the corresponding flux Jacobian matrix, 
which gives the non-isotropic characteristics to the model 
\cite{BIGA02,Turkel_Vatsa_1994}. Further details on the artificial dissipation 
model here adopted can be seen in the original paper by Turkel and Vatsa 
\cite{Turkel_Vatsa_1994} or in Ref.\ \cite{BIGA02}. Computational aspects of the 
present implementation of the model and, in particular, issues associated to 
the computation of the various terms at partition interfaces, for parallel 
implementations, are discussed in detail in 
Refs.\ \cite{Junior16,JuniorAIAA2015,jr16-aiaa}. }

\subsection{Time Marching Method}

{  The time marching method used in the present work is a 2nd-order, 
5-step Runge-Kutta scheme based on the work of Jameson and co-authors \cite{Jameson81, 
jameson_mavriplis_86}. The time integration can be written as}
\begin{equation}
	\begin{array}{ccccc}
	Q_{(i,j,k)}^{(0)}    & = & Q_{(i,j,k)}^{n} \, \mbox{,} & & \\
	Q_{(i,j,k)}^{(\ell)} & = & Q_{(i,j,k)}^{(0)} -  
	& \alpha_{\ell} \, {\Delta t} \, {RHS}_{(i,j,k)}^{(\ell - 1)} \, & 
	\hspace*{1.0 cm} \ell = 1,2 \cdots 5, \\
	Q_{(i,jk,)}^{n+1} & = & Q_{(i,jk,)}^{(5)} \, \mbox{,} & &
	\end{array}
	\label{eq:localdt}
\end{equation}
{  The process of validation of the present solver has addressed the issues 
related to the use of an explicit time integration, and it has indicated 
that the above time marching scheme is sufficiently adequate for the current 
purposes. The interested reader can find further details in such studies
in Refs.\ \cite{Junior16,JuniorAIAA2015,jr16-aiaa}. Clearly, in 
the previous equation, $\Delta t$ is the time step, and $n$ and $n+1$ 
indicate the property values at the current and at the next time step, 
respectively. The values adopted for the $\alpha_{\ell}$ parameters are}
\begin{equation}
	\begin{array}{ccccc}
		\alpha_{1} = \frac{1}{4} \hspace*{0.3 cm} \mbox{,} \hspace*{0.3 cm} & 
		\alpha_{2} = \frac{1}{6} \hspace*{0.3 cm} \mbox{,} \hspace*{0.3 cm} &
		\alpha_{3} = \frac{3}{8} \hspace*{0.3 cm} \mbox{,} \hspace*{0.3 cm} & 
		\alpha_{4} = \frac{1}{2} \hspace*{0.3 cm} \mbox{,} \hspace*{0.3 cm} & 
		\alpha_{5} = 1 \,\mbox{ ,} 
	\end{array}
\end{equation}
according to the original reference that presents this specific Runge-Kutta 
method \cite{jameson_mavriplis_86}. The time marching scheme is linearly stable for 
$CFL \leq 2\sqrt{2}$ \cite{BIGA02}.
%


\section{{ Boundary Conditions}} \label{sec:BC}

\begin{figure}[ht]
       \begin{center}
		   \subfigure[Lateral view of boundary conditions.]{
           \includegraphics[width=0.475\textwidth]
		   {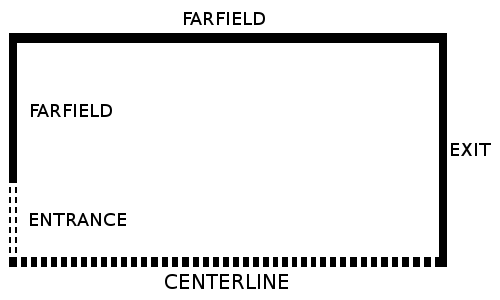} 
		   \label{fig:bc-1}
		   }
		   \subfigure[Frontal view of boundary conditions.]{
           \includegraphics[width=0.475\textwidth]
		   {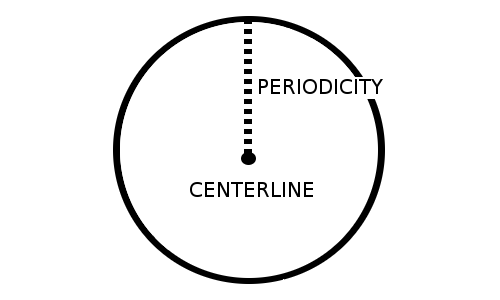} 
		   \label{fig:bc-2}
		   }
		   \caption{Lateral and frontal views of the computational domain 
		   indicating boundary conditions.}
		   \label{fig:bc}
	   \end{center}
\end{figure}

{  The geometry used in the present work presents a
cylindrical shape which is gererated by the rotation of 
a 2-D plane around a centerline. Figure \ref{fig:bc} 
presents a lateral view and a frontal view of the 
computational domain used in the present work and 
the positioning of the entrance, exit, centerline, 
far field and periodic boundary conditions. A discussion
of all boundary conditions is performed in the following 
subsections.}

\subsection{{ Far Field Boundary}}
 
{  Riemann invariants \cite{Long91} are used to implement far field boundary conditions.
They are derived from the characteristic relations for the Euler equations.
At the interface of the outer boundary, the following expressions apply}
\begin{eqnarray}
	{  \mathbf{R}^{-} = {\mathbf{R}}_{\infty}^{-}} &{   = }& {  q_{n_\infty}-\frac{2}{\gamma-1}a_\infty\, \mbox{,}}  \\
	{  \mathbf{R}^{+} = {\mathbf{R}}_{e}^{+} }&{   = }&{  q_{n_e}-\frac{2}{\gamma-1}a_e \, \mbox{,}}
 	\label{eq:R-farfield}
\end{eqnarray}
{  where $\infty$ and $e$ indexes stand for the property in the freestream and in the 
internal region, respectively. $q_n$ is the velocity component normal to the outer surface,
defined as}
\begin{equation}
	{  q_n={\bf u} \cdot \vec{n} \, \mbox{,}}
 	\label{eq:qn-farfield}
\end{equation}
{  and $\vec{n}$ is the unit outward normal vector}
\begin{equation}
	{  \vec{n} \, = \, [n_x \,\, n_y \,\, n_z ]^T \, = \,
	\frac{1}{\sqrt{\eta_{x}^2+\eta_{y}^2+\eta_{z}^2}}
	[\eta_x \ \eta_y \ \eta_z ]^T \, \mbox{.}}
 	\label{eq:norm-vec}
\end{equation}
{  Equation \eqref{eq:norm-vec} assumes that the $\eta$ direction is pointing from the jet to the 
external boundary. Solving for $q_n$ and $a$, one can obtain}
\begin{eqnarray}
	{  q_{n f} = \frac{\mathbf{R}^+ + \mathbf{R}^-}{2} \, \mbox{,}} & \ & 
	{  a_f = \frac{\gamma-1}{4}(\mathbf{R}^+ - \mathbf{R}^-) \, \mbox{.}}
 	\label{eq: qn2-farfield}
\end{eqnarray}
{  The $f$ index is linked to the property at the boundary surface and it is used to update 
the solution at this boundary. For a subsonic exit boundary, $0<q_{n_e}/a_e<1$, the 
velocity components are derived from internal properties as}
  \begin{eqnarray}
	  {  u_f}&{  =}&{  u_e+(q_{n f}-q_{n_e}) \, n_x \, \mbox{,}} \nonumber \\ 
	  {  v_f}&{  =}&{  v_e+(q_{n f}-q_{n_e}) \, n_y \, \mbox{,}} \\ 
	  {  w_f}&{  =}&{  w_e+(q_{n f}-q_{n_e}) \, n_z \, \mbox{.}} \nonumber
 	 \label{eq:vel-farfield}
  \end{eqnarray}
{  Density and pressure are obtained by extrapolating the entropy from 
the adjacent grid node,}
\begin{eqnarray}
	{  \rho_f = 
 	\left(\frac{\rho_{e}^{\gamma}a_{f}^2}{\gamma p_e} \right)^{\frac{1}{\gamma-1}}
	\, \mbox{,} } & \ & { 
	p_{f} = \frac{\rho_{f} a_{f}^2}{\gamma} \, \mbox{.}} \nonumber
 	 \label{eq:rhop-farfield}
\end{eqnarray}
{  For a subsonic entrance, $-1<q_{n_e}/a_e<0$, properties are obtained similarly 
from the freestream variables as}
 \begin{eqnarray}
	 {  u_f}&{  =}&{  u_\infty+(q_{n f}-q_{n_\infty}) \, n_x \, \mbox{,}} \nonumber \\
	 {  v_f}&{  =}&{  v_\infty+(q_{n f}-q_{n_\infty}) \, n_y \, \mbox{,}}\\
	 {  w_f}&{  =}&{  w_\infty+(q_{n f}-q_{n_\infty}) \, n_z \, \mbox{,}}\nonumber
 	\label{eq:vel2-farfield}
 \end{eqnarray}
 \begin{equation}
	 {  \rho_f = 
 	\left(\frac{\rho_{\infty}^{\gamma}a_{f}^2}{\gamma p_\infty} \right)^{\frac{1}{\gamma-1}}
	 \, \mbox{.}}
 	\label{eq:rhop2-farfield}
\end{equation}
{  For a supersonic exit boundary, $q_{n_e}/a_e>1$, the properties are extrapolated 
from the interior of the domain as}
\begin{eqnarray}
	{  \rho_f}&{  =}&{  \rho_e \, \mbox{,}} \nonumber\\
	{  u_f}&{  =}&{  u_e \, \mbox{,} }\nonumber\\
	{  v_f}&{  =}&{  v_e \, \mbox{,} }\\
	{  w_f}&{  =}&{  w_e \, \mbox{,} }\nonumber\\
	{  e_f}&{  =}&{  e_e \, \mbox{,} }\nonumber   
 	\label{eq:supso-farfield}
\end{eqnarray}
{  and for a supersonic entrance, $q_{n_e}/a_e<-1$, the properties are extrapolated 
from the freestream variables as}
 \begin{eqnarray}
	 {  \rho_f}&{ =}&{ \rho_\infty \, \mbox{,}}  \nonumber\\
	 {  u_f}&{ =}&{ u_\infty \, \mbox{,}}  \nonumber\\
	 {  v_f}&{ =}&{ v_\infty \, \mbox{,}} \\
	 {  w_f}&{ =}&{ w_\infty \, \mbox{,}} \nonumber\\
	 {  e_f}&{ =}&{ e_\infty \, \mbox{.}} \nonumber
 	\label{eq:supso2-farfield}
\end{eqnarray}
 
\subsection{{  Entrance Boundary}}
 
{  For a jet-like configuration, the entrance boundary is divided in two areas: the
jet and the area above it. The jet entrance boundary condition is implemented through 
the use of the 1-D characteristic relations for the 3-D Euler equations for a flat
velocity profile. The set of properties, then, determined is computed from within and 
from outside the computational domain. For the subsonic entrance, the $v$ and $w$ components
of the velocity are extrapolated by a zero-order extrapolation from inside the 
computational domain and the angle of flow entrance is assumed fixed. The remaining properties 
are obtained as a function of the jet Mach number, which is a known variable.} 
\begin{eqnarray}
	{  \left( u \right)_{1,j,k}} &{  =} &{   u_{j} \, \mbox{,}} \nonumber \\
	{  \left( v \right)_{1,j,k}}&{  =} &{   \left( v \right)_{2,j,k} \,\mbox{,}} \\
	{  \left( w \right)_{1,j,k}}&{  =} &{   \left( w \right)_{2,j,k} \, \mbox{.}} \nonumber
 	\label{eq:vel-entry}
\end{eqnarray}
{  The dimensionless total temperature and total pressure are defined with the isentropic relations:}
\begin{eqnarray}
	{  T_t = 1+\frac{1}{2}(\gamma-1)M_{j}^{2} \,} &{  \mbox{and}} & 
	{  P_t = \frac{1}{\gamma}(T_t)^{\frac{\gamma}{\gamma-1}} \, \mbox{.}}
 	\label{eq:Tot-entry}
\end{eqnarray}
{  The dimensionless static temperature and pressure are deduced from Eq.\ \eqref{eq:Tot-entry},
resulting in}
\begin{eqnarray}
	{  \left( T \right)_{1,j,k}=\frac{T_t}{1+\frac{1}{2}(\gamma-1)(u^2+v^2+w^2)_{1,j,k}} \,} 
	& {  \mbox{and}} & 
	{  \left( p \right)_{1,j,k}=\frac{1}{\gamma}(T)_{1,j,k}^{\frac{\gamma}{\gamma-1}} \, \mbox{.}}
 	\label{eq:Stat-entry}
\end{eqnarray}
{  For the supersonic case, all conserved variables receive the jet property values.
Such entrance boundary conditions do not include any disturbance to the 
inlet velocity. The present approach simply generates a ``top hat'' velocity profile 
at the jet entrance. The calculations performed in the context of this work have indicated, at 
least for the cases addressed here, that numerical disturbances already present in the solution 
process are sufficient to destabilize the flow and induce turbulence transition in the jet.
 
The far field boundary conditions are implemented outside of the jet area in order to correctly
propagate information coming from the inner domain of the flow to the outer regions of 
the simulation. However, in the present case, $\xi$, instead of $\eta$, as presented in 
the previous subsection, is the normal direction used to define the Riemann invariants.}
 
\newpage
\FloatBarrier

\subsection{{  Exit Boundary Conditions}}
 
{  At the exit plane, the same reasoning of the jet entrance boundary is applied. In this case, 
for a subsonic exit, the pressure is obtained from the outside, {\em i.e.}, it is assumed given, 
and all other variables are extrapolated from the interior of the computational domain by a 
zero-order extrapolation. The conserved variables are obtained as}
\begin{eqnarray}
	{  (\rho)_{I_{MAX},j,k}} &{  =}& 
	{  \frac{(p)_{I_{MAX},j,k}}{(\gamma-1)(e_i)_{I_{MAX}-1,j,k}}
	\mbox{,}} \\
	{  (\vec{u})_{I_{MAX},j,k}} &{  =}& 
	{  (\vec{u})_{I_{MAX}-1,j,k}\mbox{,} }\\
	{  (e)_{I_{MAX},j,k} }&{  =}& 
	{  (\rho)_{I_{MAX},j,k}\left[ (e_i)_{I_{MAX}-1,j,k}+
	\frac{1}{2}(\vec{u})_{I_{MAX},j,k}\cdot(\vec{u})_{I_{MAX},j,k} \right]
	\, \mbox{,}}
 	\label{eq:exit}
\end{eqnarray}
{  in which $I_{MAX}$ stands for the last point of the mesh in the axial direction. For 
the supersonic exit, all properties are extrapolated from the interior domain.}
 
\subsection{{  Centerline Boundary Conditions}}
 
\begin{figure}[ht]
        \begin{center}
        {\includegraphics[width=0.5\textwidth]
		{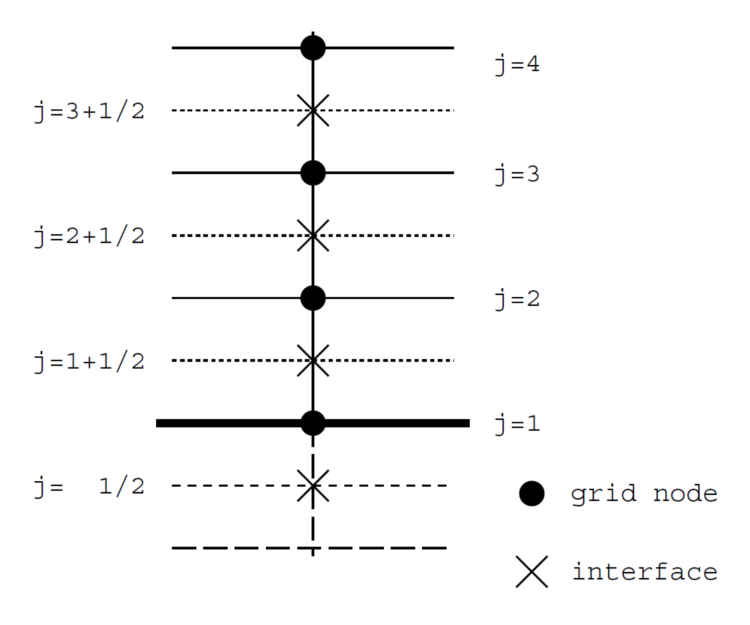}}\\
 	    \caption{Boundary point dissipation operator at centerline \cite{BIGA02}.}
 	    \label{fig:centerline}
        \end{center}
\end{figure}

{  The centerline boundary is a singularity of the coordinate transformation and, hence, 
an adequate treatment of this boundary must be provided. The conserved properties 
are extrapolated from the adjacent longitudinal plane and they are averaged in the azimuthal 
direction in order to define the updated properties at the centerline of the jet.
 
The fourth-difference terms of the artificial dissipation scheme, used in the present 
work, are carefully treated in order to avoid the five-point difference stencils at 
the centerline singularity. 
If one considers the flux balance at one grid point near the centerline boundary in 
a certain coordinate direction, let $w_{j}$ denote a component of the $\mathcal{W}$ 
vector from Eq.\ \eqref{eq:W_dissip} and $\mathbf{d}_{j}$ denote the corresponding artificial
dissipation term at the $j$-th mesh point. In the present example, 
$\left(\Delta w\right)_{j+\frac{1}{2}}$ stands for the difference between the solution
at the interface for the points $j+1$ and $j$. The fourth-difference of the dissipative
fluxes from Eq.\ \eqref{eq:dissip_term} can be written as}
\begin{equation}
	{  \mathbf{d}_{j+\frac{1}{2}} = \left( \Delta w \right)_{j+\frac{3}{2}} 
 	- 2 \left( \Delta w \right)_{j+\frac{1}{2}}
	+ \left( \Delta w \right)_{j-\frac{1}{2}} \, \mbox{.}}
\end{equation}
{  Considering the centerline and the point $j=1$, as presented in 
Fig.\ \ref{fig:centerline}, the calculation of $\mathbf{d}_{1+\frac{1}{2}}$ demands the 
$\left( \Delta w \right)_{\frac{1}{2}}$ term, which is unknown since it is outside the
computation domain. In the present work a extrapolation is performed and given by}
\begin{equation}
	{  \left( \Delta w \right)_{\frac{1}{2}} =
	- \left( \Delta w \right)_{1+\frac{1}{2}} \, \mbox{.}}
\end{equation}
{  This extrapolation modifies the calculation of $\mathbf{d}_{1+\frac{1}{2}}$ 
that can be written as}
\begin{equation}
	{  \mathbf{d}_{j+\frac{1}{2}} = \left( \Delta w \right)_{j+\frac{3}{2}} 
	- 3 \left( \Delta w \right)_{j+\frac{1}{2}} \, \mbox{.}}
\end{equation}
{  The approach is plausible since the centerline region is smooth and does 
not have high gradient of properties.}
 
\subsection{{ Periodic Boundary Conditions}}
 
{  A periodic condition is implemented between the first ($K=1$) and the last point in the 
azimuthal direction ($K=K_{MAX}$) in order to close the 3-D computational domain. There 
are no boundaries in this direction, since all the points are inside the domain. The first
and the last points, in the azimuthal direction, are superposed in order to facilitate
the boundary condition implementation which is given by}
\begin{eqnarray}
	{
	  (\rho)_{i,j,K_{MAX}}
	} &{  =}&{   (\rho)_{i,j,1} \, \mbox{,}} \nonumber\\
	{
	  (u)_{i,j,K_{MAX}}
	} &{  =}&{   (u)_{i,j,1} \, \mbox{,}} \nonumber\\
	{
	  (v)_{i,j,K_{MAX}}
	} &{  =}&{   (v)_{i,j,1} \, \mbox{,}} \\
	{
	  (w)_{i,j,K_{MAX}}
	} &{  =}&{   (w)_{i,j,1} \, \mbox{,}} \nonumber\\
	{
	  (e)_{i,j,K_{MAX}}
	} &{  =}&{   (e)_{i,j,1} \, \mbox{.}} \nonumber
 	\label{eq:periodicity}
\end{eqnarray}
%


\section{Study of Supersonic Jet Flow}

Four {  test cases} are {  addressed} in the present {  work} in 
order to study the use of 2nd-order spatial discretization on large 
eddy simulations of a perfectly expanded jet flow configuration.
{  These test cases compare the effects of mesh refinement and SGS models on the 
results.} Two different meshes are created for the grid refinement study. Results for the three 
SGS models implemented in the code, namely, classic Smagorinsky, dynamic Smagorinsky and 
Vreman models, are compared in the current section. The present results are compared with 
analytical, numerical and experimental data from the 
literature \cite{bridges2008turbulence,Mendez10,Mendez12}. 


\newpage
\FloatBarrier

\subsection{Geometry Characteristics} 

Two different {  computational domain} geometries are created for the {  jet} 
simulations discussed in the current work. One geometry presents a cylindrical shape 
and the other one presents a divergent conical shape. For the sake of simplicity, the 
{  cylindrical} geometry is named geometry A and the other one is named geometry 
B in the present paper.
The computational domains are created in two steps. First, a 2-D 
region is generated. In the sequence, this region is rotated {  about the jet axis} in 
order to generate a fully 3-D geometry. An in-house code is used 
for the generation of the 2-D domain of geometry A. The commercial 
mesh generator ANSYS\textsuperscript{\textregistered} ICEM CFD 
\cite{ICEM} is used for the 2-D domain of geometry B. 

\begin{figure}[htb!]
       \begin{center}
		   \subfigure[3-D view of two XZ {  planes} of geometry A.]{
           \includegraphics[trim= 5mm 5mm 5mm 5mm, clip, width=0.475\textwidth]
		   {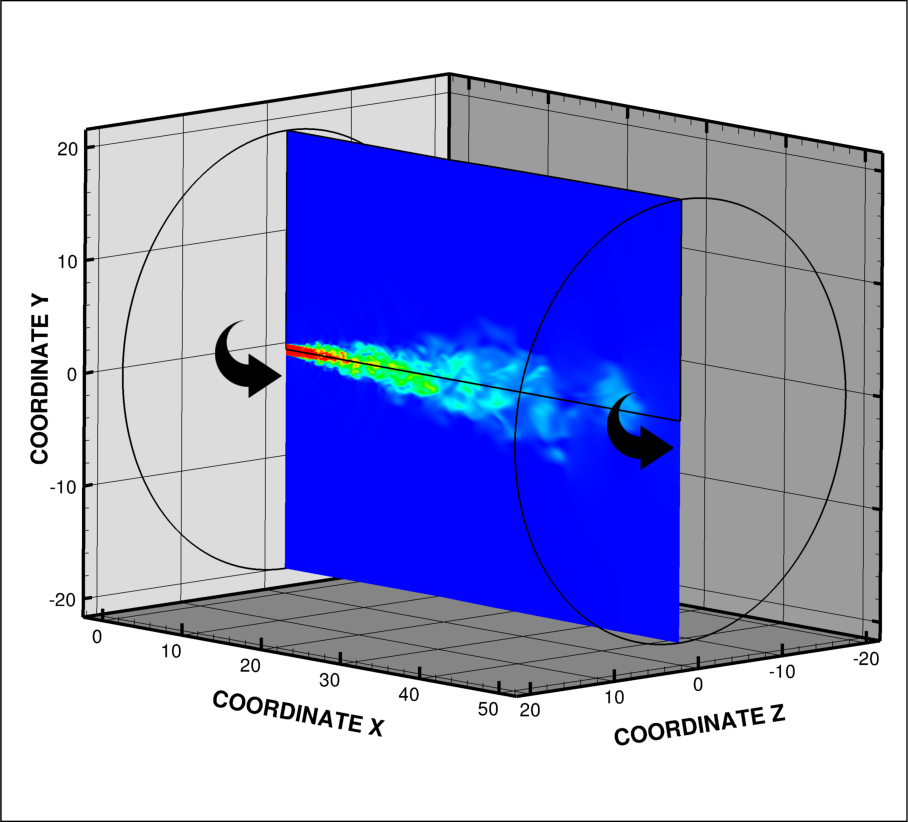} 
		   \label{fig:3d-geom-14m}
		   }
		   \subfigure[3-D view of two XZ {  planes} of geometry B.]{
           \includegraphics[trim= 5mm 5mm 5mm 5mm, clip, width=0.475\textwidth]
		   {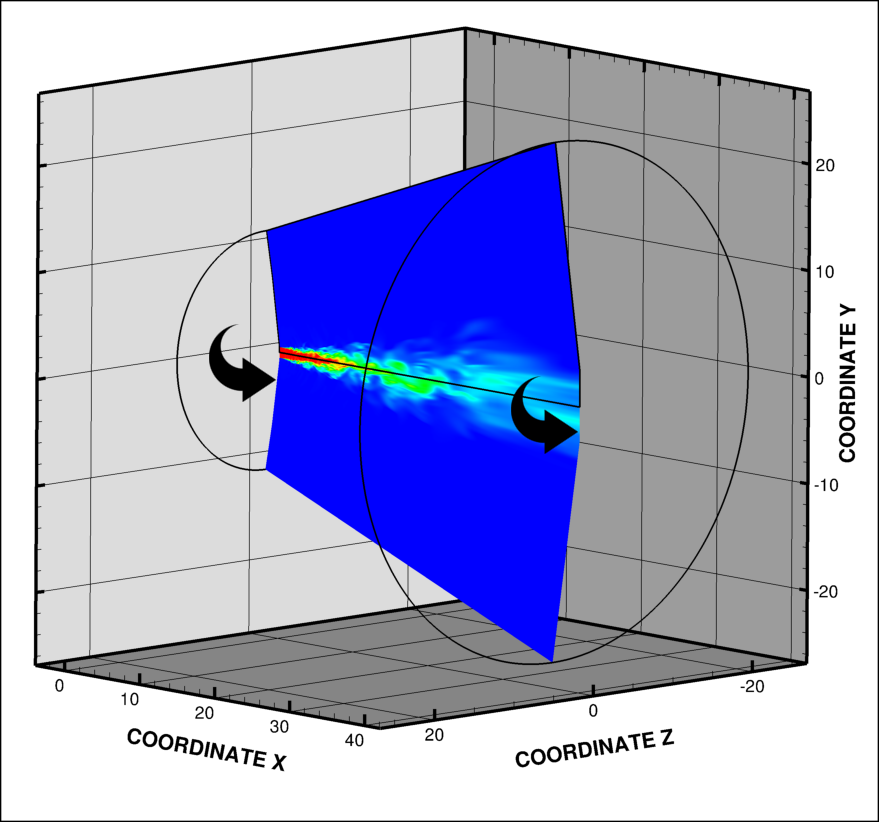} 
		   \label{fig:3d-geom-50m}
		   }
		   \caption{3-D view of {  computational domain} geometries used for the LES 
		   {  calculations}.}
		   \label{fig:geom}
	   \end{center}
\end{figure}

Geometry A is a cylindrical domain with radius of $20D$ and a length of $50D$.
Geometry B presents a divergent form whose axis length is $40D$\@. The minimum 
and maximum heights of geometry B are $\approx 16D$ and $25D$, respectively. 
{  Geometry B is created based on results from simulations using geometry A in 
order to refine the mesh in the shear layer region of the jet flow. The 2-D coordinates 
of the zones created for geometry B and further details of the geometry can be found in 
Ref.\ \cite{Junior16}.}
Geometries A and B are illustrated in Fig.\ \ref{fig:geom} which 
presents a 3-D view of the two computational domains used in the 
current work. The geometries are colored by {  an instantaneous visualization of the} 
solution for the axial component of the flow velocity. 
%


\subsection{Mesh Configurations}

\begin{figure}[htb!]
       \begin{center}
		   \subfigure[2-D view of mesh A in the XZ {  plane}.]{
           \includegraphics[trim= 5mm 5mm 5mm 5mm, clip, width=0.475\textwidth]
		   {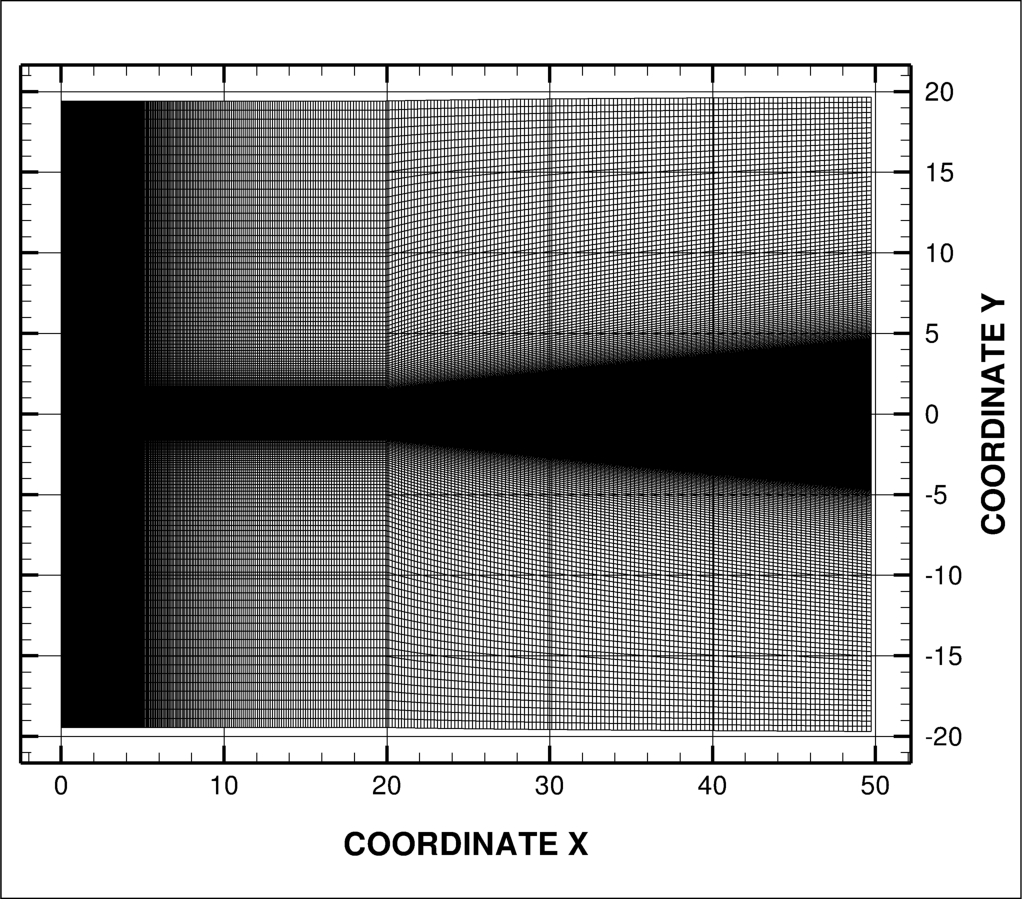} 
		   \label{fig:2d-mesh-14m}
		   }
		   \subfigure[2-D view of mesh B in the XZ {  plane}.]{
           \includegraphics[trim= 5mm 5mm 5mm 5mm, clip, width=0.475\textwidth]
		   {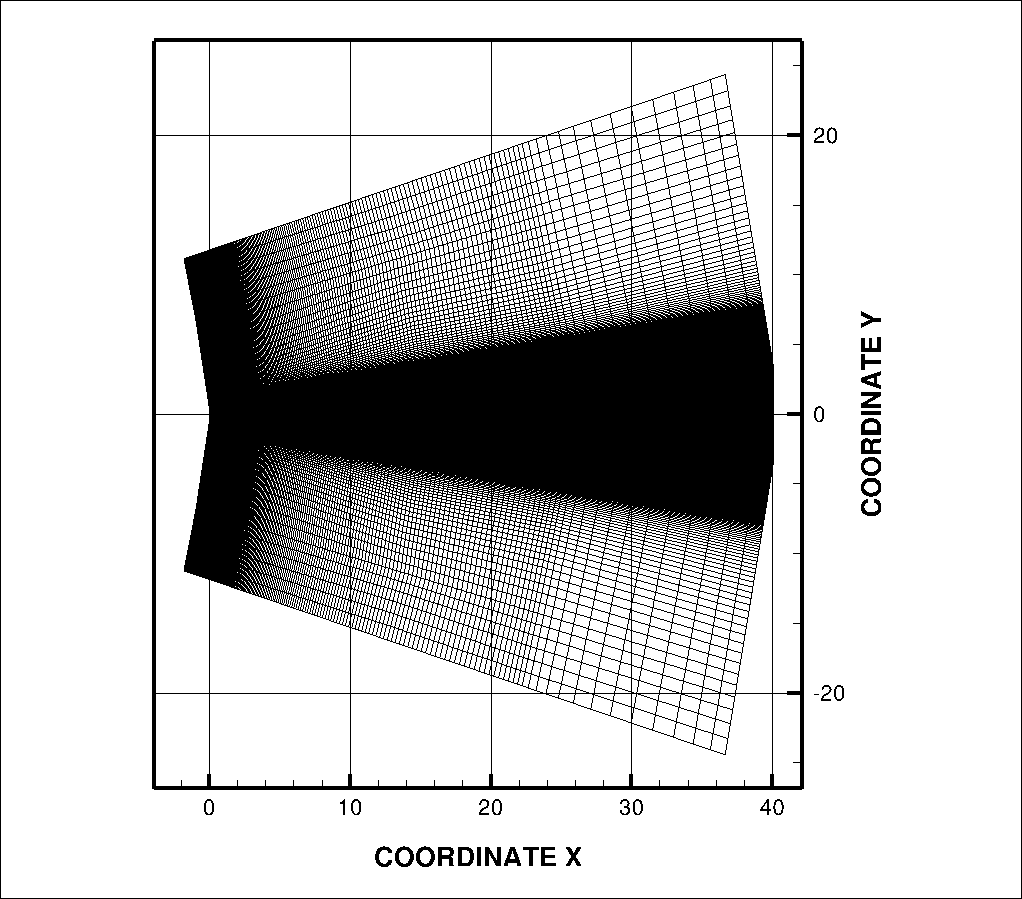} 
		   \label{fig:2d-mesh-50m}
		   }
		   \caption{2-D view of the computational meshes used in the present work.}
		   \label{fig:mesh-ab}
	   \end{center}
\end{figure}

One grid is generated for each geometry used in the present study. These 
computational grids are named mesh A and mesh B. 
{  An illustration of the computational grids is presented in Fig.\ \ref{fig:mesh-ab}.} 
Mesh A is created using a mesh generator developed by the research group for the 
cylindrical shape configuration, {\em i.e.}, geometry A\@. This computational mesh 
is composed of 400 points in the axial direction, 200 points in the radial direction 
and 180 points in the azimuthal direction, which yields a total of 14.4 million grid 
points. Hyperbolic tangent functions are used for the point distributions in the 
radial and axial directions. Grid points are clustered near the shear layer of the jet. 
The mesh is coarsened towards the outer regions of the domain in order to dissipate 
properties of the flow far from the jet. Such mesh refinement approach can avoid 
{  the reflection of waves back} 
into the domain. 

The radial and longitudinal dimensions of the smallest distance 
between mesh points of grid A are given by 
$(\Delta \underline{r})_{min}=0.002D$ and $(\Delta \underline{x})_{min}=0.0126D$, 
respectively. This minimal spacing occurs at the shear layer of the jet and at the 
entrance of the computational domain. {  Mesh A characteristics are chosen 
based on data provided by the work of Mendez {\em et al.} \cite{Mendez10,Mendez12}, 
who have also used a second-order accurate spatial discretization scheme for the 
same jet flow configuration.}
{  Simulations were initially performed using mesh A\@. This particular 
calculation has used the static Smagorinsky SGS model, and details of the calculation 
are discussed in the forthcoming sections. The results indicated that further grid refinement 
was necessary and they also helped in the decision of which regions of the domain should be 
refined. Therefore, the authors created a second computational grid, mesh B, which also adopted 
a somewhat different topology for the computational domain, geometry B, as previously discussed.} 

\begin{figure}[htb!]
       \begin{center}
           \includegraphics[width=0.5\textwidth]
		   {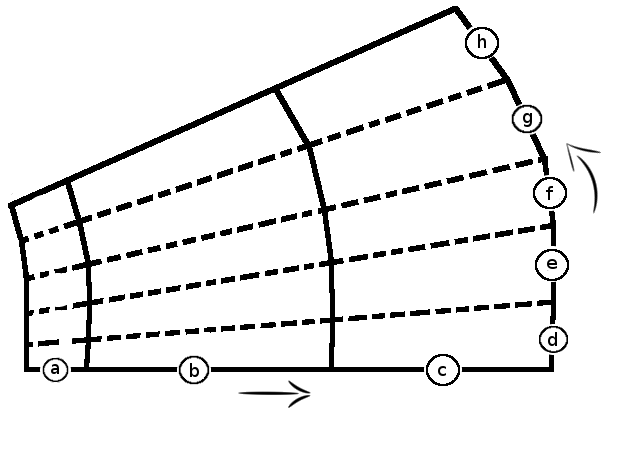} 
		   \caption{Auxiliary edges used for the generation of mesh B.}
		   \label{fig:mesh-alloc}
	   \end{center}
\end{figure}

{  The more refined computational grid, mesh B,} is composed of 343 
points in the axial direction, 398 points in the radial direction and 360 
points in the azimuthal direction. This yields a mesh with approximately 50 million grid 
points. The 2-D mesh is generated with ANSYS\textsuperscript{\textregistered} 
ICEM CFD \cite{ICEM}. The points are allocated using different 
distributions in eight edges of the 2-D domain. The same coarsening 
approach used for mesh A is also applied for mesh B\@. The distance 
between mesh points increases towards the outer region of the domain. 
This procedure forces the dissipation of properties far from the jet 
in order to avoid {  reflection of waves} 
into the domain. {  The grid coarsening can be understood as an 
implicit damping which can smooth out properties far from the jet, in the region 
where the mesh is no longer refined. Figure \ref{fig:mesh-alloc} illustrates 
the edges used to generate the point distribution and the direction of the 
mesh coarsening. Table \ref{tab:mesh-b} presents the number of points, the 
growth factor and the smallest grid spacing for all auxiliary edges 
used to generate the mesh.}
\begin{table}[htbp]
\begin{center}
	\caption{Characteristics of mesh B.}
\label{tab:mesh-b}
\begin{tabular}{|c|c|c|c|}
\hline
Edge & Nb. of points & Growth Factor & Smallest Spacing \\
\hline
a & 300 & 1.01912 & $1.000 \times 10^{-3}$ D    \\
b & 30  & 1.02739 & $2.8231 \times 10^{-1}$ D   \\
c & 15  & 1.30370 & $3.4482 \times 10^{-1}$ D   \\
d & 45  & 1.00    & $3.40909 \times 10^{-3}$ D  \\ 
e & 155 & 1.03724 & $3.40909 \times 10^{-3}$ D  \\
f & 155 & 1.01034 & $1.18655 \times 10^{-2}$ D  \\
g & 40  & 1.07161 & $5.73327 \times 10^{-2}$ D  \\
h & 7   & 1.08206 & $7.93887 \times 10^{-1}$ D  \\
\hline
\end{tabular}
\end{center}
\end{table}
%

\subsection{Flow Configuration and Boundary Conditions} 

An unheated perfectly expanded jet flow is chosen to {  perform the present studies 
with} the LES tool. The jet entrance Mach number is $1.4$\@. The pressure ratio, $PR=P_{j}/P_\infty$, 
and the temperature ratio, $TR=T_{j}/T_\infty$, between the jet entrance and the ambient freestream 
conditions, are equal to one, {\em i.e.}, $PR = 1$ and $TR=1$. The Reynolds number of the jet is 
$Re = 1.57 \times 10^{6}$, based on the jet entrance diameter, D\@. This flow configuration is 
chosen due to the absence of strong shocks waves. Strong discontinuities {  clearly create 
additional numerical difficulties, that the authors did not want to add to the present study.} 
Moreover, numerical and experimental data are available in the literature for this flow configuration, 
such as the work of Mendez {\it et al.} \cite{Mendez10,Mendez12} and the work of 
Bridges and Wernet \cite{bridges2008turbulence}. 

The boundary conditions discussed in Sect.\ \ref{sec:BC} are used 
in the simulations performed in the {  present paper}. 
Figure \ref{fig:bc} presented a lateral view and a frontal view of the 
computational domain used for the simulations, indicating the positioning 
of each boundary condition. A {  top hat} velocity profile, 
with $M=1.4$, is used at the entrance boundary. Riemann invariants are 
used at the {  stagnated ambient regions}. A special singularity 
treatment is performed at the centerline. Periodicity is imposed in the 
azimuthal direction. 
Properties of flow at the inlet and at the farfield regions have to be 
provided to the code in order to impose the boundary conditions. Density, 
$\rho$, temperature, $T$, velocity, $U$, Reynolds number, $Re$, and specific 
heat at constant volume, $C_{v}$, are provided in the dimensionless form 
to the simulation. These properties are given by
\begin{eqnarray}
	\rho_{j} = 1.00 \, \mbox{,} & \hspace*{1.0 cm} & 
	\rho_{\infty} = 1.00 \, \mbox{,} \nonumber \\
	T_{j} = 1.00 \, \mbox{,} & \hspace*{1.0 cm} & 
	T_{\infty} = 1.00 \, \mbox{,} \\
	U_{j} = 1.4 \, \mbox{,} & \hspace*{1.0 cm} & 
	U_{\infty} = 0.00 \nonumber \, \mbox{,} \\
	Re_{j}=1.57 \times 10^{6} \, \mbox{,}& \hspace*{1.0 cm} & 
	C_{v} = 1.786 \, \mbox{,} \nonumber
\end{eqnarray}
where the $j$ subscript {  identifies the properties at} the jet entrance and the 
$\infty$ subscript stands for properties at the farfield region. 
%

\subsection{Large Eddy Simulations} 

Four simulations are performed in the present {  work}. The objective is to 
study the effects of mesh refinement and to evaluate the three 
different SGS models included into the code. The calculations are 
performed in two steps. First, a preliminary simulation is performed 
in order to achieve a statistically steady state condition. In the 
sequence, the simulations are run for another period in order to 
collect enough data for the calculation of time averaged properties 
of the flow and their fluctuations. 

\subsubsection{{ Statistically Steady Flow Condition}} \label{sec:stat-steady-flow}

The configurations of all simulations are discussed in the current subsection
towards the description of the preliminary calculations which are performed
in order to drive the flow to a statistically steady flow condition.
{  In this preliminary stage, the goal is to achieve a starting point 
to begin the calculation of statistical data. Table \ref{tab:simu} presents 
the operating conditions of all four numerical studies performed in the current 
research. S1, S2 and S3 simulations were the test cases that admitted stagnated flow 
conditions as the initial conditions for the preliminary simulations. Differently 
from the former studies, the S4 simulation used the statistically steady state 
condition of the S2 computational study as initial condition for this preliminary stage. 
Mesh A is only used in the S1 simulation. The other calculations are performed using 
the more refined grid, Mesh B\@. S1, S2, S3 and S4 studies apply, respectively, time 
increments of $2.5 \times 10^{-5}$, $1.0 \times 10^{-4}$, $5.0 \times 10^{-5}$ and 
$1.0 \times 10^{-4}$, in dimensionless time units, as indicated in the 
4th column of Tab.\ \ref{tab:simu}}. The dimensionless time increment used 
for all configurations is the largest one which the solver can handle 
without diverging the solution. The static Smagorinsky model 
\cite{Lilly65,Lilly67,Smagorinsky63} is used in the S1 and S2 simulations. 
The dynamic Smagorinsky model \cite{Germano91,moin91} and the Vreman model 
\cite{vreman2004} are used in simulations S3 and S4, respectively.

{  The total (physical) time simulated by all numerical studies in order to achieve the 
statistically steady state condition is indicated in the 6th column of 
Tab.\ \ref{tab:simu} in flow through time (FTT) units. One flow through time 
is the necessary amount of time for a particle to cross all the domain 
considering the inlet velocity of the jet. The S1 simulation is the 
least expensive test case studied and, therefore, one could afford to let it run for a longer 
period of time.} It uses a 14 million point mesh while 
the other simulations use the 50 million point grid. 
On the other hand, simulation S3 is the most expensive 
numerical test case, because the dynamic Smagorinsky SGS model requires more 
computational time per time step when compared with the other SGS models 
implemented in the code. Moreover, the time step restrictions for the dynamic 
Smagorinsky model are much more stringent, as one can see in Tab.\ \ref{tab:simu}. 
Hence, the S3 simulation has only been run for 8.20 FTT for this study. 
\begin{table}[htbp]
\begin{center}
	\caption{Overall characteristics of the large eddy simulations performed 
	in the present work.}
\label{tab:simu}
\begin{tabular}{|c|c|c|c|c|c|}
\hline
Simulation & Mesh & SGS & $\Delta t$ & Initial Condition & FTT \\
\hline
S1 & A & Static Smagorinsky  & $2.5 \times 10^{-5}$ & \small Stagnated flow & {  52.9} \\
S2 & B & Static Smagorinsky  & $1 \times 10^{-4}$ & \small Stagnated flow   & {  14.2} \\
S3 & B & Dynamic Smagorinsky & $5 \times 10^{-5}$ & \small Stagnated flow   & {  8.20}  \\
S4 & B & Vreman              & $1 \times 10^{-4}$ & \small S2               & {  19.1} \\
\hline
\end{tabular}
\end{center}
\end{table}

\subsubsection{{  Mean Flow Property Calculations}}

{  After the statistically stationary state is achieved, the 
simulations are restarted and run for another period of time in which data of the 
flow are extracted and recorded in a fixed frequency.} The collected data are 
time averaged in order to calculate mean properties of the flow and compare them 
with the results of the numerical and experimental references.
In the present {  section}, time-averaged properties are denoted as 
$\langle \cdot \rangle$. Table \ref{tab:mean-simu} summarizes the information 
on data collecting and time averaging for the four test cases addressed here. 
The 2nd column presents the number of extractions performed during the simulations. 
Data are extracted each 0.02 dimensionless time units in the present work, which 
is equivalent to a dimensionless frequency of 50\@. The choice of this frequency 
is based on the numerical work reported in Refs.\ \cite{Mendez10,Mendez12}. 
The last column of Tab.\ \ref{tab:mean-simu} presents the total additional dimensionless 
time simulated in order to extract the data to calculate the mean properties.
\begin{table}[htbp]
\begin{center}
	\caption{Details of the data extraction for mean flow calculations for each test case.}
\label{tab:mean-simu}
\begin{tabular}{|c|c|c|c|}
\hline
Simulation & Number of Extractions & Frequency & Data Extraction Time \\
\hline
S1 & 2048 & 50 & 1.60 FTT \\
S2 & 3365 & 50 & 3.30 FTT \\
S3 & 2841 & 50 & 2.77 FTT \\
S4 & 1543 & 50 & 1.51 FTT \\
\hline
\end{tabular}
\end{center}
\end{table}
%

\subsection{Study of Mesh Refinement Effects} 
\label{sec:mesh-study}

\begin{figure}[htb!]
  \centering
    {\includegraphics[trim= 5mm 5mm 5mm 5mm, clip, width=0.7\textwidth]
	{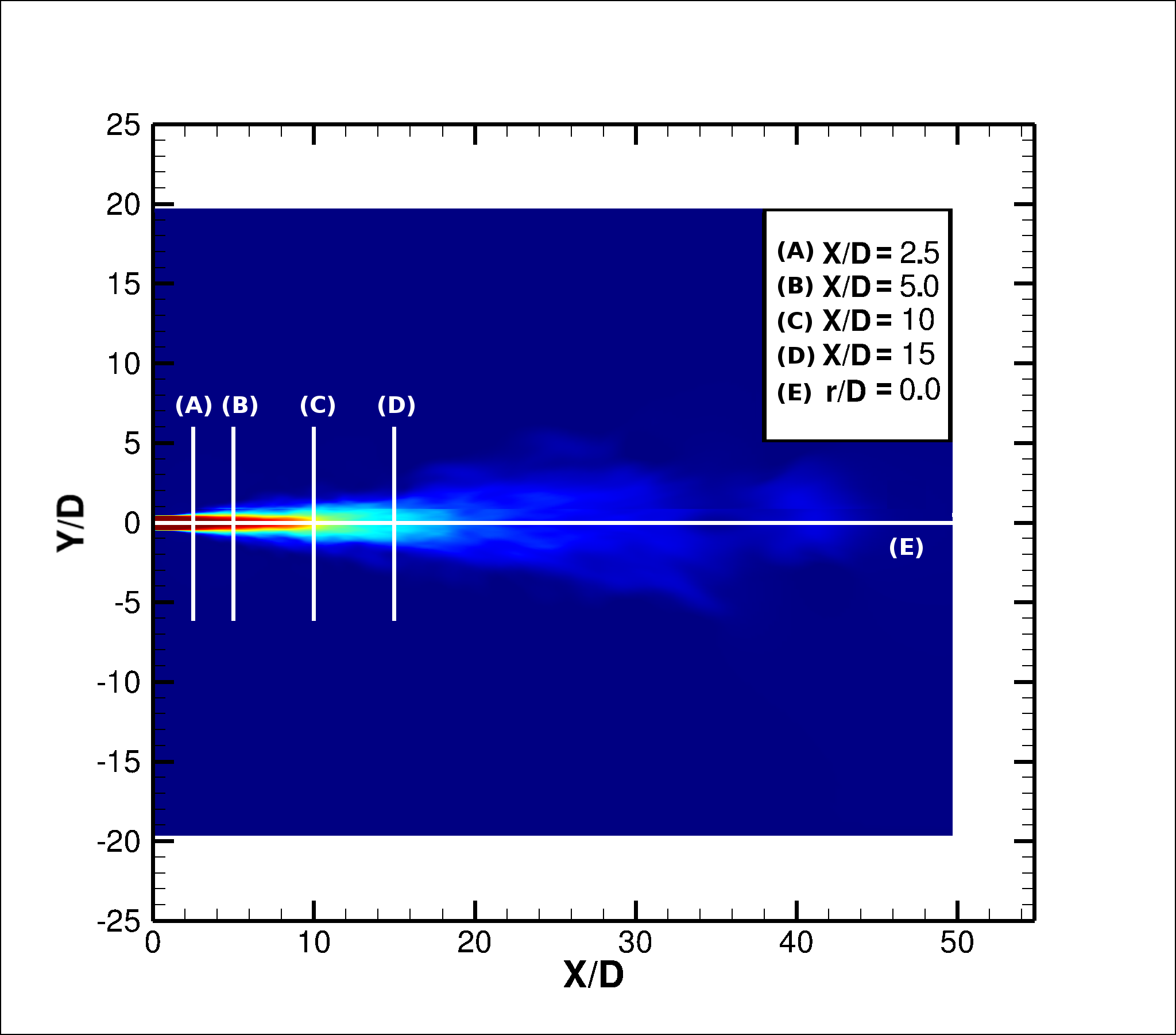}}
	\caption{ Positioning in the computational domain of surfaces 
	studied in the present work.}
	\label{fig:extract-new}
\end{figure}

Effects of mesh refinement on compressible LES using the JAZzY 
solver are discussed in the present section. 
Time-averaged 2-D distributions and profiles of the axial component of velocity
are collected from simulations S1 and S2, and they 
compared with numerical and experimental results from the 
literature \cite{bridges2008turbulence,Mendez10,Mendez12}. 
Both simulations use the same SGS model, the static Smagorinsky
model \cite{Lilly65,Lilly67,Smagorinsky63}. Mesh A is used on the S1 simulation 
while Mesh B is used on the S2 simulation. 
Figure 
\ref{fig:extract-new} illustrates the positioning 
of surfaces and profiles extracted for all simulations performed in the present 
work.
%
%
{  Cuts (A) through (D) are radial profiles at different positions 
downstream of the jet entrance. }{{  An average in the azimuthal 
direction is performed when the radial profiles are calculated. }
{  The last region indicated in Fig.\ \ref{fig:extract-new}, Cut (E), 
represents the jet axis.}

\subsubsection{{ Time Averaged Axial Component of Velocity}}

One important characteristic of round jet flow configurations is 
the potential core length, $\delta_{j}^{95\%}$. The potential 
core is defined as the region in which the axial velocity component, $U_{j}^{95\%}$, 
{  is at least} $95\%$ of the velocity of the jet at the inlet, {\em i.e.},
\begin{equation}
	U_{j}^{95\%} = 0.95 \,\, U_{j}
	\, \mbox{.}
	\label{eq:pot-core}
\end{equation}
Therefore, the {  potential core length} can be defined by the point in which 
the mean axial velocity component reaches $U_{j}^{95\%}$ along the centerline.

Lateral views of $\langle U \rangle$ for the S1 and S2 simulations are presented 
in Fig.\ \ref{fig:lat-u-av-mesh}. In these contour plots, $U_{j}^{95\%}$ is indicated by the 
black solid line. 
%
%
\begin{figure}[htb!]
  \centering
  \subfigure[Lateral view of $\langle U \rangle$ for S1 simulation.]
    {\includegraphics[trim= 5mm 5mm 5mm 5mm, clip, width=0.495\textwidth]
	{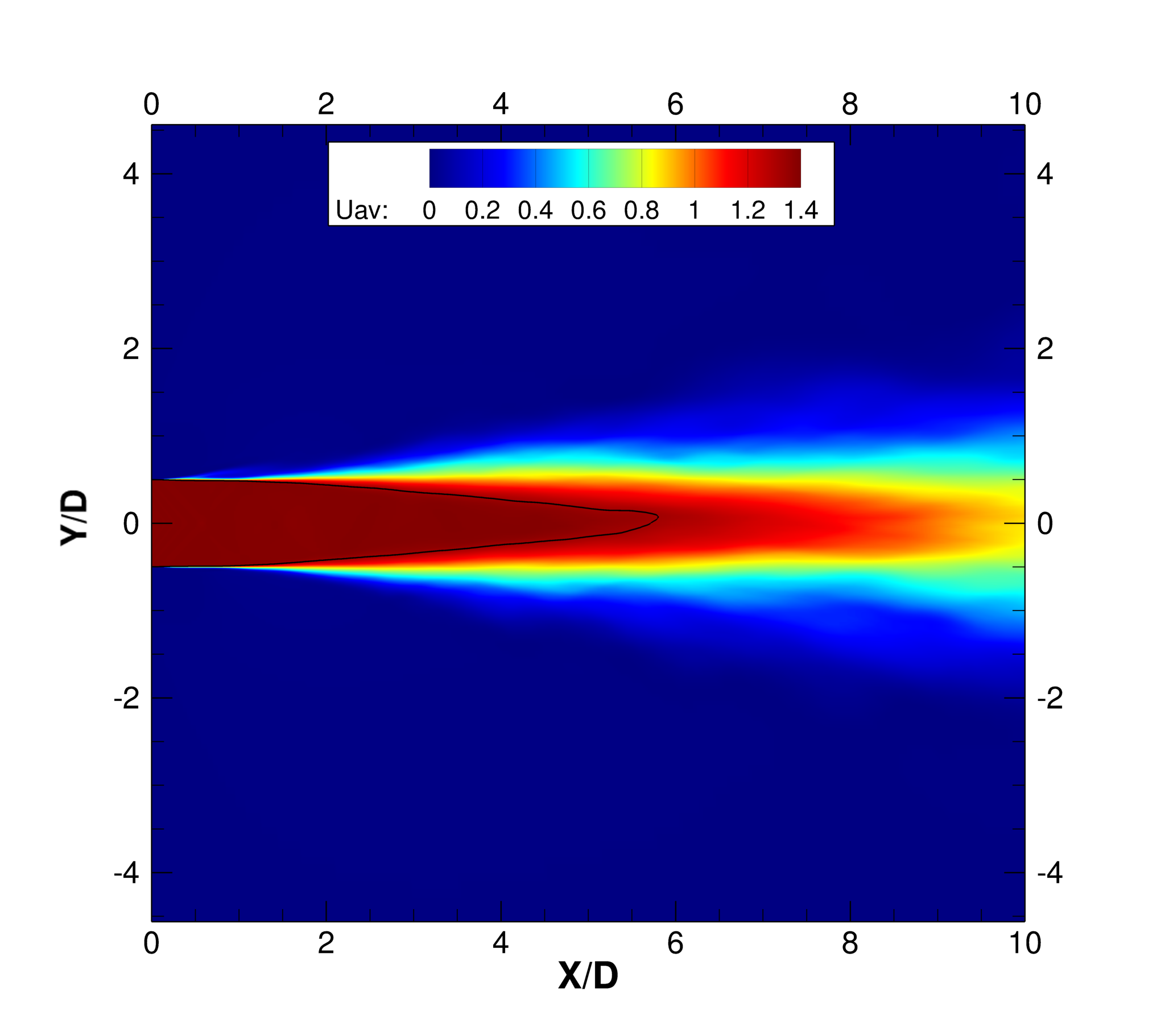}}
  \subfigure[Lateral view of $\langle U \rangle$ for S2 simulation.]
    {\includegraphics[trim= 5mm 5mm 5mm 5mm, clip, width=0.495\textwidth]
	{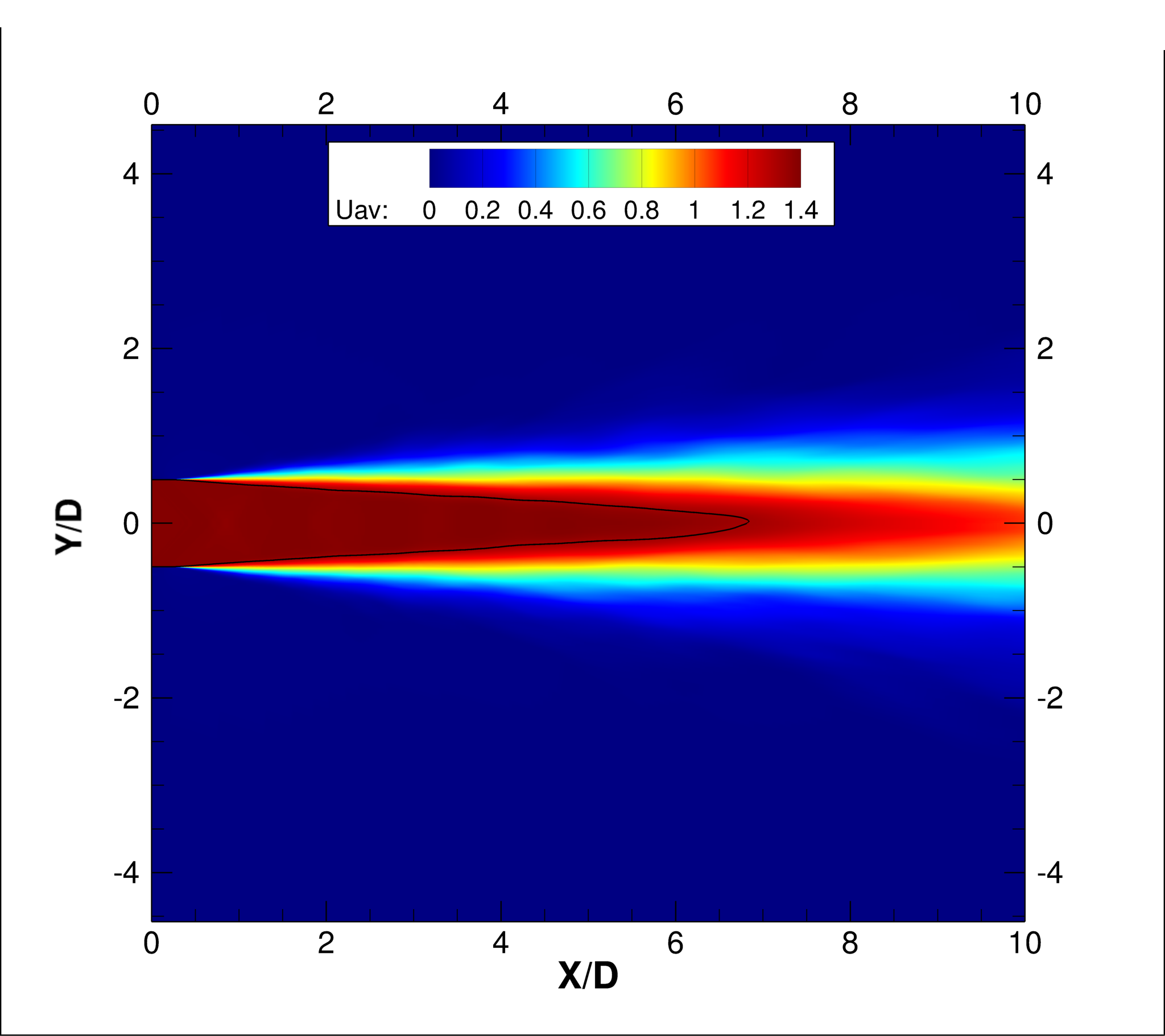}}
	\caption{Lateral view of the averaged axial component of 
	velocity, $\langle U \rangle$, for the S1 and S2 simulations. 
	The solid black line (\textbf{--}) indicates the 
	potential core of the jet, $U_{j}^{95\%}$.} 
	\label{fig:lat-u-av-mesh}
\end{figure}
Table \ref{tab:core-mesh} presents the potential core length for the S1 and S2 simulations. 
The present calculations are compared to the numerical results from Refs.\ \cite{Mendez10,Mendez12}, 
both in terms of the actual values of $\delta_{j}^{95\%}$ as well as in terms of the relative error 
with respect to the experimental data \cite{bridges2008turbulence}.
\begin{table}[htbp]
\begin{center}
  \caption{Potential core length and relative error with respect to the experimental data \cite{bridges2008turbulence}.}
  \label{tab:core-mesh}
  \begin{tabular}{|c|c|c|}
  \hline
  Simulation & $\delta_{j}^{95\%}$ & Relative error\\
  \hline
  S1 & 5.57 & 40\%\\
  S2 & 6.84 & 26\%\\
  Mendez {\it et al.} & 8.35 & 8\%\\
  \hline
  \end{tabular}
\end{center}
\end{table}
There are significant differences between the results for the S1 and S2 test cases, as well as 
between those and the computational results from Ref.\ \cite{Mendez12}. The results for the S1 
simulation present a smaller $\delta_{j}^{95\%}$, when compared to the results for the S2 
simulation, {\em i.e.}, 5,57 and 6.84, respectively. The S1 solution is over dissipative when 
compared to the S2 results and, hence, the jet {  decays} earlier in the S1 simulation. 
As previously discussed, the mesh used in the S1 test case is very coarse when compared to the 
grid used for the S2 simulation. This lack of spatial resolution can generate very dissipative 
solutions which yield the under prediction of the potential core length. The mesh refinement 
reduced in 14\% the relative error of the S2 simulation when compared to the experimental data. 

Profiles of $\langle U \rangle$ along the mainstream direction, and the evolution of 
$\langle U \rangle$ along the centerline 
are compared with numerical and experimental results in Figs.\ 
\ref{fig:prof-u-av-mesh-new} and \ref{fig:prof-u-av-mesh-centerline}, respectively. 
The centerline 
{  is} indicated as the (E) 
{  cut} in Fig.\ \ref{fig:extract-new}. The dash-point line and the solid line stand for 
the results of the S1 and S2 test cases, respectively, in 
{  Figs.\ \ref{fig:prof-u-av-mesh-new} and \ref{fig:prof-u-av-mesh-centerline}}. The square 
symbols are the LES results of Mendez {\it et al.} \cite{Mendez10,Mendez12}, while the triangular 
symbols indicate the experimental data of Bridges and Wernet \cite{bridges2008turbulence}.
%
%
\begin{figure}[htb!]
  \centering
  \subfigure[X=2.5D ; $-1.5D\leq Y\leq 1.5D$]
    {\includegraphics[trim= 5mm 5mm 5mm 5mm, clip, width=0.42\textwidth]
	{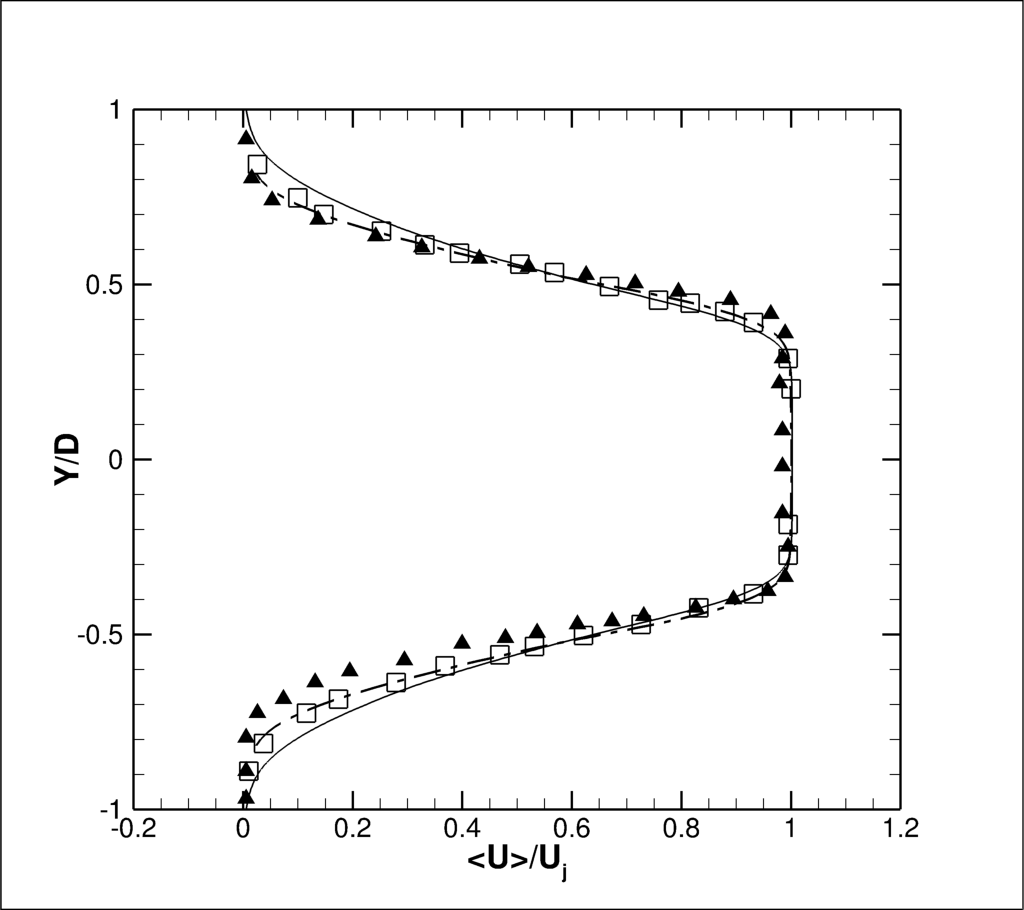}\label{fig:u-2-5-mesh-new}}
  \subfigure[X=5.0D ; $-1.5D\leq Y\leq 1.5D$]
    {\includegraphics[trim= 5mm 5mm 5mm 5mm, clip, width=0.42\textwidth]
	{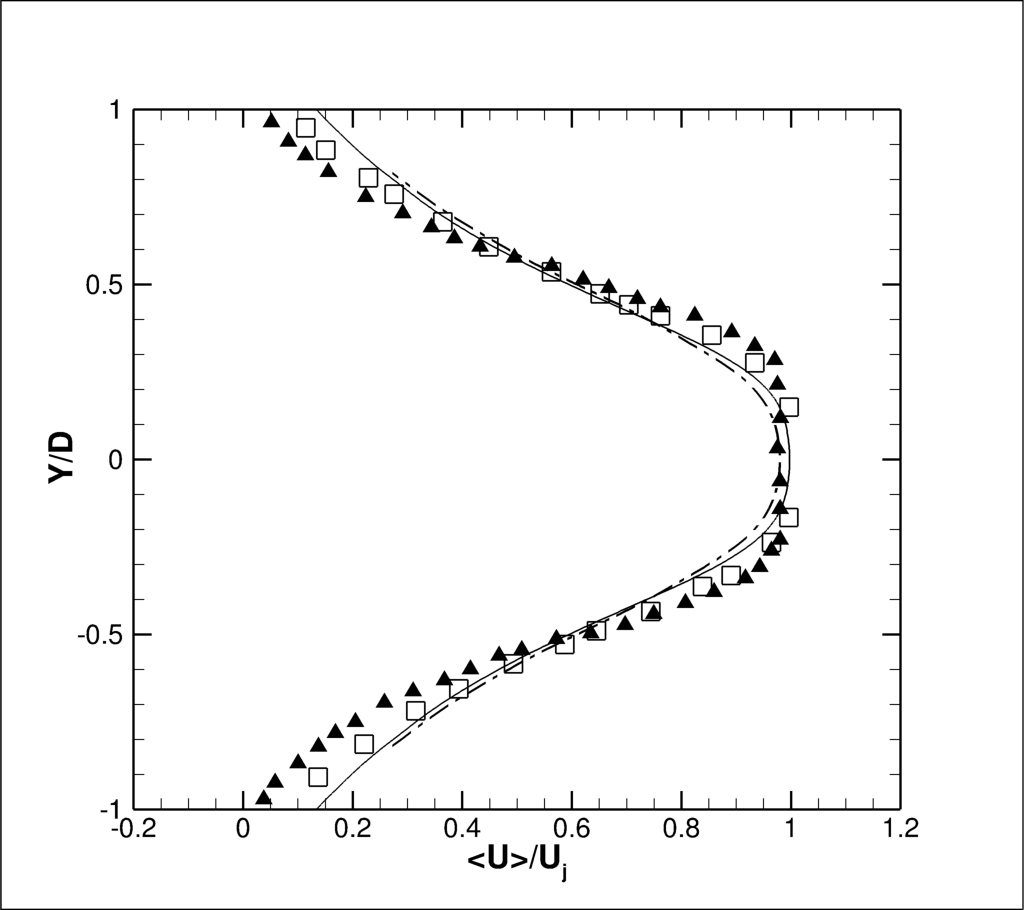}\label{fig:u-5-0-mesh-new}}
  \subfigure[X=10D ; $-1.5D\leq Y\leq 1.5D$]
    {\includegraphics[trim= 5mm 5mm 5mm 5mm, clip, width=0.42\textwidth]
	{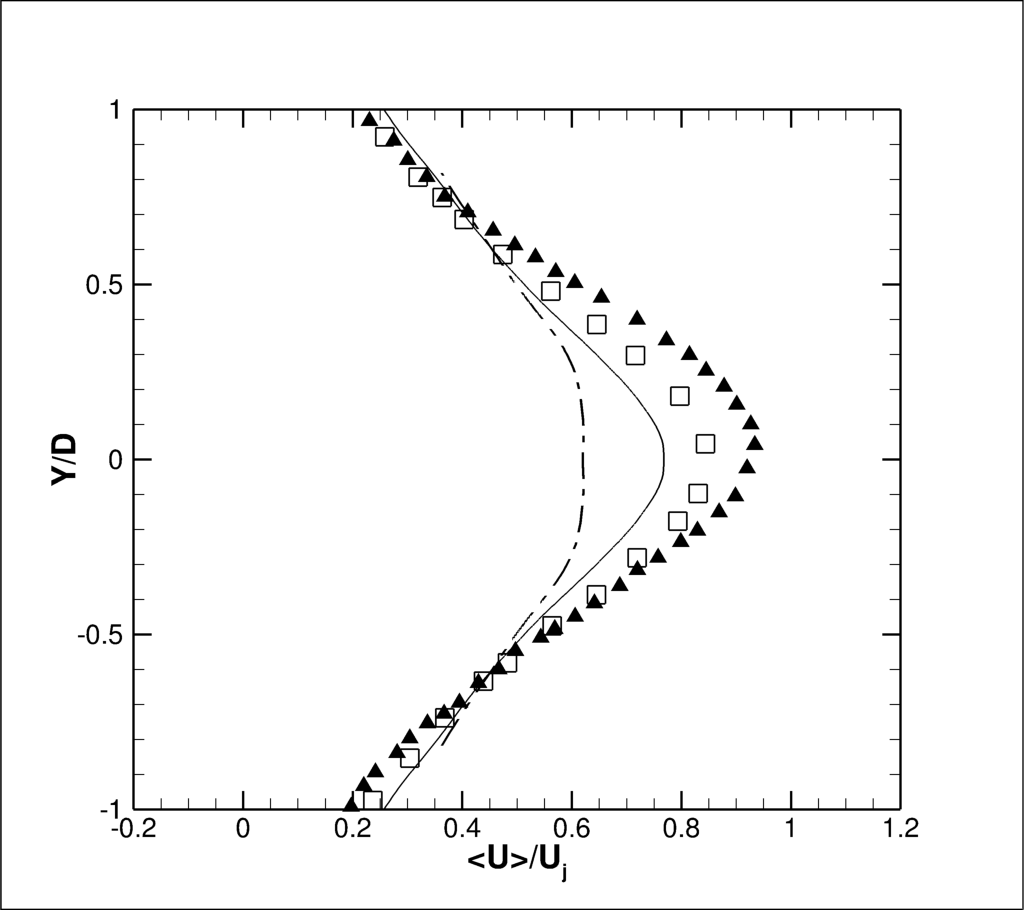}\label{fig:u-10-0-mesh-new}}
  \subfigure[X=15D ; $-1.5D\leq Y\leq 1.5D$]
    {\includegraphics[trim= 5mm 5mm 5mm 5mm, clip, width=0.42\textwidth]
	{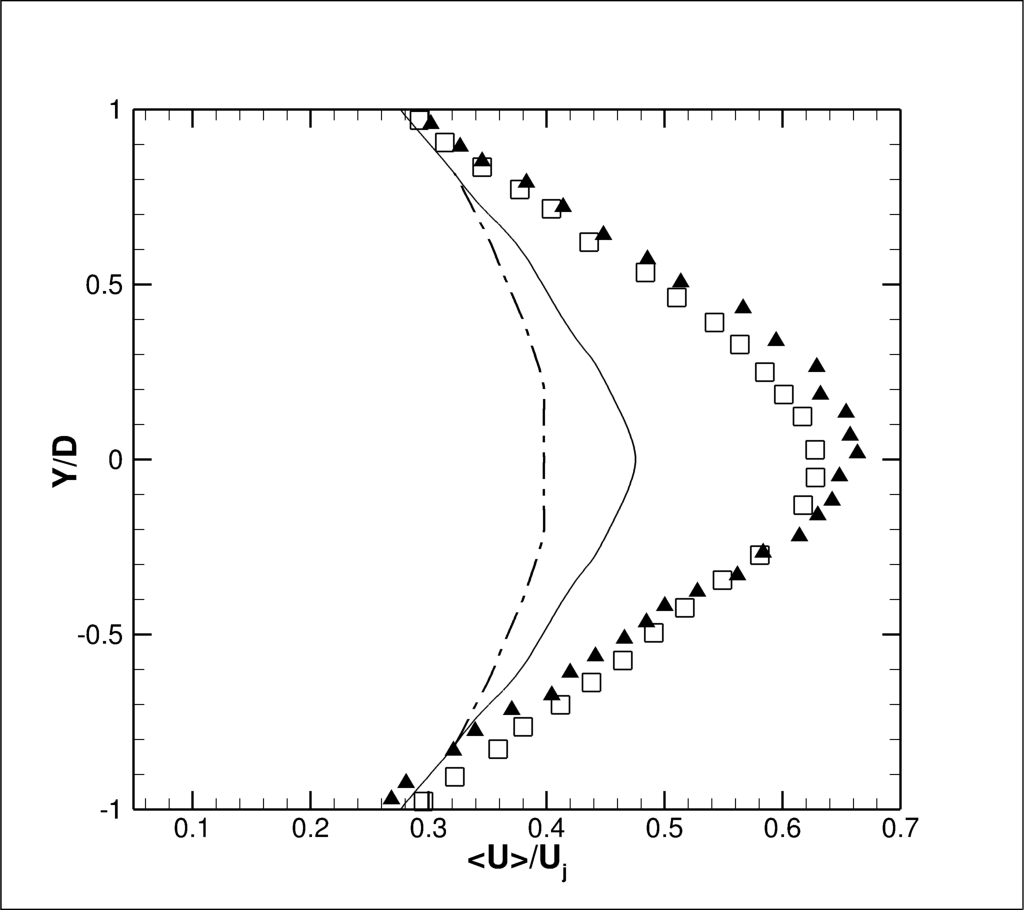}\label{fig:u-15-0-mesh-new}}
	\caption{{ Profiles of the averaged axial component of velocity, $\langle U \rangle$, at 
	different positions within the computational domain: (\textbf{-}$\cdot$\textbf{-}) S1 simulation;
	(\textbf{--}) S2 simulation; ($\square$) numerical data; ($\blacktriangle$) experimental data.}}
	\label{fig:prof-u-av-mesh-new}
\end{figure}
\begin{figure}[htb!]
  \centering
    {\includegraphics[trim= 5mm 5mm 5mm 5mm, clip, width=0.6\textwidth]
	{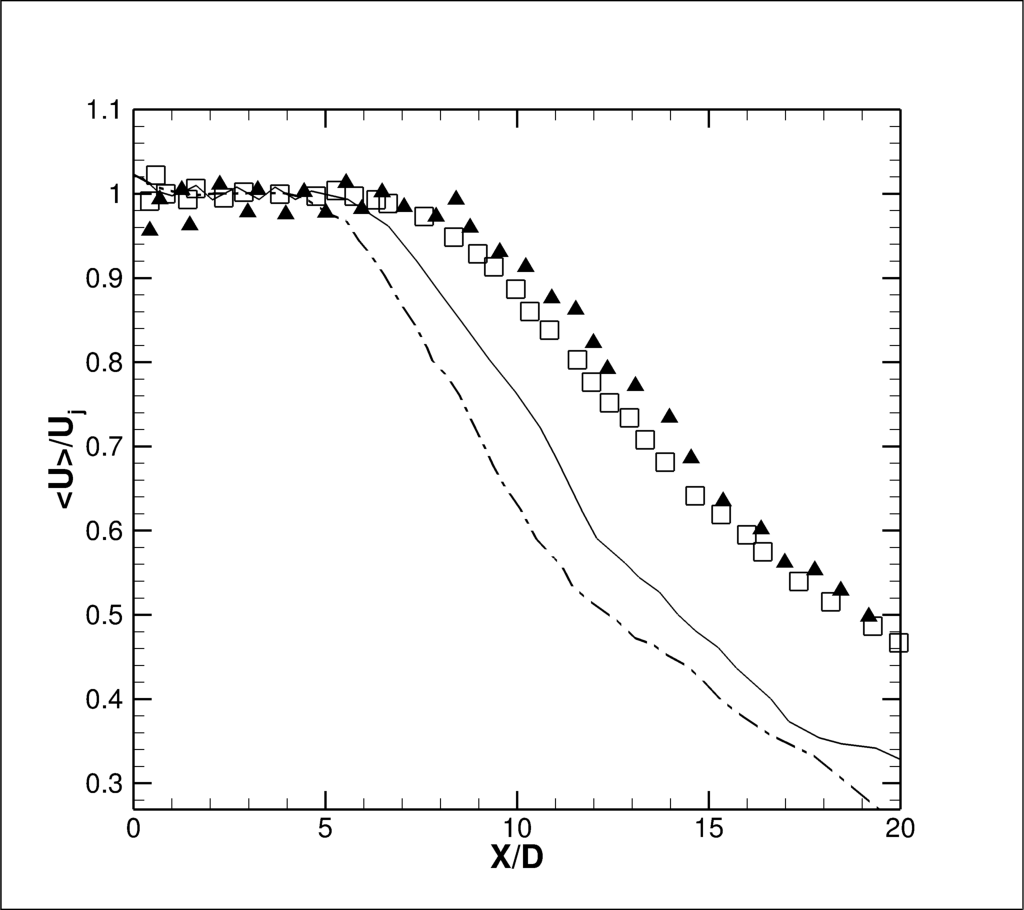}}
	\caption{{ Evolution of the averaged axial component of velocity, $\langle U \rangle$, 
	along the centerline (Y=0); $0\leq X \leq 20D$. (\textbf{-}$\cdot$\textbf{-}) S1 simulation; 
	(\textbf{--}) S2 simulation; ($\square$) numerical data; ($\blacktriangle$) experimental data.}}
	\label{fig:prof-u-av-mesh-centerline}
\end{figure}

{  In compressible jet flow analisys it is very important the study 
the evolution of flow properties along the lipline which is defined by the 
region where $r/D=0.5$. The four points at the lipline from the profiles 
presented in Figs.\ \ref{fig:u-2-5-mesh-new}, \ref{fig:u-5-0-mesh-new}, 
\ref{fig:u-10-0-mesh-new} and \ref{fig:u-15-0-mesh-new} are illustrated in 
Fig.\ \ref{fig:prof-u-av-mesh-lipline} as an evolution the averaged axial 
component of velocity, $\langle U \rangle$, along the lipline. A spline 
is used to create the curve using the four points extracted from the 
$\langle U \rangle$ profiles at $X/D=2.5$, $X/D=5.0$, $X/D=10$
and $X/D=15$. The black line stands for the numerical data, the red line for 
the experimental, the magenta line for the S1 simulation and the green line 
for the S2 simulation.}
\begin{figure}[htb!]
  \centering
    {\includegraphics[trim= 5mm 5mm 5mm 5mm, clip, width=0.6\textwidth]
	{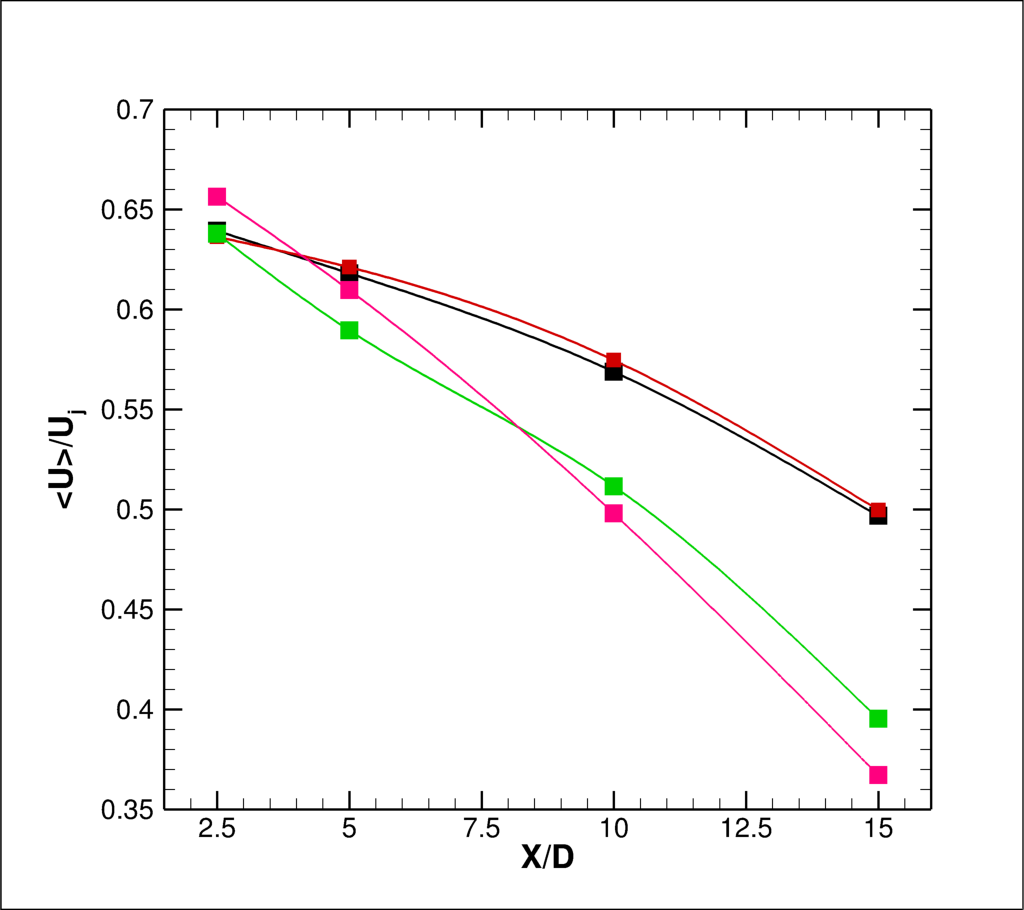}}
	\caption{{ Extractions of the averaged axial component of velocity, 
	$\langle U \rangle$, along the lipline (Y/D=0.5); $0\leq X \leq 20D$.
	({\color{black}$\blacksquare$}) numerical data; 
	({ $\blacksquare$}) experimental data;
	({\color{magenta}$\blacksquare$}) S1 simulation;
	({\color{green}$\blacksquare$}) S2 simulation.}}
	\label{fig:prof-u-av-mesh-lipline}
\end{figure}

The comparison of profiles indicates that the distributions of $\langle U \rangle$, 
calculated in the present work, correlate well with the reference data up to $X=5.0D$. 
$\langle U \rangle$ profile, calculated in the S2 case, at $X=10.0D$ is under predicted 
when compared with the reference profiles. However, the comparison is still quite a bit better 
than the one obtained for the S1 results at this axial position. One can observe that the 
$\langle U \rangle$ distributions along the centerline, for the S1 and S2 cases, correlate 
well with the reference data in the region in which there is good grid resolution. 
{  The collection of $\langle U \rangle$ magnitude from the profiles along the 
lipline for the S2 simulation correlates well with the reference and numerical data until 
$X=5.0D$. The averaged axial component of velocity calculated in S1 simulation is always 
understimated when compared to the same property calculated with S2 computation. However,
as the mesh spacing increases, the time-averaged axial velocity component computed using
S1 and S2 numerical studies starts to correlate poorly with the references.}

{  Figure \ref{fig:vort-mesh} presents a lateral view of an instantaneous snapshot 
of the pressure contours, in grey scale, superimposed by vorticity magnitude contours, in 
color, for both S1 and S2 numerical studies. It is clear that the finer grid results, that 
is, the S2 simulation, provide greater details of the pressure and vorticity fields.} 
\begin{figure}[htb!]
  \centering
  \subfigure[S1 simulation.]
    {\includegraphics[width=0.9\textwidth]
	{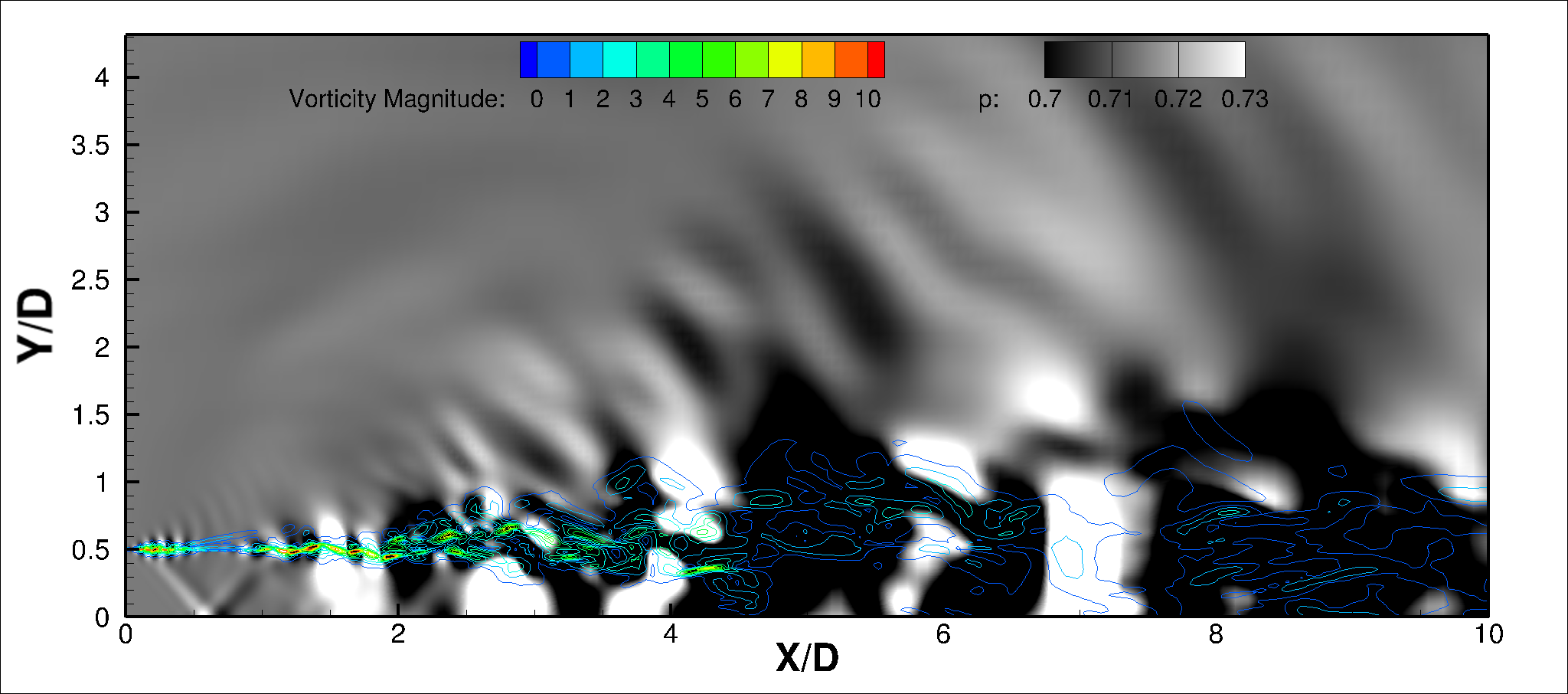}}
  \subfigure[S2 simulation.]
    {\includegraphics[width=0.9\textwidth]
	{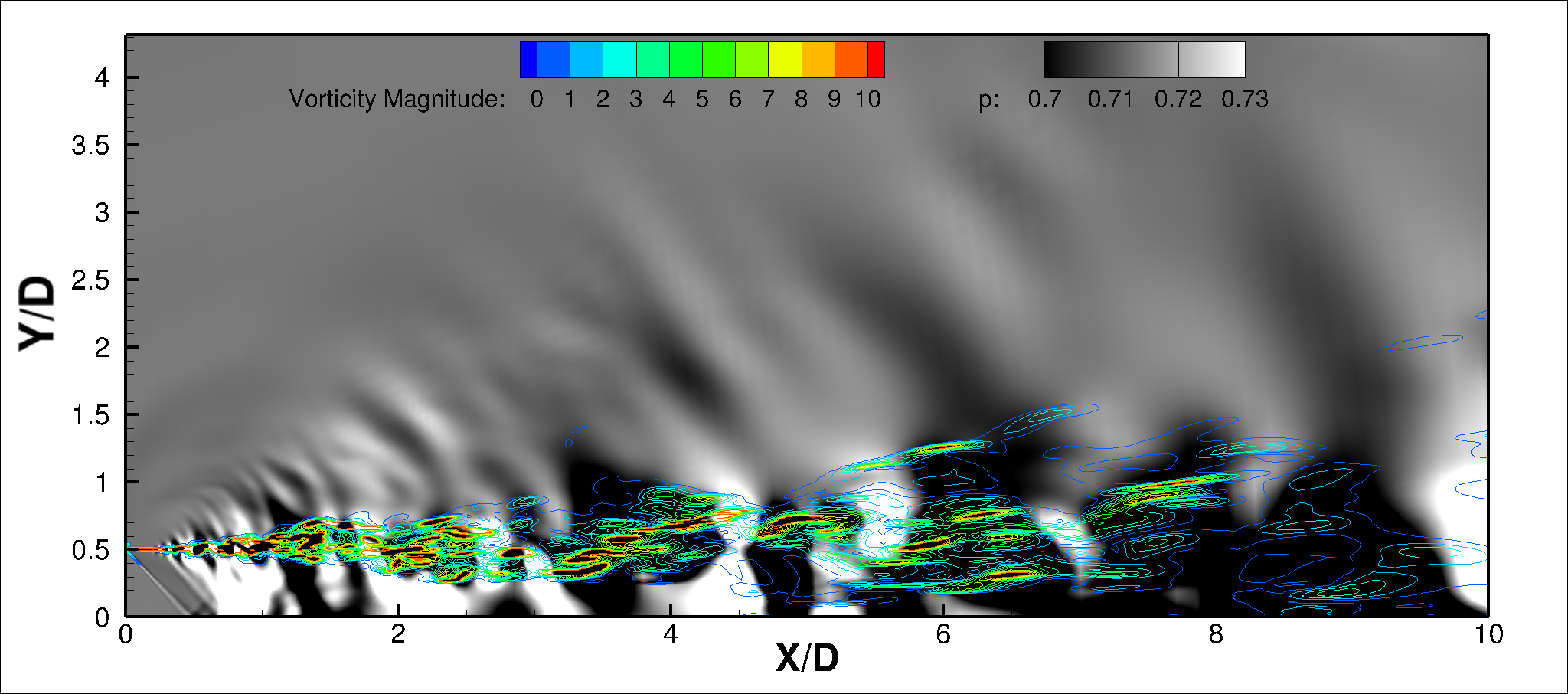}}
	\caption{{  Lateral view of instantaneous pressure contours, in grey scale, superimposed 
	by vorticity magnitude contours, in color, of the jet flow computed in the S1 and S2 simulations.}}
	\label{fig:vort-mesh}
\end{figure}
{  A detailed visualization of the jet entrance is shown in Fig.\ \ref{fig:vort-zoom-mesh}, 
for the two test cases. As before, an instantaneous snapshot of one longitudinal plane of the 
jet flow is shown. Pressure contours are shown as a background visualization in grey scale, and this 
is superimposed by vorticity magnitude contours, in color. The better resolution of flow features 
obtained for the S2 test case is even more evident in this detailed plot of the jet entrance. In 
particular, compression waves generated at the shear layer, and their reflections at the jet axis, 
are much more clearly visible in the S2 results. At the end, this simulation provides a much 
richer visualization of flow features.}
\begin{figure}[htb!]
  \centering
  \subfigure[S1 simulation.]
    {\includegraphics[width=0.9\textwidth]
	{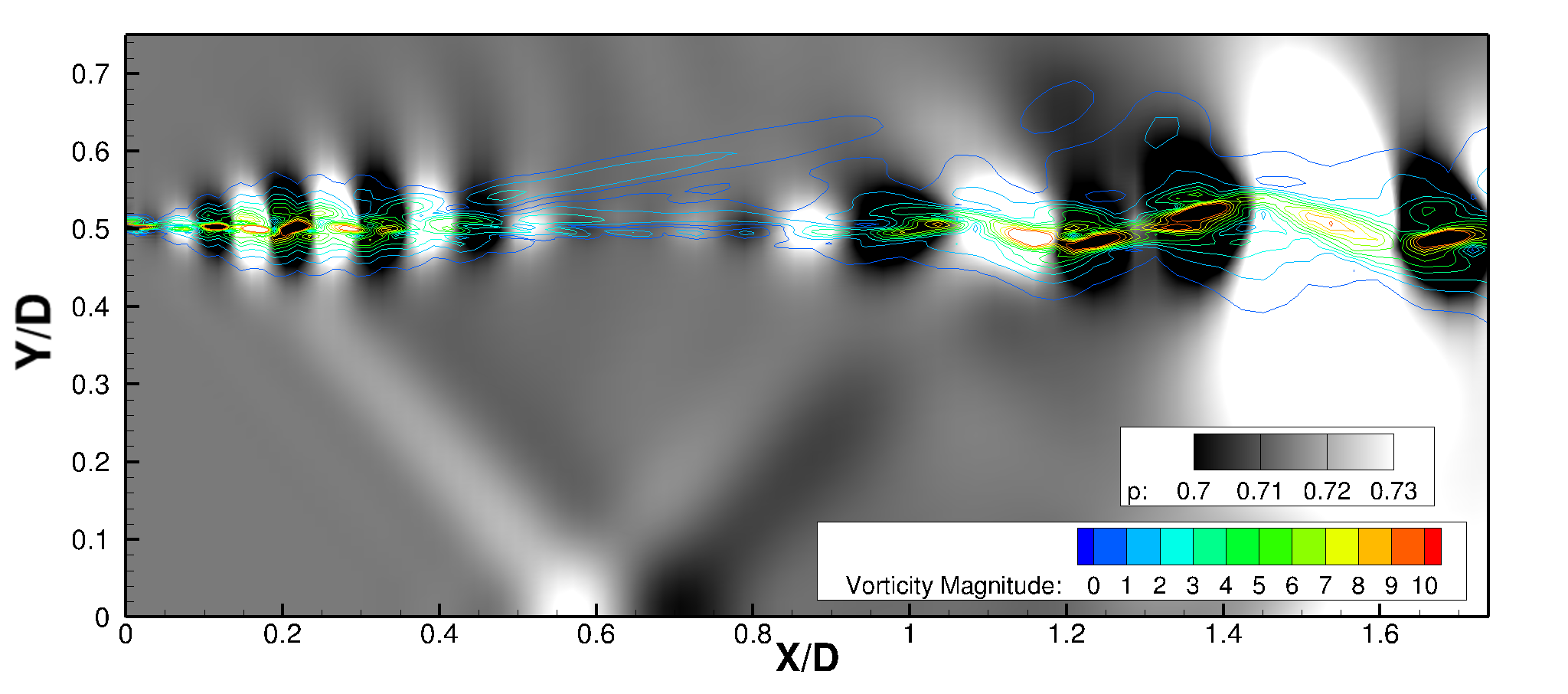}}
  \subfigure[S2 simulation.]
    {\includegraphics[width=0.9\textwidth]
	{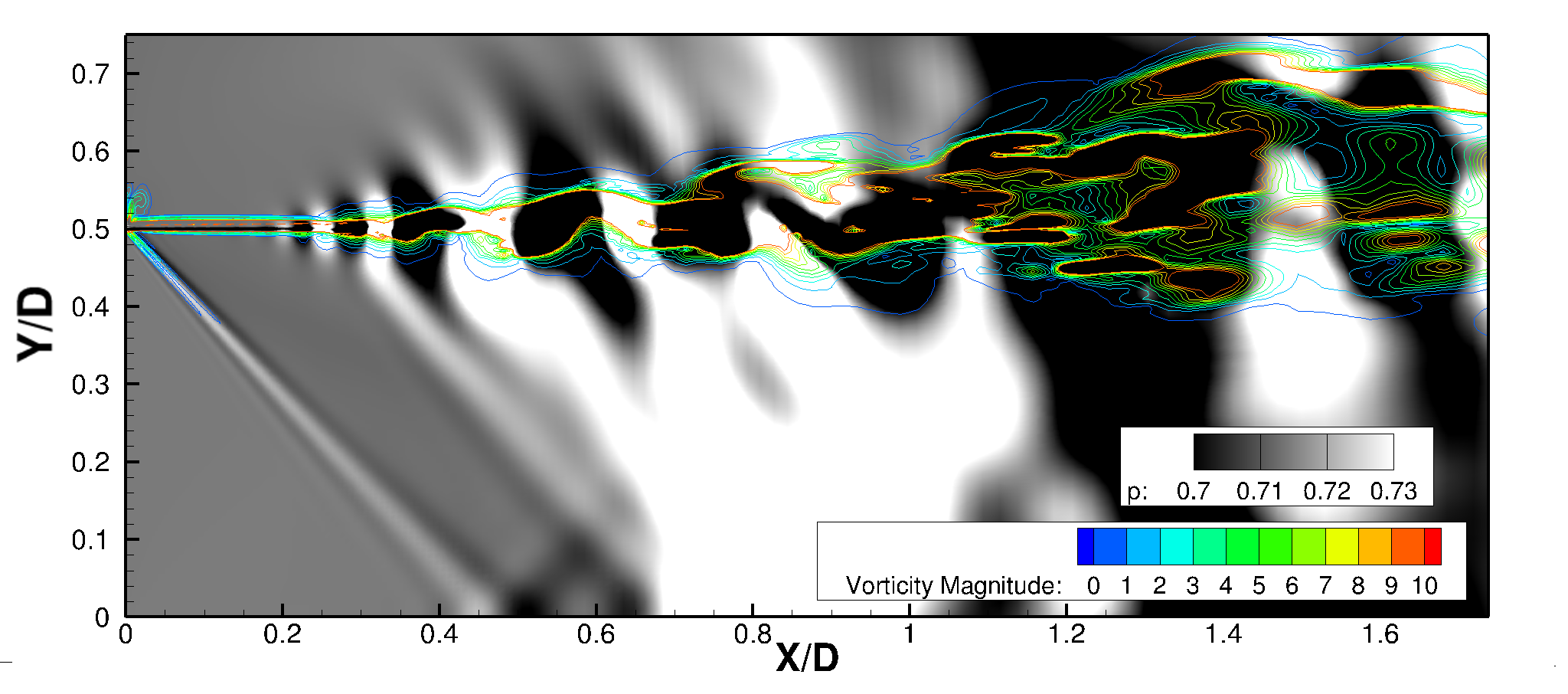}}
	\caption{  Detailed view near the jet entrance of pressure contours, in grey scale, superimposed 
	by vorticity magnitude contours, in color, for the S1 and S2 simulations.}
	\label{fig:vort-zoom-mesh}
\end{figure}

\subsubsection{{  RMS Distributions of Time Fluctuations 
of Axial Velocity Component}}

{  The time fluctuating part of the flow is also important to be 
studied. Therefore, the analysis of the effects of mesh refinement is also performed for the 
fluctuating part of the axial velocity component using the root mean square approach. A 
lateral view, {\em i.e.}, a complete longitudinal plane, and a detailed view of the region near the 
jet entrance the root mean square (RMS) values of the fluctuating part of the axial velocity component, 
$u^{*}_{RMS}$, computed in the S1 and S2 simulations, are presented in Fig.\ \ref{fig:lat-u-rms-mesh}.}
\begin{figure}[htb!]
  \centering
  \subfigure[$u^{*}_{RMS}$ for S1 simulation.]
    {\includegraphics[trim= 5mm 5mm 5mm 5mm, clip, width=0.495\textwidth]
	{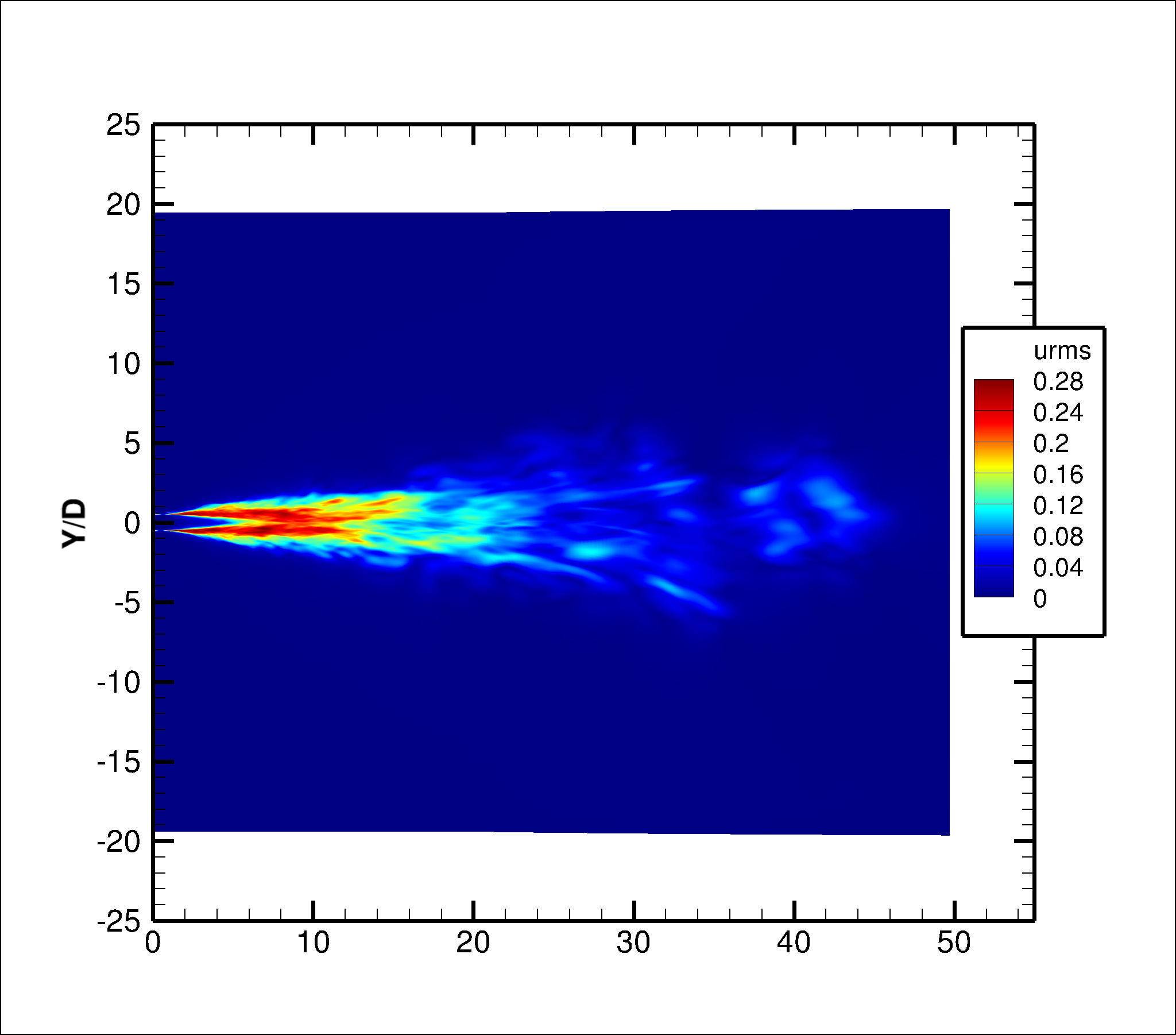}\label{subfig:urms-lat-s1}}
  \subfigure[$u^{*}_{RMS}$ for S2 simulation.]
    {\includegraphics[trim= 5mm 5mm 5mm 5mm, clip, width=0.495\textwidth]
	{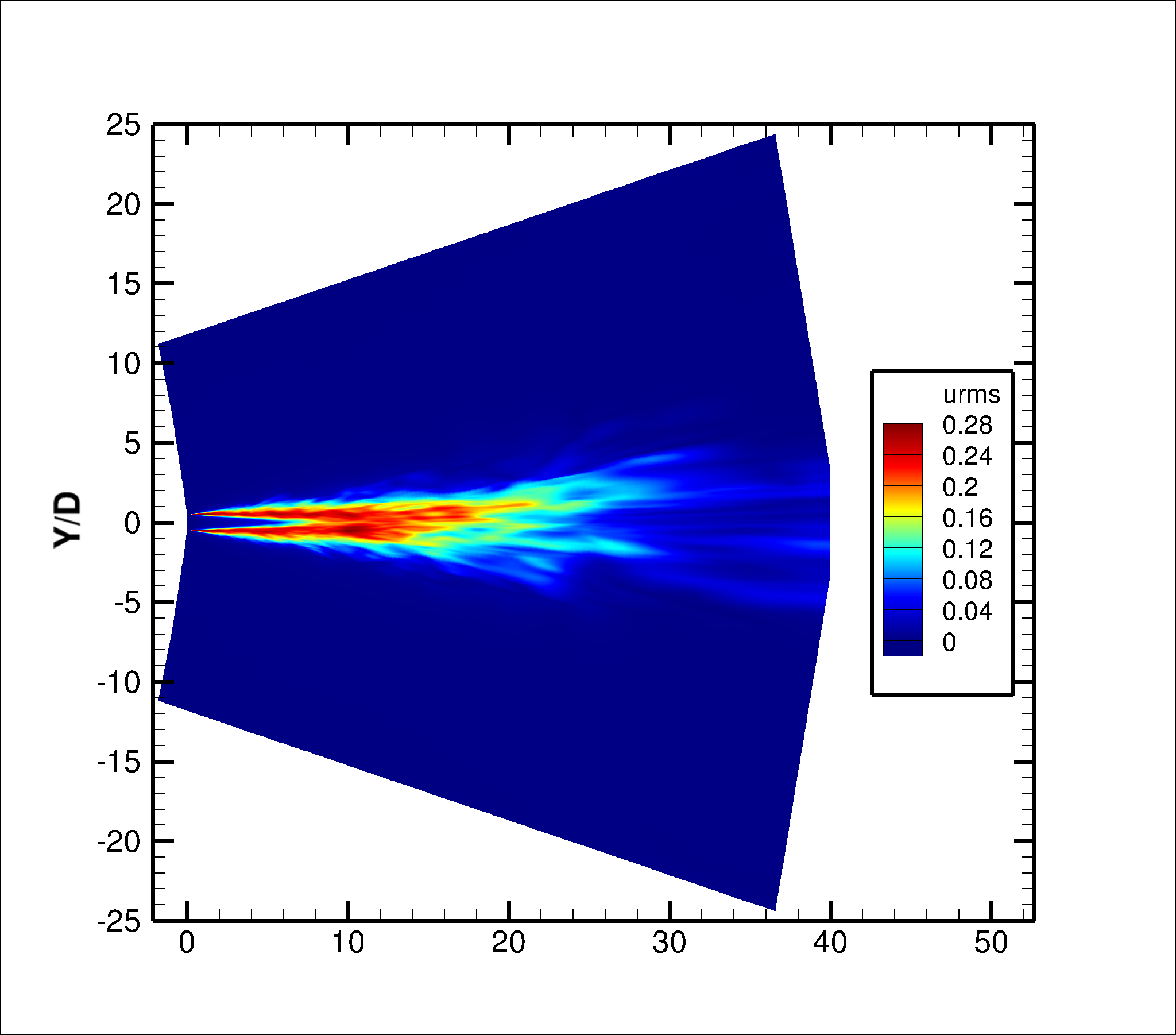}\label{subfig:urms-lat-s2}}
  \subfigure[$u^{*}_{RMS}$ for S1 simulation.]
    {\includegraphics[trim= 5mm 5mm 5mm 5mm, clip, width=0.495\textwidth]
	{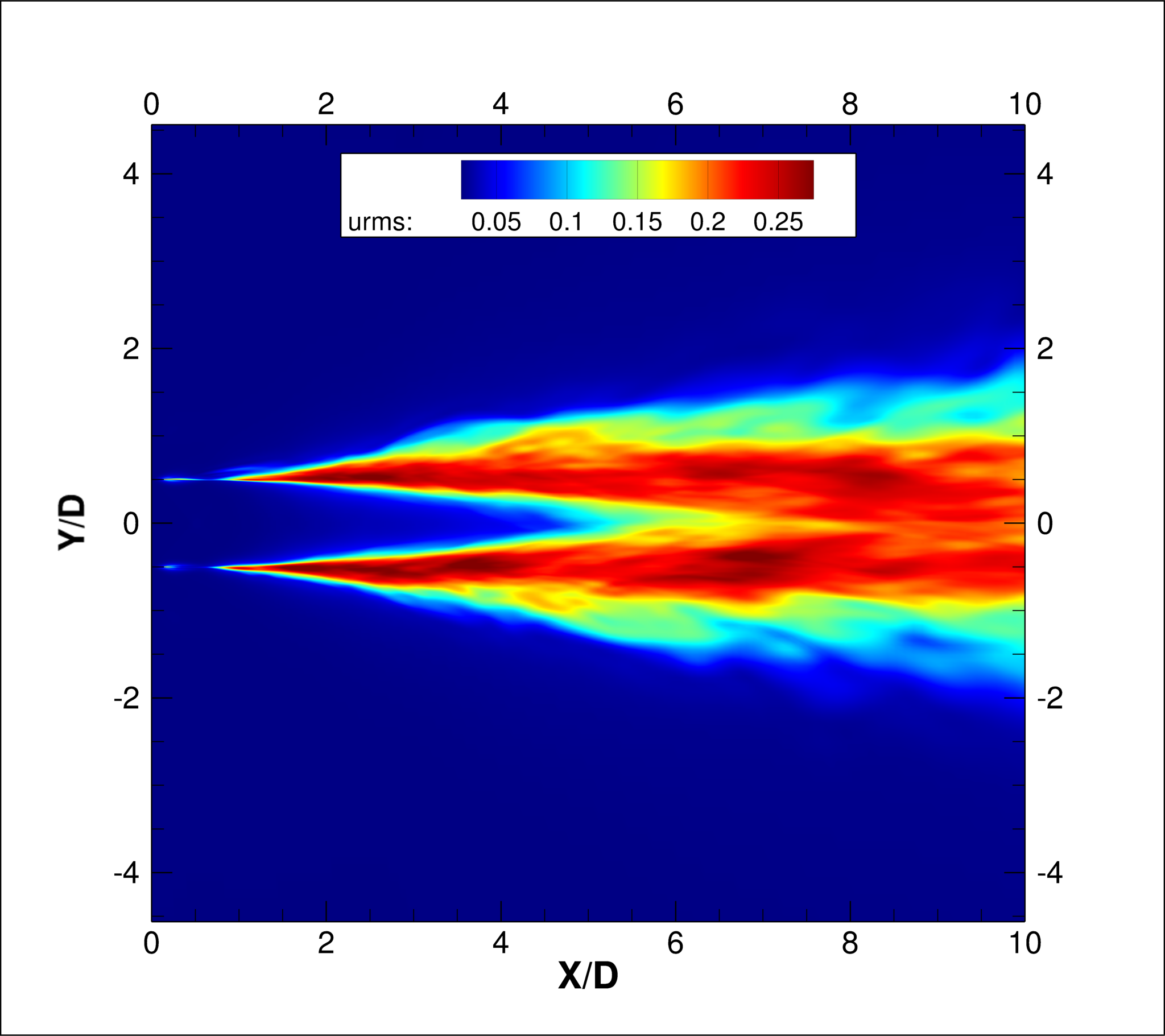}\label{subfig:urms-det-s1}}
  \subfigure[$u^{*}_{RMS}$ for S2 simulation.]
    {\includegraphics[trim= 5mm 5mm 5mm 5mm, clip, width=0.495\textwidth]
	{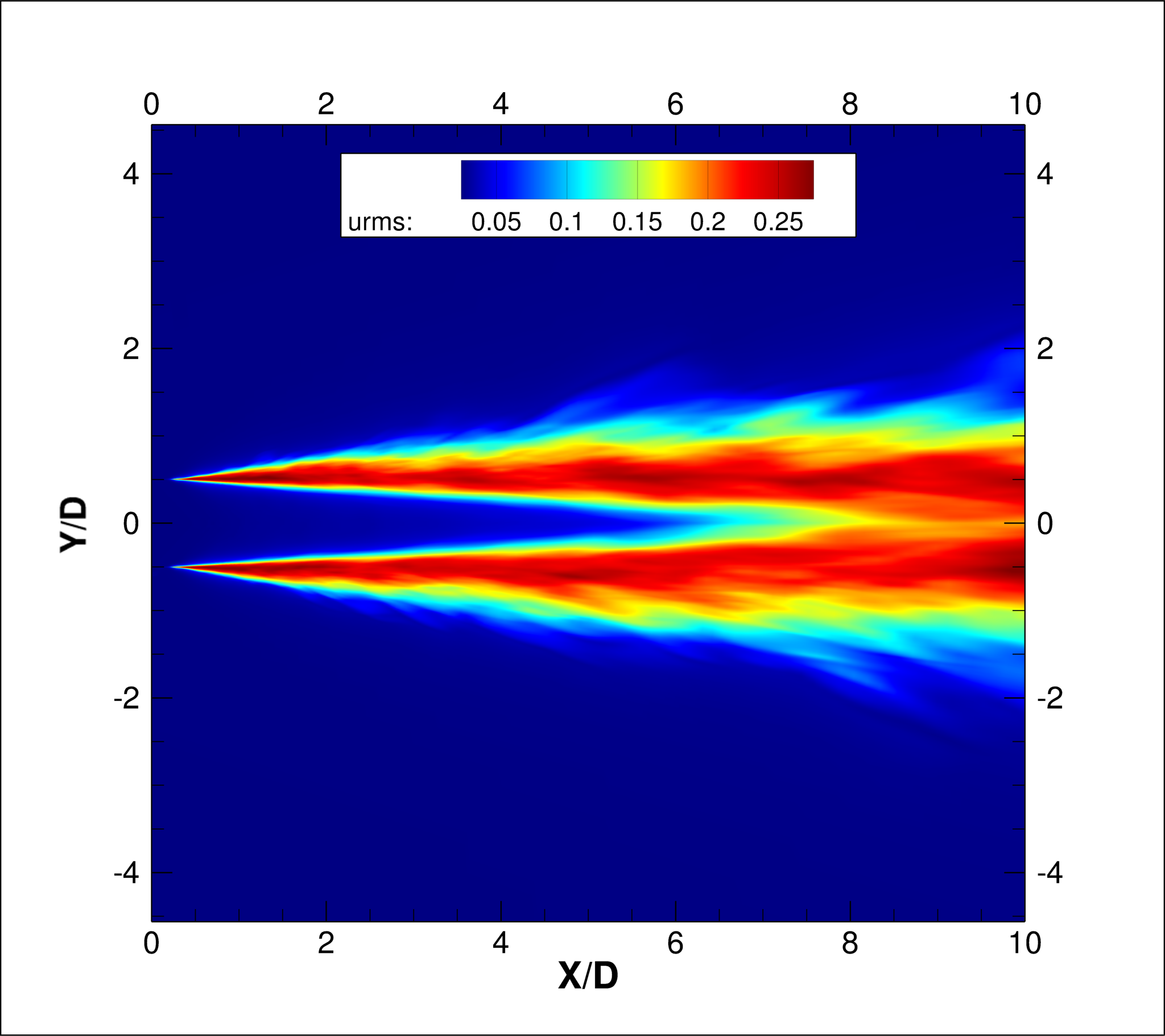}\label{subfig:urms-det-s2}}
	\caption{{ Lateral and detailed views of the RMS values of the time 
	fluctuation of the axial component of velocity, $u_{RMS}^{*}$, for S1 and 
	S2 simulations.}} 
	\label{fig:lat-u-rms-mesh}
\end{figure}

{  One can observe, for instance, in Figs.\ \ref{subfig:urms-lat-s1} and 
\ref{subfig:urms-lat-s2}, that the mesh coarsening in the streamwise direction, towards the farfield, 
is working as designed. In other words, the mesh stretching destroys structures of the flow towards 
the exit boundary and, therefore, there are no wave reflections back into the computational domain. 
As previously discussed, coarse meshes implicitly add dissipation to the solution and, as such, they 
work as sponge zones in the vicinity of undisturbed flow regions.} 
{  The detailed views of $u^{*}_{RMS}$, {\em i.e.}, Figs.\ \ref{subfig:urms-det-s1} and 
\ref{subfig:urms-det-s2}, indicate that the properties calculated in the S2 study provide a 
better definition of the potential core and of the near jet entrance region. The results for the 
S1 test case are more spread and indicating a much reduced potential core length. The same effect can 
be observed on crossflow plane visualizations of $u^{*}_{RMS}$ contours, which, however, are not 
shown here for the sake of brevity.} 

The same strategy used to compare the mean profiles of velocity is 
used for the study of $u^{*}_{RMS}$. 
{  Figure 
\ref{fig:prof-u-rms-mesh-new}} {  presents the comparison of RMS 
profiles of $u^{*}$, from the S1 and S2 simulations, with reference results. 
}
{  The profile of $u^{*}_{RMS}$ calculated using the numerical 
approach fits very well the experimental reference profile at $X=2.5D$ near
the centerline region. However, all numerical calculations overpredict the 
peaks of $u^{*}_{RMS}$ at the proximity of $Y=0.5D$. The profile obtained 
with the S2 calculation, at the same position, presents a good correlation 
with numerical data whereas the $u^{*}_{RMS}$ profile calculated in S1 
simulation cannot correctly represent the two peaks of the profile.}
{  At $X=5.0D$, the $u^{*}_{RMS}$ profile calculated using 
S2 simulation is close to the profiles achieved by the references. 
S1 study overstimates the RMS of $u^{*}$ near the centerline region,
{\it i.e.} $r/D=0.0$. The same beahavior can be observed}
{  at $X=10.0D$}, {  where} {  the profiles of 
the fluctuating part of the axial velocity component obtained in the S1 
and S2 calculations start to diverge from the reference results}
{  at $-0.5D<Y<0.5D$ vicinity.}
{  At $X=15.0D$, the $u^{*}_{RMS}$ profile for the S1 
calculation is completely underpredicted when compared with the profiles
of $u^{*}_{RMS}$ obtained from numerical and experimental references. 
At the same position, the fluctuation profile from the S2 computation 
presents similiar magnitudes shape when compared with the numerical and 
references profiles of $u^{*}_{RMS}$. However, the shape of the same
profile achieved using S2 computation does not match with the reference 
data at $X=15.0D$.}
%
%
\begin{figure}[htb!]
  \centering
  \subfigure[X=2.5D ; $-1.5D\leq R\leq 1.5D$.]
    {\includegraphics[trim= 5mm 5mm 5mm 5mm, clip, width=0.42\textwidth]
	{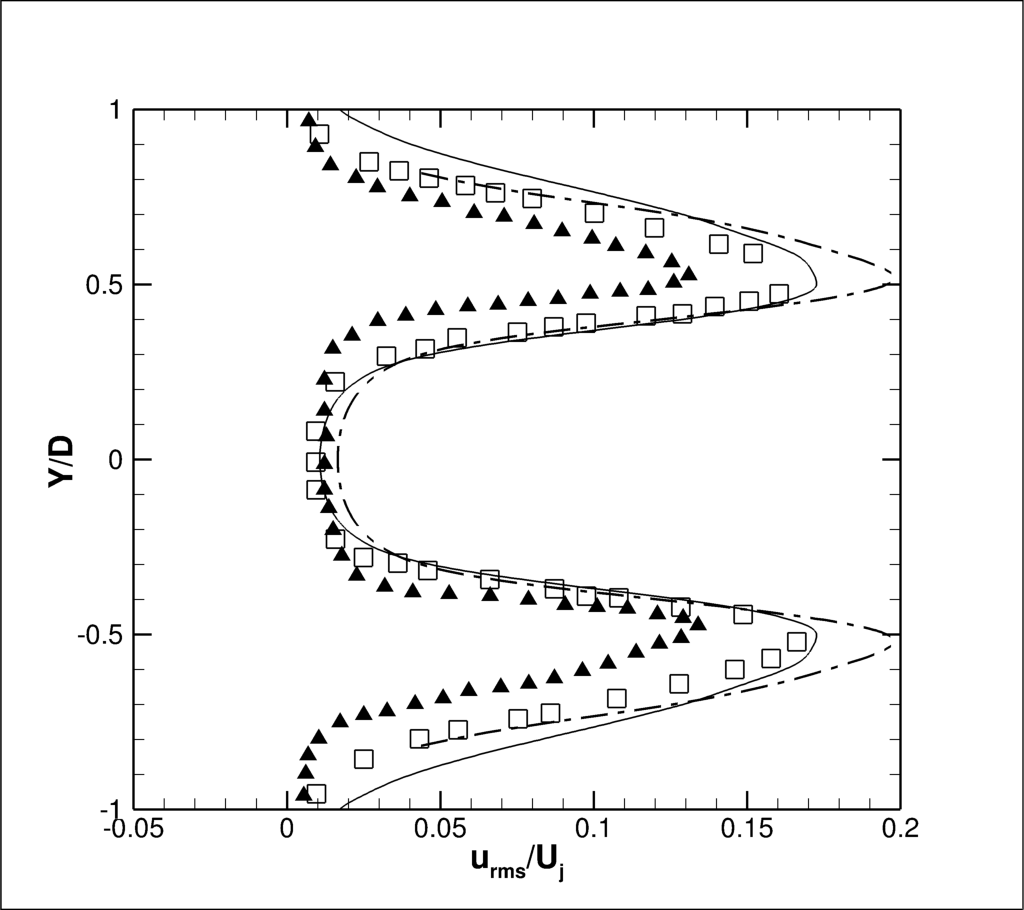}\label{fig:urms-2-5-mesh-new}}
  \subfigure[X=5.0D ; $-1.5D\leq R\leq 1.5D$.]
    {\includegraphics[trim= 5mm 5mm 5mm 5mm, clip, width=0.42\textwidth]
	{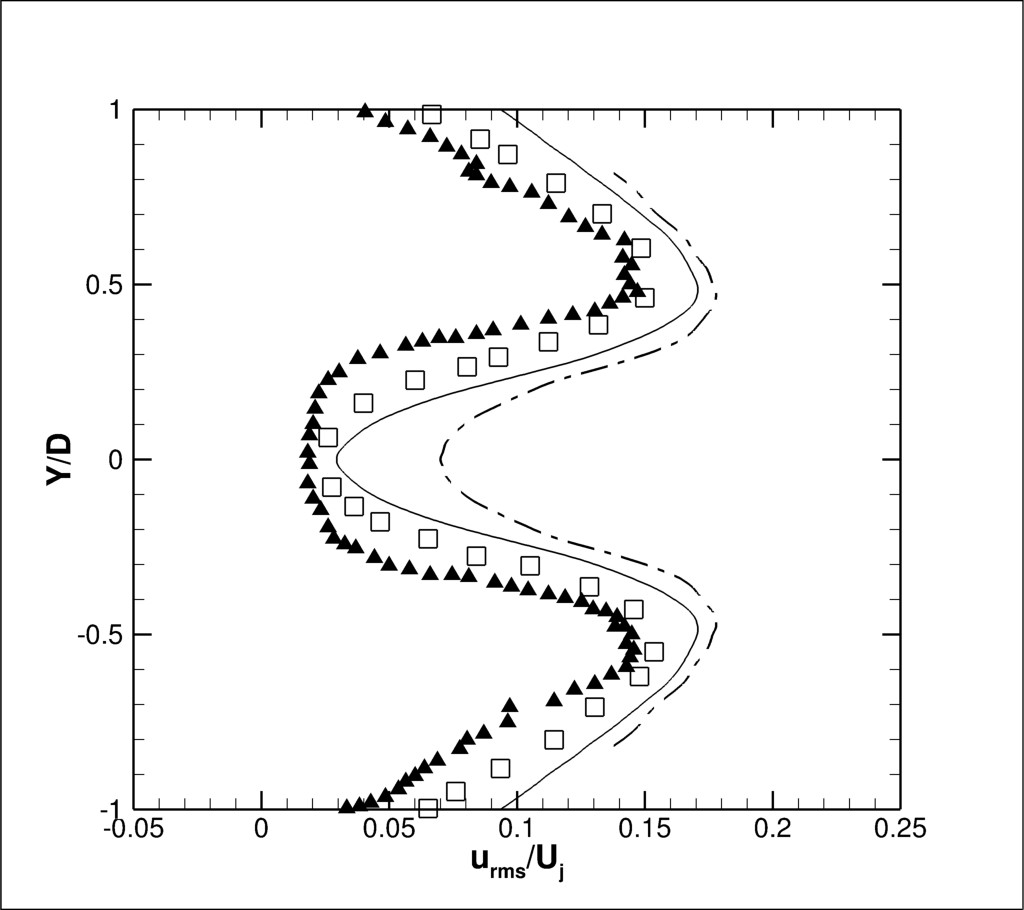}\label{fig:urms-5-0-mesh-new}}
  \subfigure[X=10D ; $-1.5D\leq R\leq 1.5D$.]
    {\includegraphics[trim= 5mm 5mm 5mm 5mm, clip, width=0.42\textwidth]
	{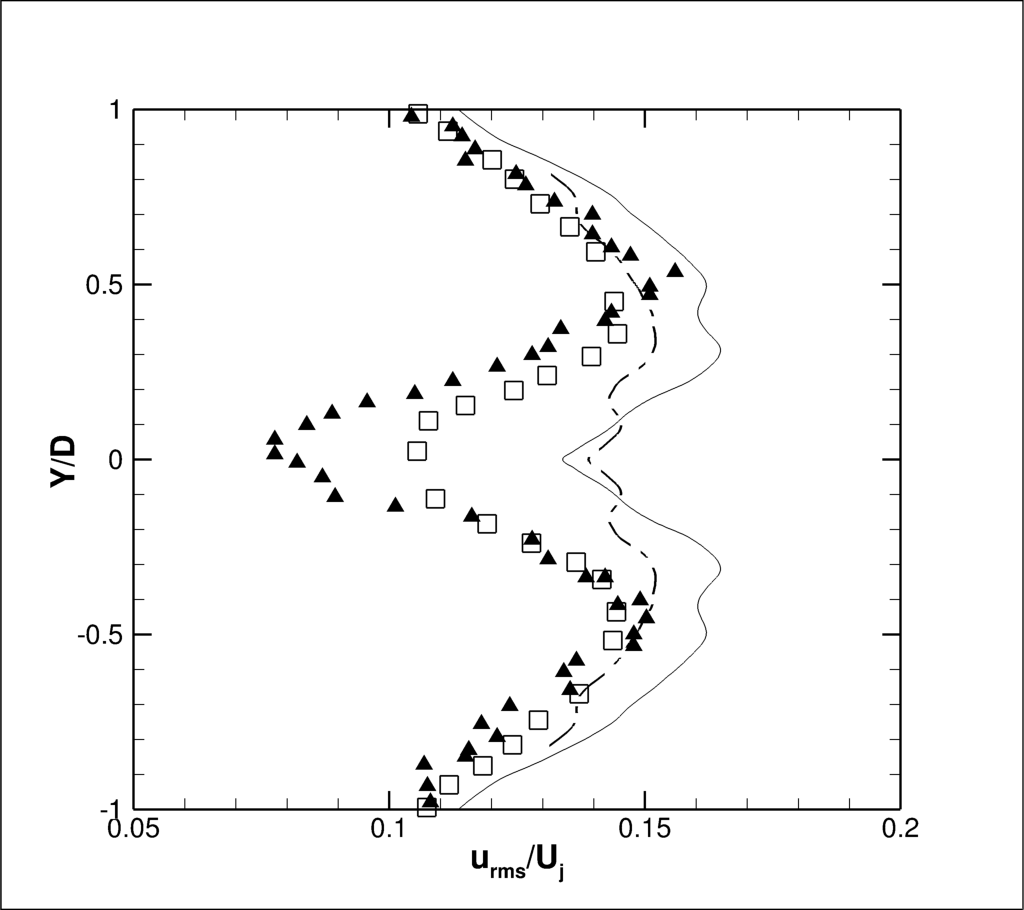}\label{fig:urms-10-0-mesh-new}}
  \subfigure[X=15D ; $-1.5D\leq R\leq 1.5D$.]
    {\includegraphics[trim= 5mm 5mm 5mm 5mm, clip, width=0.42\textwidth]
	{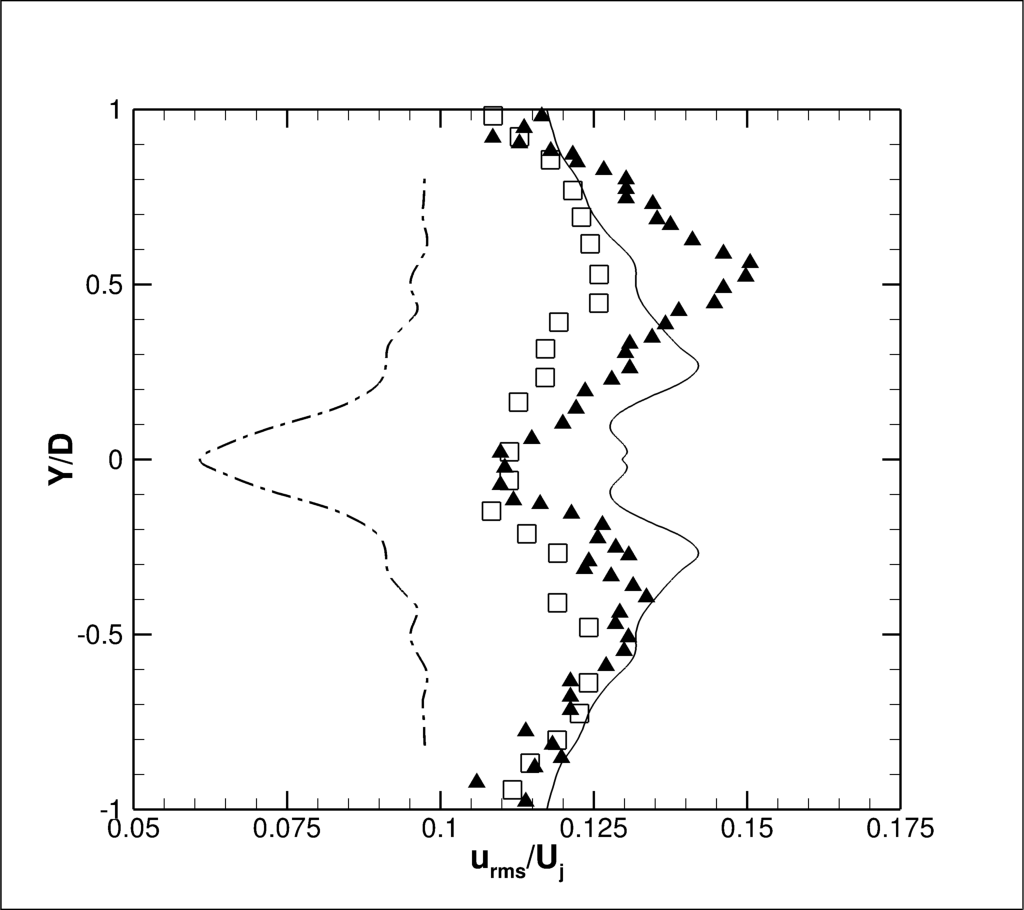}\label{fig:urms-15-0-mesh-new}}
	\caption{{  Profiles of RMS values of time fluctuation of axial 
	component of velocity, $u_{RMS}^{*}$, for the S1 and S2 simulations,
	at different positions within the computational domain:
	(\LARGE \textbf{-}$\cdot$\textbf{-}\small) S1;
	(\LARGE \textbf{--}\small) S2; 
	($\square$) numerical data; ($\blacktriangle$) experimental data.}}
	\label{fig:prof-u-rms-mesh-new}
\end{figure}


{  Figure \ref{fig:prof-u-rms-mesh-centerline} presents the 
distribution of $u^{*}_{RMS}$ along the centerline of the jet. The
S2 calculation presents overpredicted values of $u^{*}_{RMS}$ along
the centerline when compared with reference data. Nevertheless, the 
evolution of the same property along the centerline presents a similar shape 
of the evolution of $u^{*}_{RMS}$ achieved by the numerical and experimental
references. One can notice that the distribution of $u^{*}_{RMS}$ along 
the centerline achieved by S1 calculation presents the same behavior 
of the distribution obtained from S2 study for $X<10D$. Both results are 
overpredicted when compared with the refence. However, in particular
for S1 simulation, the value of $u^{*}_{RMS}$ decreases quickly in the
streamwise direction for $X>10D$. This behavior of S1 calculation
generates a very underpredicted distribution of $u^{*}_{RMS}$ when 
compared to the results of S2 simulation and refence data. Even the
shape of the evolution of the property calculated by S1 simulation, 
for $X>10D$, along the centerline, is completely different from the 
evolution presented by the reference data. Therefore, it is possible 
to notice a significant improvement of the solution with the refinement 
of the mesh performed for S2 study.}
\begin{figure}[htb!]
  \centering
    {\includegraphics[trim= 5mm 5mm 5mm 5mm, clip, width=0.6\textwidth]
	{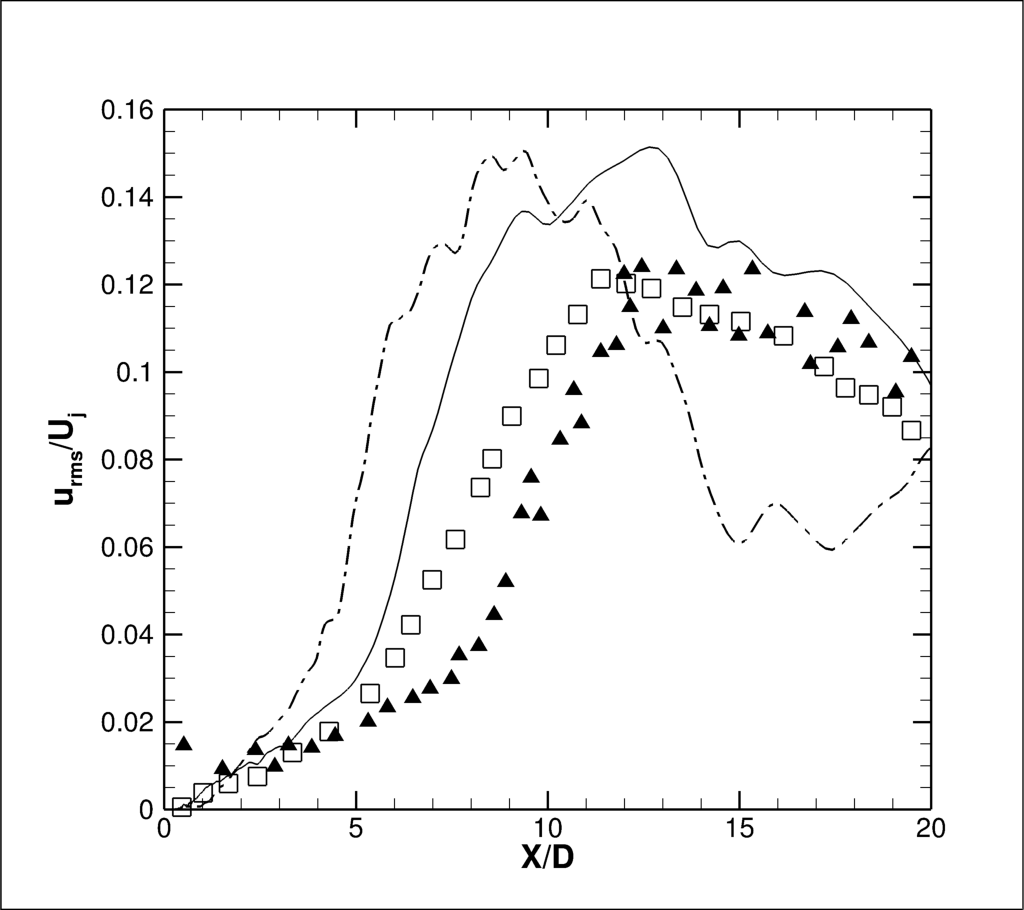}}
	\caption{{ Evolution of RMS values of time fluctuation of axial 
	component of velocity, $u_{RMS}^{*}$, along the centerline (Y=0); 
	$0\leq X \leq 20D$. (\textbf{-}$\cdot$\textbf{-}) S1 simulation; 
	(\textbf{--}) S2 simulation; ($\square$) numerical data; 
	($\blacktriangle$) experimental data.}}
	\label{fig:prof-u-rms-mesh-centerline}
\end{figure}

{  Figure \ref{fig:prof-u-rms-mesh-lipline} presents the 
evolution of $u_{RMS}^{*}$ along the lipline using four points extracted
from Figs.\ \ref{fig:urms-2-5-mesh-new}, \ref{fig:urms-5-0-mesh-new}, 
\ref{fig:urms-10-0-mesh-new} and \ref{fig:urms-15-0-mesh-new}.
A spline is used to create the curve using the four points extracted from 
the $u_{RMS}^{*}$ profiles at $X/D=2.5$, $X/D=5.0$, $X/D=10$ and $X/D=15$. 
The black line stands for the numerical data, the red line for the 
experimental, the magenta line for the S1 simulation and the green line for 
the S2 simulation. The peaks of $u_{RMS}^{*}$ are located near the lipline
region. Figure \ref{fig:prof-u-rms-mesh-lipline} indicates that the results
obtained from the numerical simulations performed in the present work and
the numerical reference are overpredicted when compared with the experimental
data at $X/D=2.5$ and $X/D=5.0$. The values of $u_{RMS}^{*}$ obtained
by S2 calculation and the numerical reference are close to the results 
presented in the experimental refence at $X/D=10.0$ and $X/D=15.0$.
One can notice that the results achieved using the less refined mesh in the 
present work fails to predict the peaks of $u_{RMS}^{*}$ at the four
points compared along the lipline.}
\begin{figure}[htb!]
  \centering
    {\includegraphics[trim= 5mm 5mm 5mm 5mm, clip, width=0.6\textwidth]
	{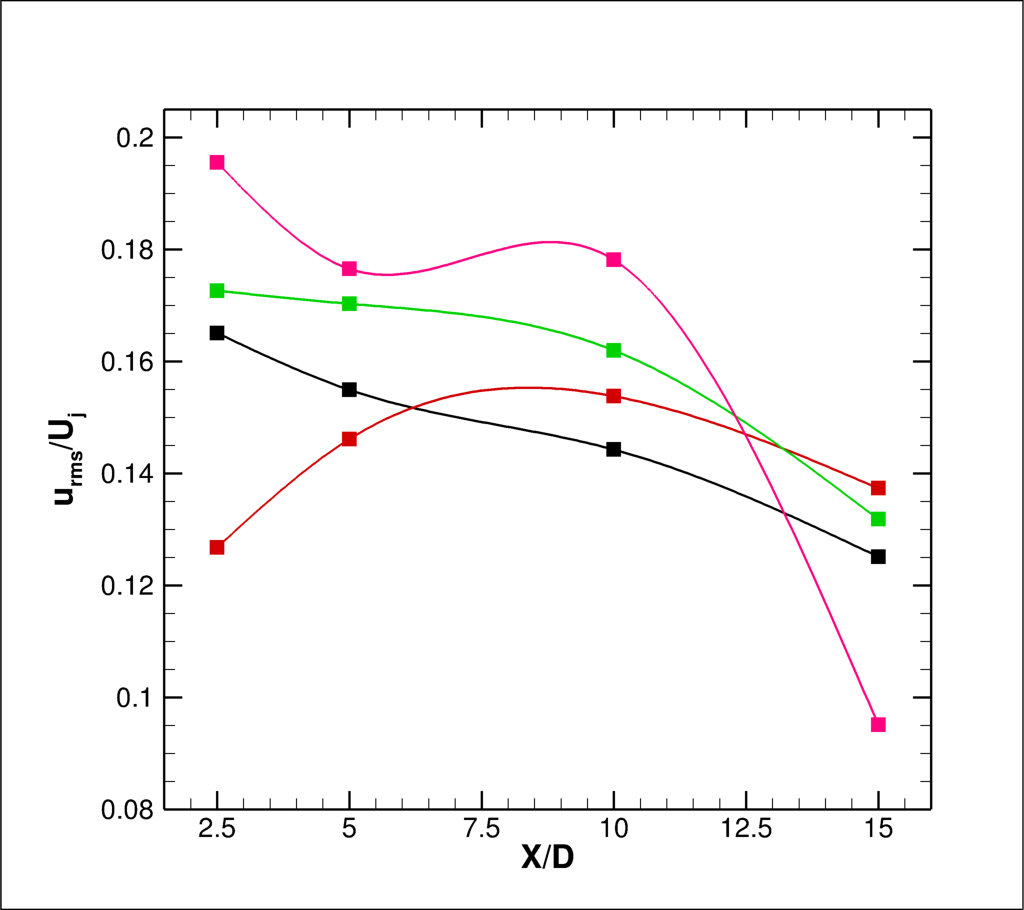}}
	\caption{{ Extractions of RMS values of time fluctuation 
	of axial component of velocity, $u_{RMS}^{*}$, along the lipline 
	(Y/D=0.5); $0\leq X \leq 20D$.
	({\color{black}$\blacksquare$}) numerical data; 
	({ $\blacksquare$}) experimental data;
	({\color{magenta}$\blacksquare$}) S1 simulation;
	({\color{green}$\blacksquare$}) S2 simulation.}}
	\label{fig:prof-u-rms-mesh-lipline}
\end{figure}
%


\subsection{Subgrid Scale Modeling Study}

After the mesh refinement study, the three SGS models added
to the solver are compared. S2, S3 and S4 simulations
are performed using the static Smagorinsky model 
\cite{Smagorinsky63,Lilly65,Lilly67}, the dynamic Smagorinsky 
model \cite{germano90,moin91} and the Vreman model 
\cite{vreman2004}, respectively. The same mesh with 50 
million points is used for all three simulations.
The stagnated flow condition is used as intial condition
for S2 and S3 simulations. A restart of S2 simulation is used as initial 
condition for the S4 simulation. The configuration of the numerical 
studies is presented at Tab.\ \ref{tab:simu}.
An extensive comparison study is perfomed in this subsection.
Time-averaged distributions of the axial and radial
velocity components and eddy viscosity are presented in 
the subsection along with the RMS distribution of axial and radial 
components of velocity and distributions of the $\langle u^{*}v^{*}
\rangle$ component of the Reynolds stress tensor.

\subsubsection{Time Averaged Axial Component of Velocity}

Effects of the SGS modeling on the time averaged results of the 
axial component of velocity are presented in the subsection. A 
lateral view of $\langle U \rangle$ for S2, S3 and S4 simulations, side by 
side, are presented in Fig.\ \ref{fig:lat-u-av-sgs}, where 
$U_{j}^{95\%}$ is indicated by the solid line. 
\begin{figure}[htb!]
  \centering
  \subfigure[Lateral view of $\langle U \rangle$ for simulation S2.]
    {\includegraphics[trim= 5mm 5mm 5mm 5mm, clip, width=0.32\textwidth]
	{./stat-cs-XY-zoom-u-av.png}}
  \subfigure[Lateral view of $\langle U \rangle$ for simulation S3.]
    {\includegraphics[trim= 5mm 5mm 5mm 5mm, clip, width=0.32\textwidth]
	{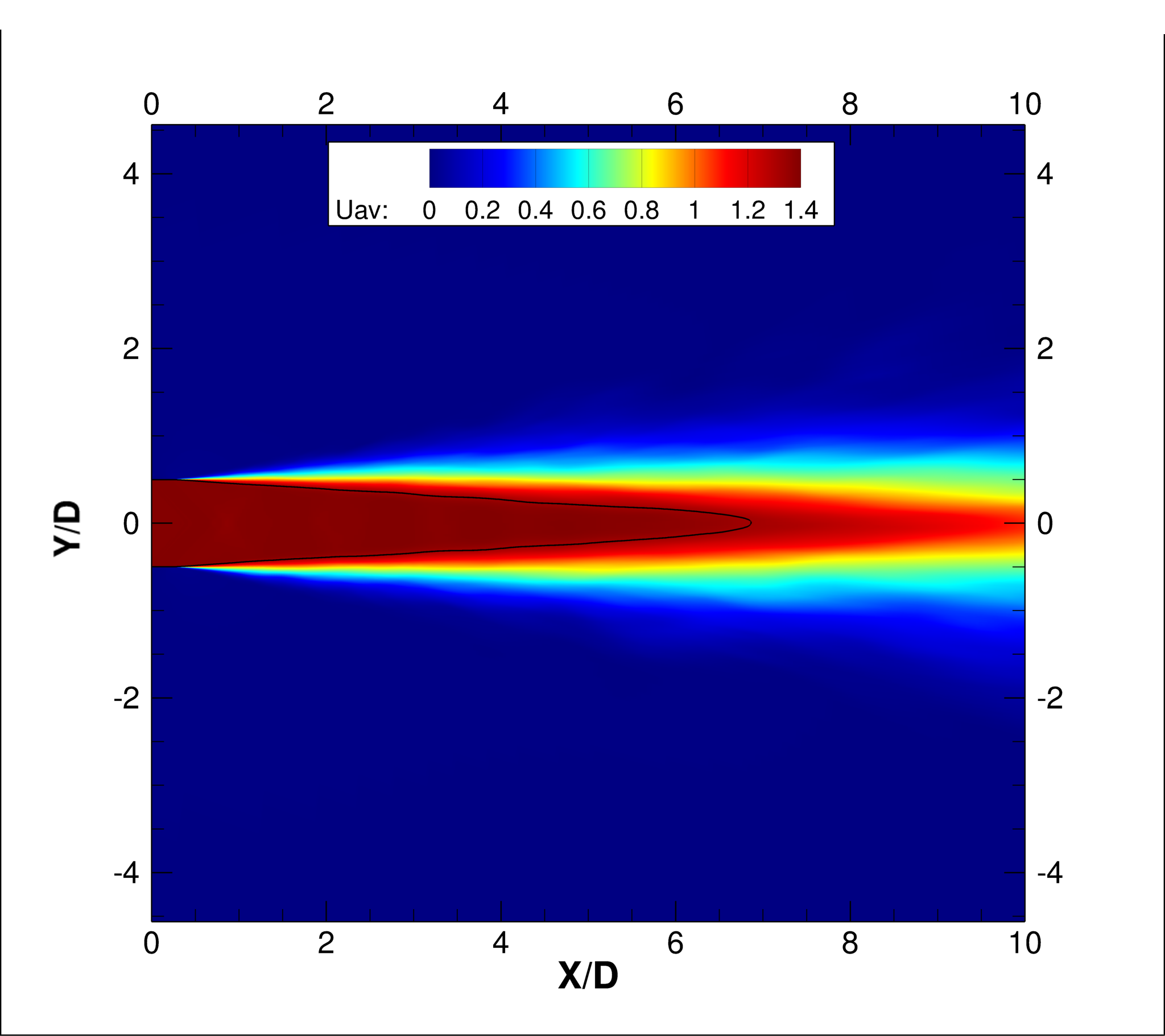}}
  \subfigure[Lateral view of $\langle U \rangle$ for simulation S4.]
    {\includegraphics[trim= 5mm 5mm 5mm 5mm, clip, width=0.32\textwidth]
	{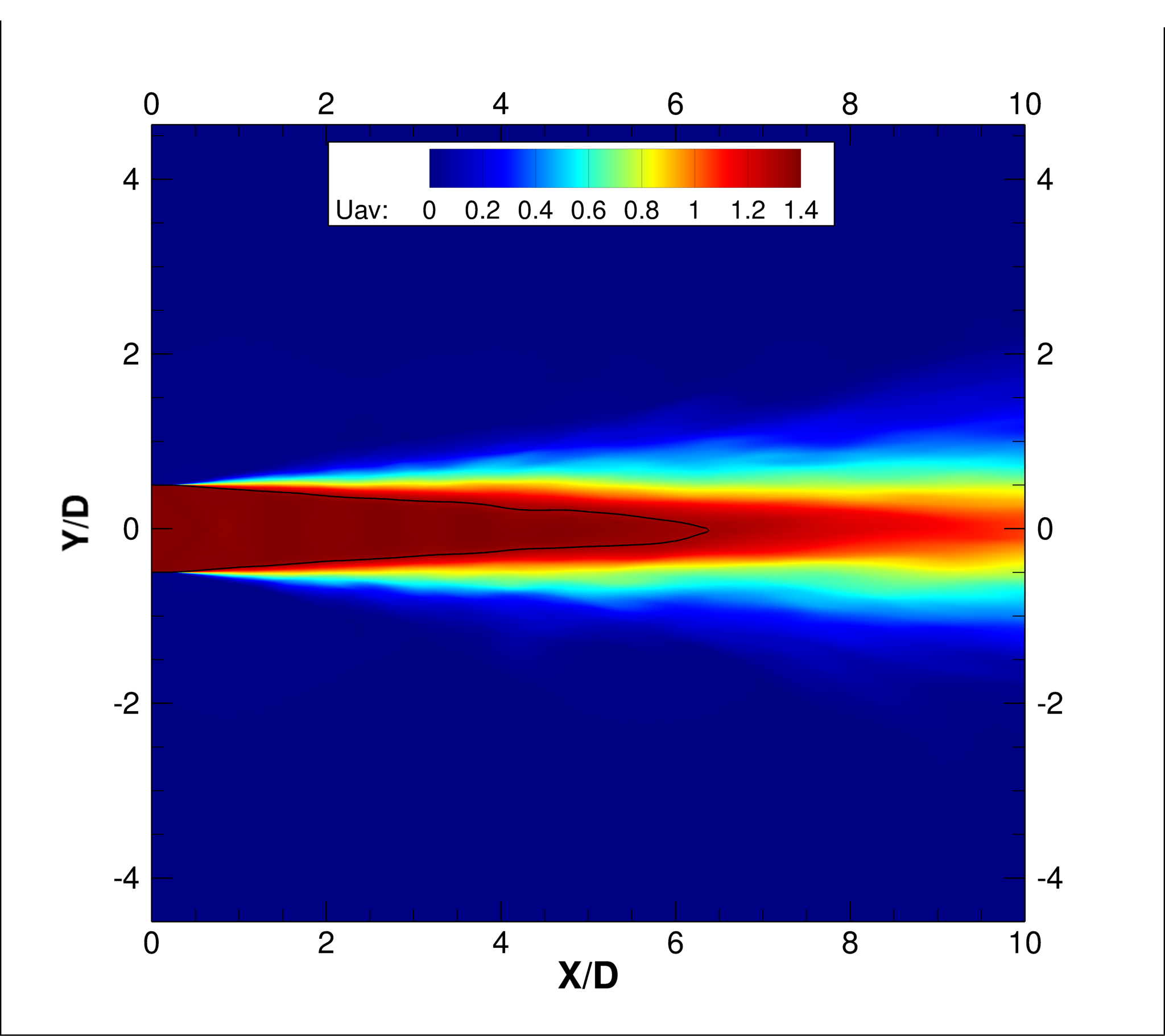}}
	\caption{Lateral view of the averaged axial component 
	of velocity, $\langle U \rangle$, for S2, S3 and S4 simulations.
	(\textbf{--}) indicates the 
	potential core of the jet, $U_{j}^{95\%}$.} 
	\label{fig:lat-u-av-sgs}
\end{figure}
Table \ref{tab:core-sgs} presents the size of 
the potential core of S2, S3 and S4 simulations and the numerical reference 
\cite{Mendez10,Mendez12} along with the relative error compared 
with the experimental data \cite{bridges2008turbulence}.
\begin{table}[htb!]
\begin{center}
  \caption{Potential core length and relative error of S2, S3 and S4 
	simulations.}
  \label{tab:core-sgs}
  \begin{tabular}{|c|c|c|}
  \hline
  Simulation & $\delta_{j}^{95\%}$ & Relative error\\
  \hline
  S2 & 6.84 & 26\%\\
  S3 & 6.84 & 26\%\\
  S4 & 6.28 & 32\%\\
  Mendez {\it et al.} & 8.35 & 8\%\\
  \hline
  \end{tabular}
\end{center}
\end{table}

Comparing the results, one cannot observe significant differences 
on the potential core length between S2, S3 and S4 simulations. The distribution
of $\langle U \rangle$ calculated using the dynamic Smagorinky model
has shown to be slightly more concentrated at the centerline region.
S2 and S4 simulations time averaged distribution of $U$ are, on some small 
scale, more spread than the distribution obtained by S3 simulation.
{  Figure \ref{fig:vort-zoom-sgs} presentes instantaneous fields of
pressure, in grey scale, colored by the vorticity magnitude for the
numerical simulations performed with the more refined grid, Mesh B.
Although the comparison is only qualitative, one cannot notice a 
significative effect of the SGS on the pressure fields and vorticity
magnitude of the flow. }
\begin{figure}[htb!]
  \centering
  \subfigure[S2 simulation.]
    {\includegraphics[width=0.9\textwidth]
	{./stat-cs-press-vort-snapshot-1.png}}
  \subfigure[S3 simulation.]
    {\includegraphics[width=0.9\textwidth]
	{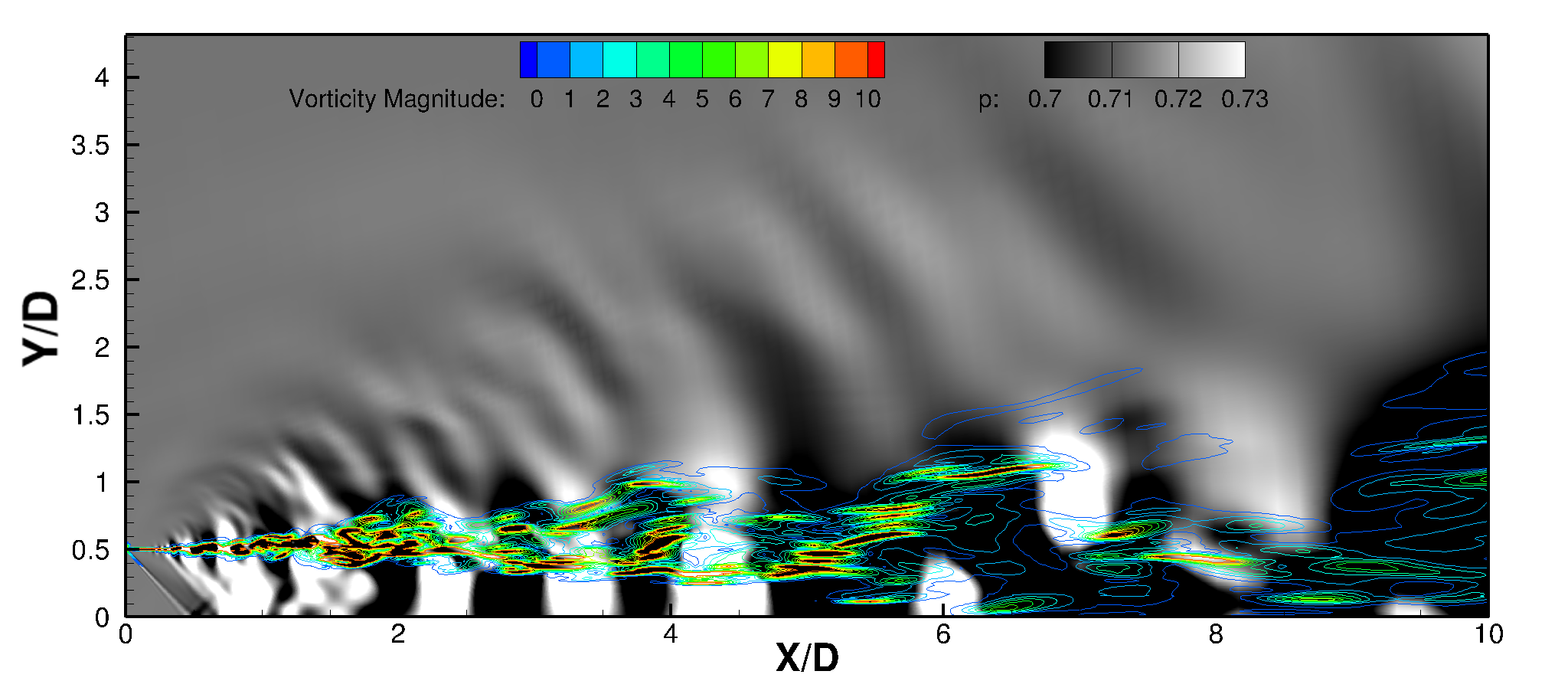}}
  \subfigure[S4 simulation.]
    {\includegraphics[width=0.9\textwidth]
	{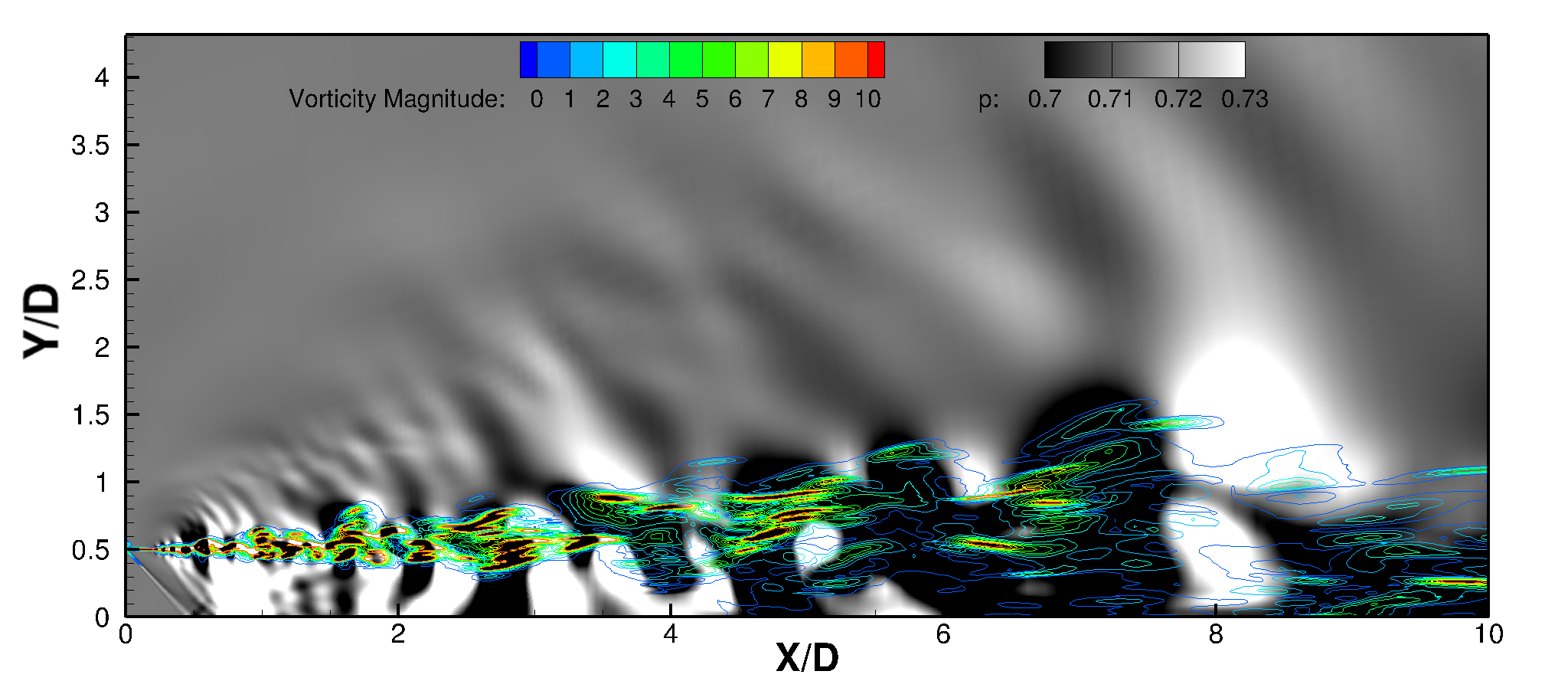}}
	\caption{{ Lateral view of field of snapshot of pressure,
	in grey scale, colored by the vorticity magnitude of 
	the jet flow for S2, S3 and S4 simulations.}}
	\label{fig:vort-zoom-sgs}
\end{figure}

Profiles of $\langle U \rangle$ from S2, S3 and S4 simulations, along the
mainstream direction 
are compared 
with numerical and experimental results in Fig.\ 
{ \ref{fig:prof-u-av-sgs-new}.
The evolution of $\langle U \rangle$ along the centerline is illustrated
in Fig.\ \ref{fig:prof-u-av-sgs-centerline}.}
The solid line, the dashed line and the circular symbol stand for the 
profiles of $\langle U \rangle$ computed by S2, S3 and S4 simulations, 
respectively. The reference data are represented by the same symbols 
presented in the mesh refinement study. 
%
%
\begin{figure}[htb!]
  \centering
  \subfigure[$\langle U \rangle$ - X=2.5D ; $-1.5D\leq Y\leq 1.5D$]
    {\includegraphics[trim= 5mm 5mm 5mm 5mm, clip, width=0.42\textwidth]
	{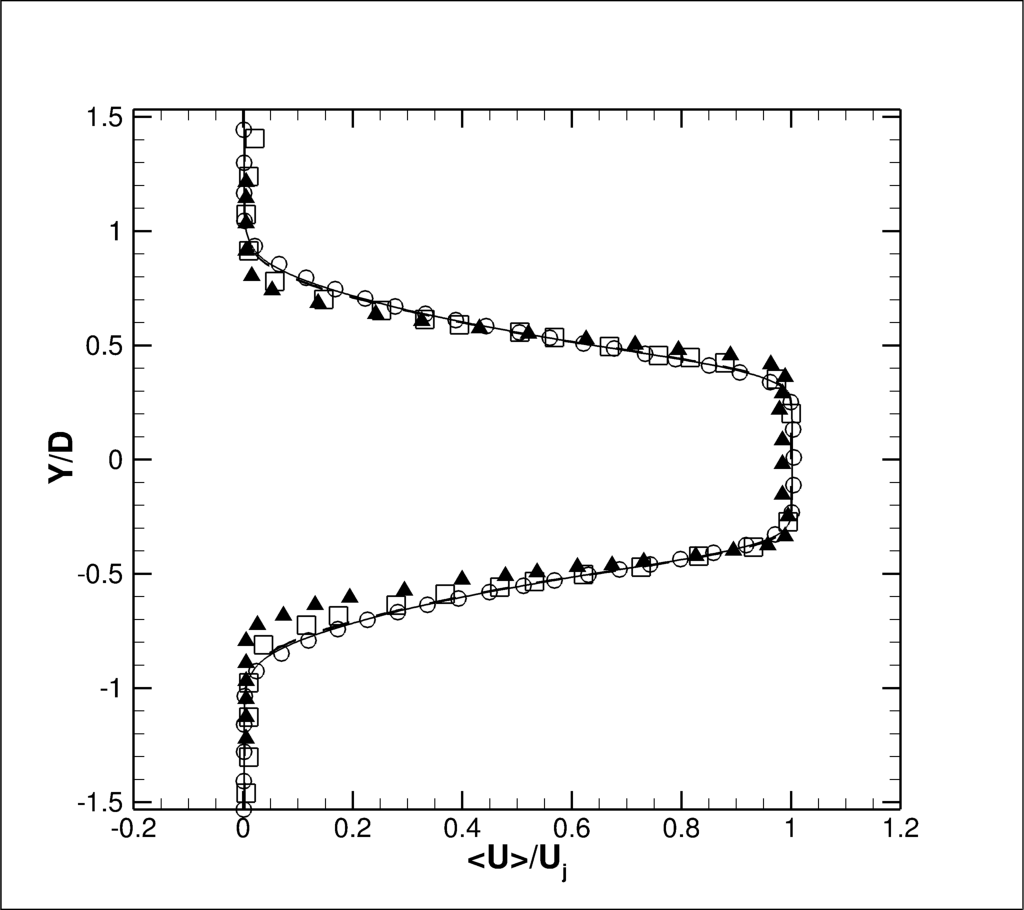}\label{fig:u-2-5-sgs-new}}
  \subfigure[$\langle U \rangle$ - X=5.0D ; $-1.5D\leq Y\leq 1.5D$]
    {\includegraphics[trim= 5mm 5mm 5mm 5mm, clip, width=0.42\textwidth]
	{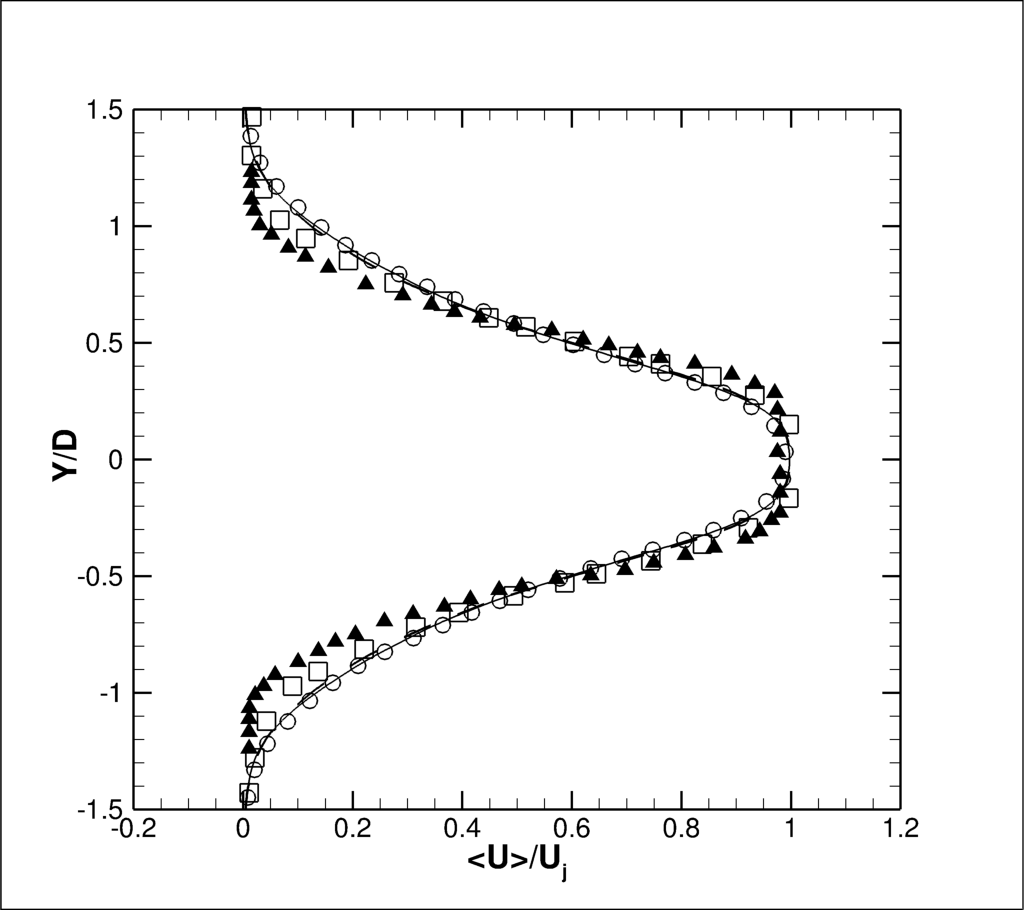}\label{fig:u-5-0-sgs-new}}
  \subfigure[$\langle U \rangle$ - X=10D ; $-1.5D\leq Y\leq 1.5D$]
    {\includegraphics[trim= 5mm 5mm 5mm 5mm, clip, width=0.42\textwidth]
	{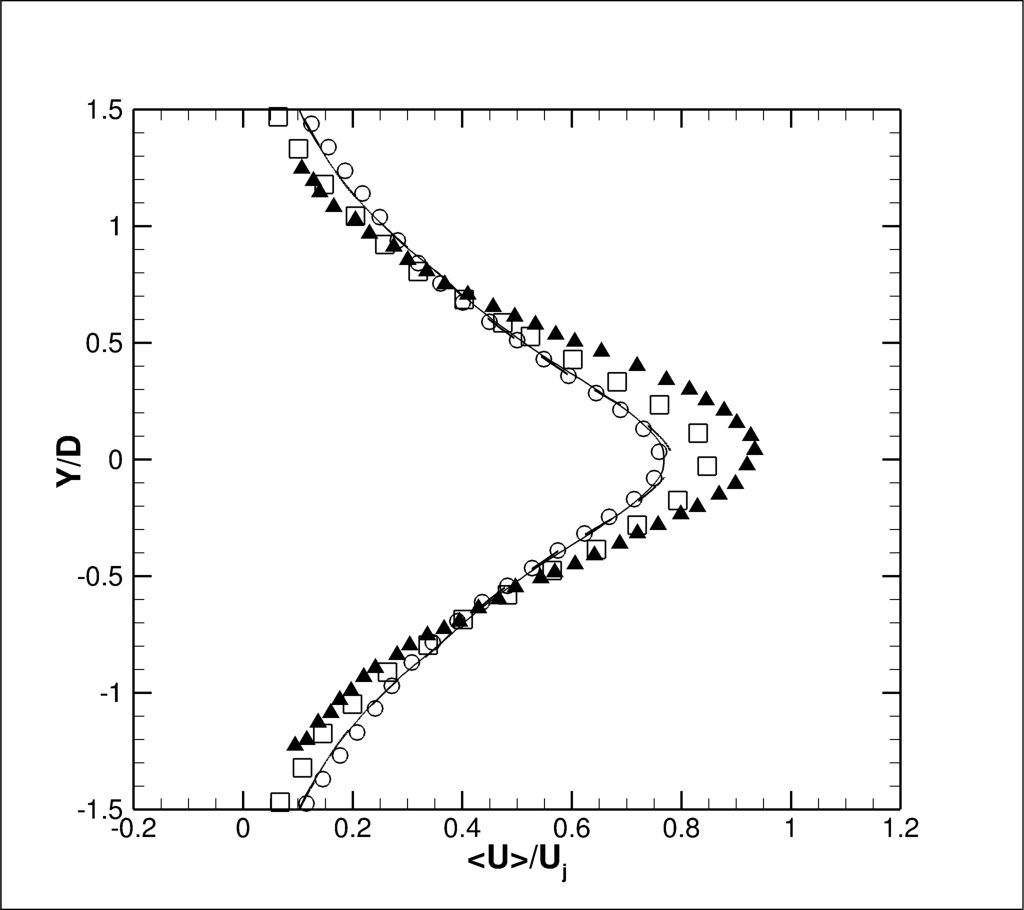}\label{fig:u-10-0-sgs-new}}
  \subfigure[$\langle U \rangle$ - X=15D ; $-1.5D\leq Y\leq 1.5D$]
    {\includegraphics[trim= 5mm 5mm 5mm 5mm, clip, width=0.42\textwidth]
	{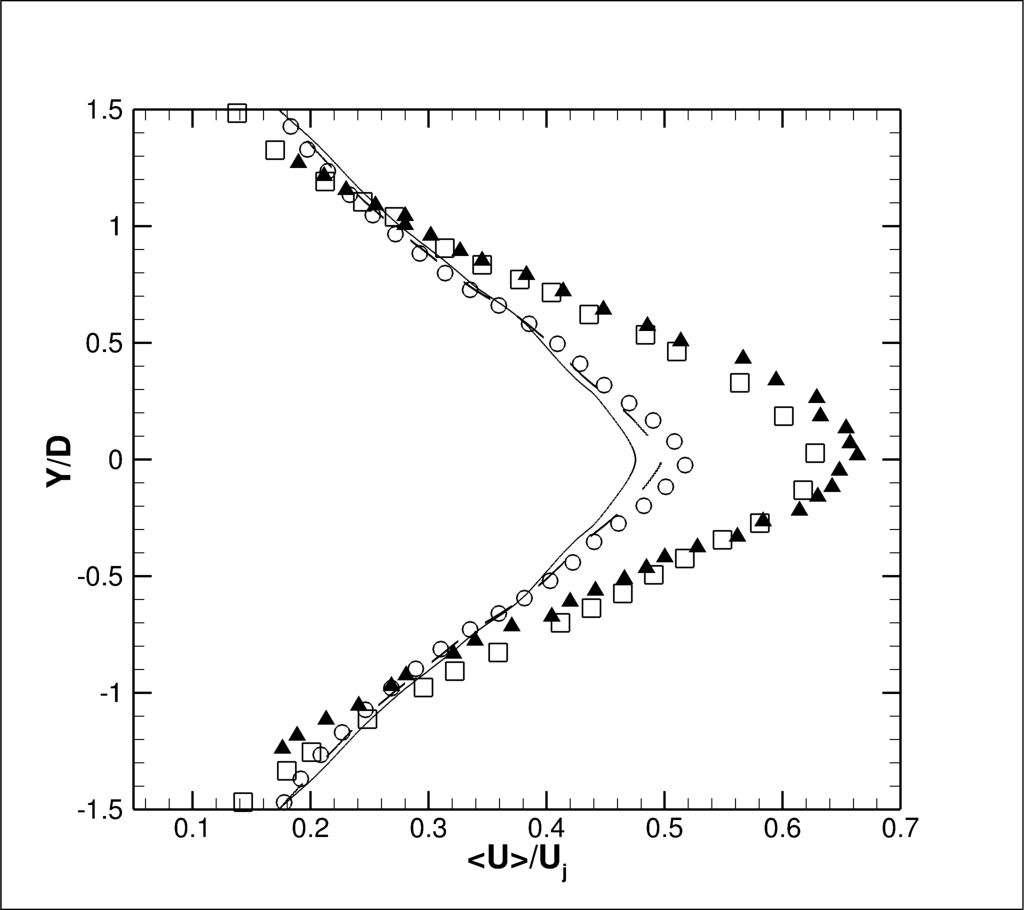}\label{fig:u-15-0-sgs-new}}
	\caption{{  Profiles of averaged axial component of velocity at 
	different positions within the computational domain.
	(\textbf{--}), S2 simulation; 
	(\textbf{-}\textbf{-}), S3 simulation;
	($\bigcirc$), S4 simulation; ($\square$), numerical data; 
	($\blacktriangle$), experimental data.}}
	\label{fig:prof-u-av-sgs-new}
\end{figure}
\begin{figure}[htb!]
  \centering
    {\includegraphics[trim= 5mm 5mm 5mm 5mm, clip, width=0.6\textwidth]
	{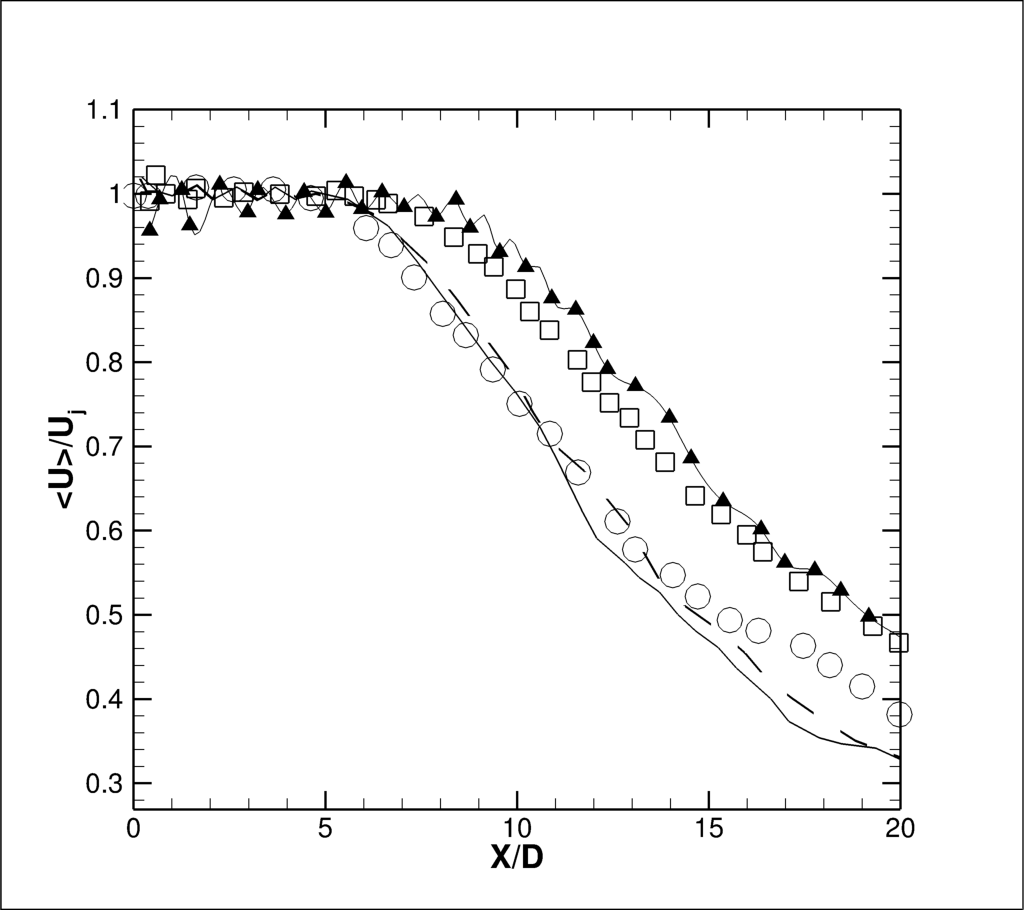}}
	\caption{{ Evolution of the averaged axial component of 
	velocity, $\langle U \rangle$, 	along the centerline (Y=0); 
	(\textbf{--}), S2 simulation; 
	(\textbf{-}\textbf{-}), S3 simulation;
	($\bigcirc$), S4 simulation; ($\square$), numerical data; 
	($\blacktriangle$), experimental data.}}
	\label{fig:prof-u-av-sgs-centerline}
\end{figure}
{  The four points at the lipline from the profiles 
presented in Figs.\ \ref{fig:u-2-5-sgs-new}, \ref{fig:u-5-0-sgs-new}, 
\ref{fig:u-10-0-sgs-new} and \ref{fig:u-15-0-sgs-new} are illustrated in 
Fig.\ \ref{fig:prof-u-av-sgs-lipline} as an evolution the averaged axial 
component of velocity, $\langle U \rangle$, along the lipline. A spline 
is used to create the curve using the four points extracted from the 
$\langle U \rangle$ profiles at $X/D=2.5$, $X/D=5.0$, $X/D=10$
and $X/D=15$. The black line stands for the numerical data, the red line for 
the experimental, the green line for the S2 simulation, the red line for
the S3 simulation and orange for the S4 simulation.}
\begin{figure}[htb!]
  \centering
    {\includegraphics[trim= 5mm 5mm 5mm 5mm, clip, width=0.6\textwidth]
	{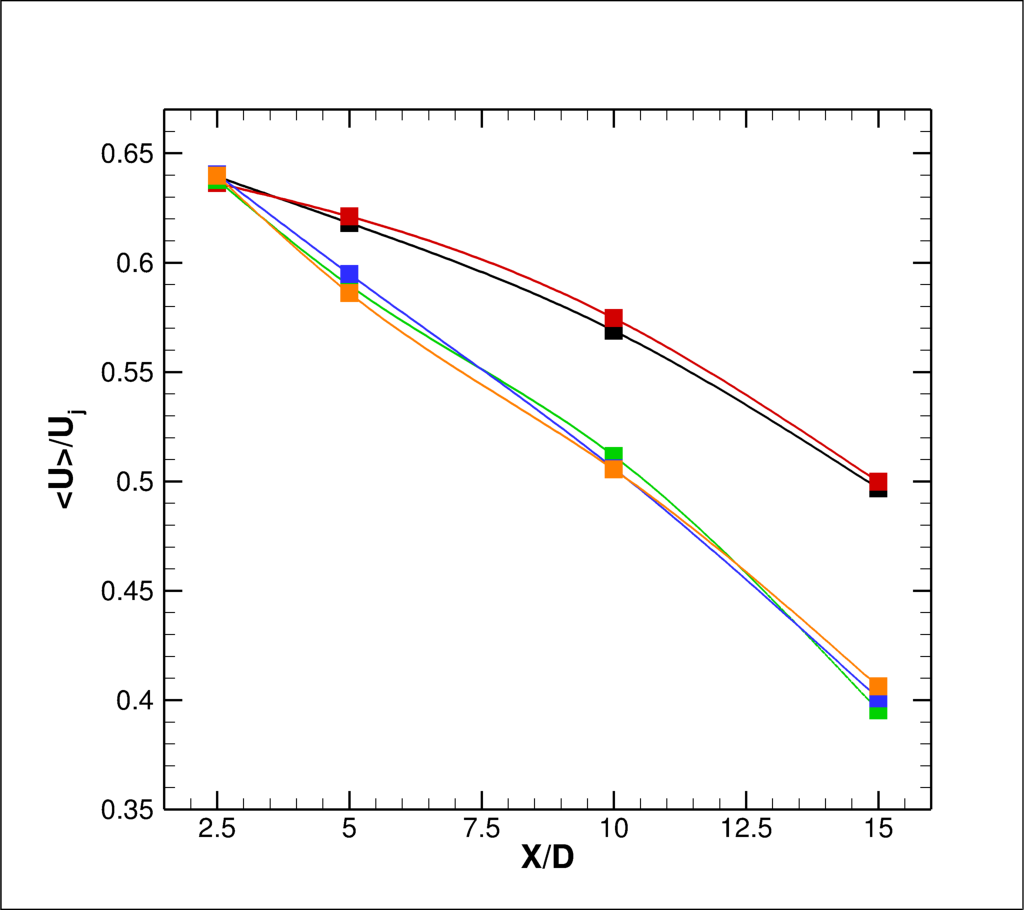}}
	\caption{{ Extractions of the averaged axial component of 
	velocity, 
	$\langle U \rangle$, along the lipline (Y/D=0.5); $0\leq X \leq 20D$.
	({\color{black}$\blacksquare$}) numerical data; 
	({ $\blacksquare$}) experimental data;
	({\color{green}$\blacksquare$}) S2 simulation.
	({\color{blue}$\blacksquare$}) S3 simulation.
	({\color{orange}$\blacksquare$}) S4 simulation.}}
	\label{fig:prof-u-av-sgs-lipline}
\end{figure}

The comparison of profiles indicates that distributions of 
$\langle U \rangle$ calculated on S2, S3 and S4 simulations correlates 
well with the references until $X=5.0D$. For $X>10.0D$ 
all SGS models understimate the magnitude of $\langle U\rangle$
at the $-0.5D<Y<0.5D$ region when the results are compared with 
the reference data. One can notice that the evolution of 
$\langle U\rangle$ along the centerline, calculated by all 
three simulations, are in good agreement with the numerical 
and experimental reference data at the region where the mesh 
presents a good resolution. Moreover, the three distributions 
calculated using different SGS closures have presented a very 
similar behavior. {  Results from S2, S3 and S4 
simulations along the lipline, presented in Fig.\ 
\ref{fig:prof-u-av-sgs-lipline}, also indicates a good correlation 
with the references data in the region where the mesh is more
refined. The difference of magnitude of $\langle U\rangle$ at 
the lipline is about $20\%$ at $X=15.0D$.}
%

\subsubsection{Root Mean Square Distribution of Time Fluctuations 
of Axial Velocity Component}

A lateral view of $u^{*}_{RMS}$ computed by S2, S3 and S4 simulations 
are presented in Figs.\ \ref{subfig:urms-sgs-s2}, \ref{subfig:urms-sgs-s3}
and \ref{subfig:urms-sgs-s4}, respectively. 
\begin{figure}[htb!]
  \centering
  \subfigure[Lateral view of $u^{*}_{RMS}$ for S2 simulation.]
    {\includegraphics[trim= 5mm 5mm 5mm 5mm, clip, width=0.32\textwidth]
	{./stat-cs-XY-zoom-u-rms.png}\label{subfig:urms-sgs-s2}}
  \subfigure[Lateral view of $u^{*}_{RMS}$ for S3 simulation.]
    {\includegraphics[trim= 5mm 5mm 5mm 5mm, clip, width=0.32\textwidth]
	{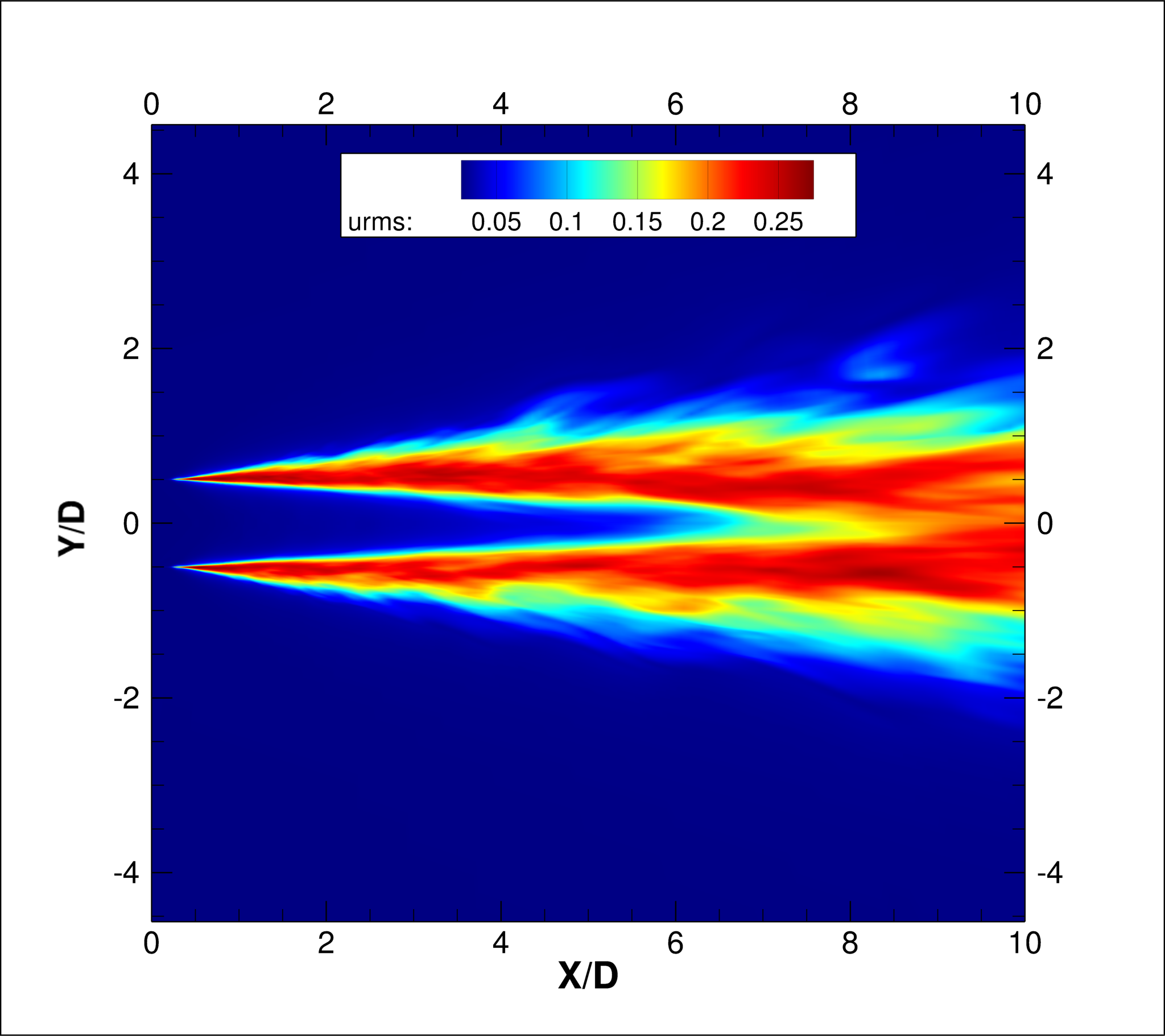}\label{subfig:urms-sgs-s3}}
  \subfigure[Lateral view of $u^{*}_{RMS}$ for S4 simulation.]
    {\includegraphics[trim= 5mm 5mm 5mm 5mm, clip, width=0.32\textwidth]
	{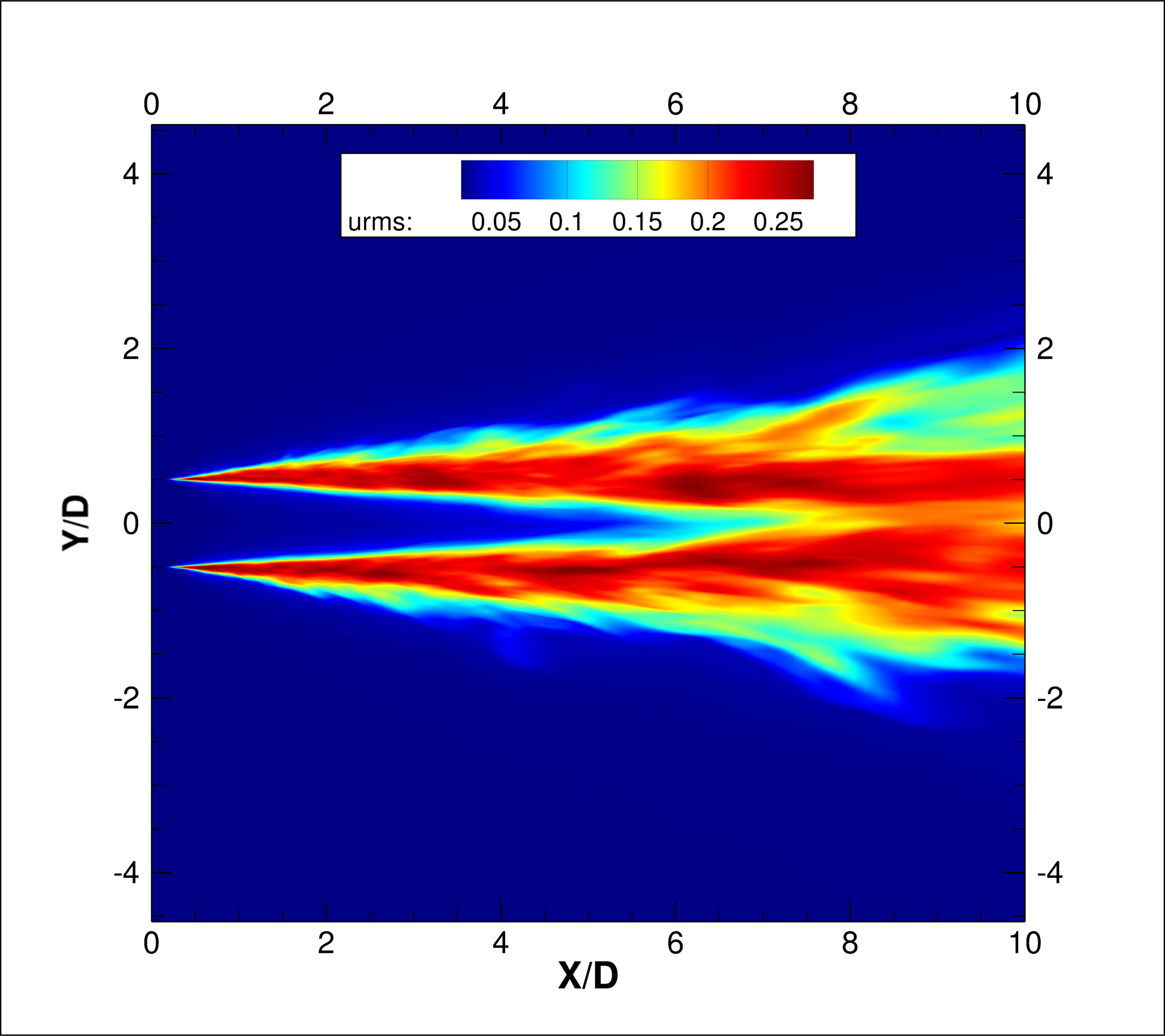}\label{subfig:urms-sgs-s4}}
  \subfigure[Lateral view of $v^{*}_{RMS}$ for S2 simulation.]
    {\includegraphics[trim= 5mm 5mm 5mm 5mm, clip, width=0.32\textwidth]
	{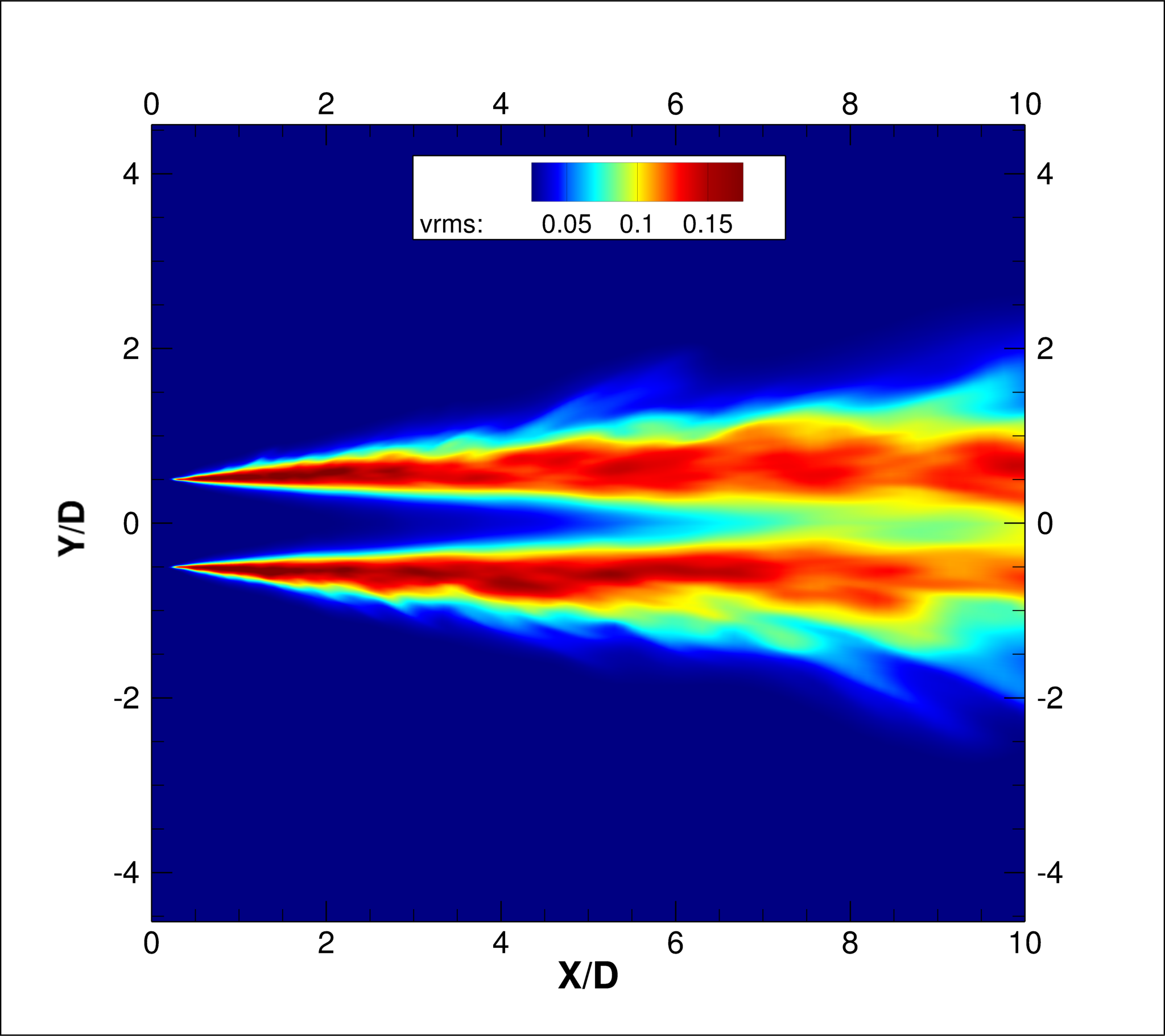}\label{subfig:vrms-sgs-s2}}
  \subfigure[Lateral view of $v^{*}_{RMS}$ for S3 simulation.]
    {\includegraphics[trim= 5mm 5mm 5mm 5mm, clip, width=0.32\textwidth]
	{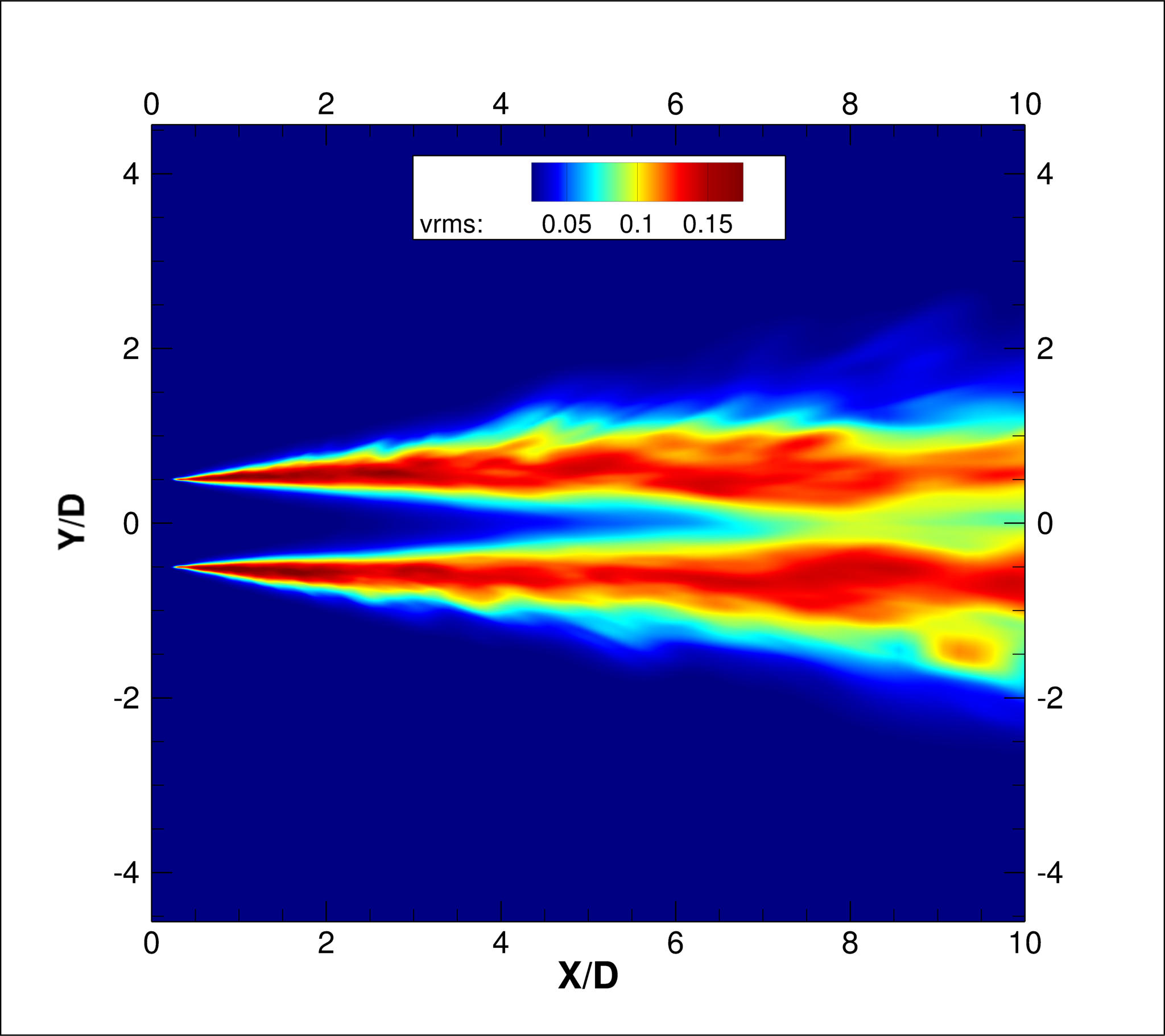}\label{subfig:vrms-sgs-s3}}
  \subfigure[Lateral view of $v^{*}_{RMS}$ for S4 simulation.]
    {\includegraphics[trim= 5mm 5mm 5mm 5mm, clip, width=0.32\textwidth]
	{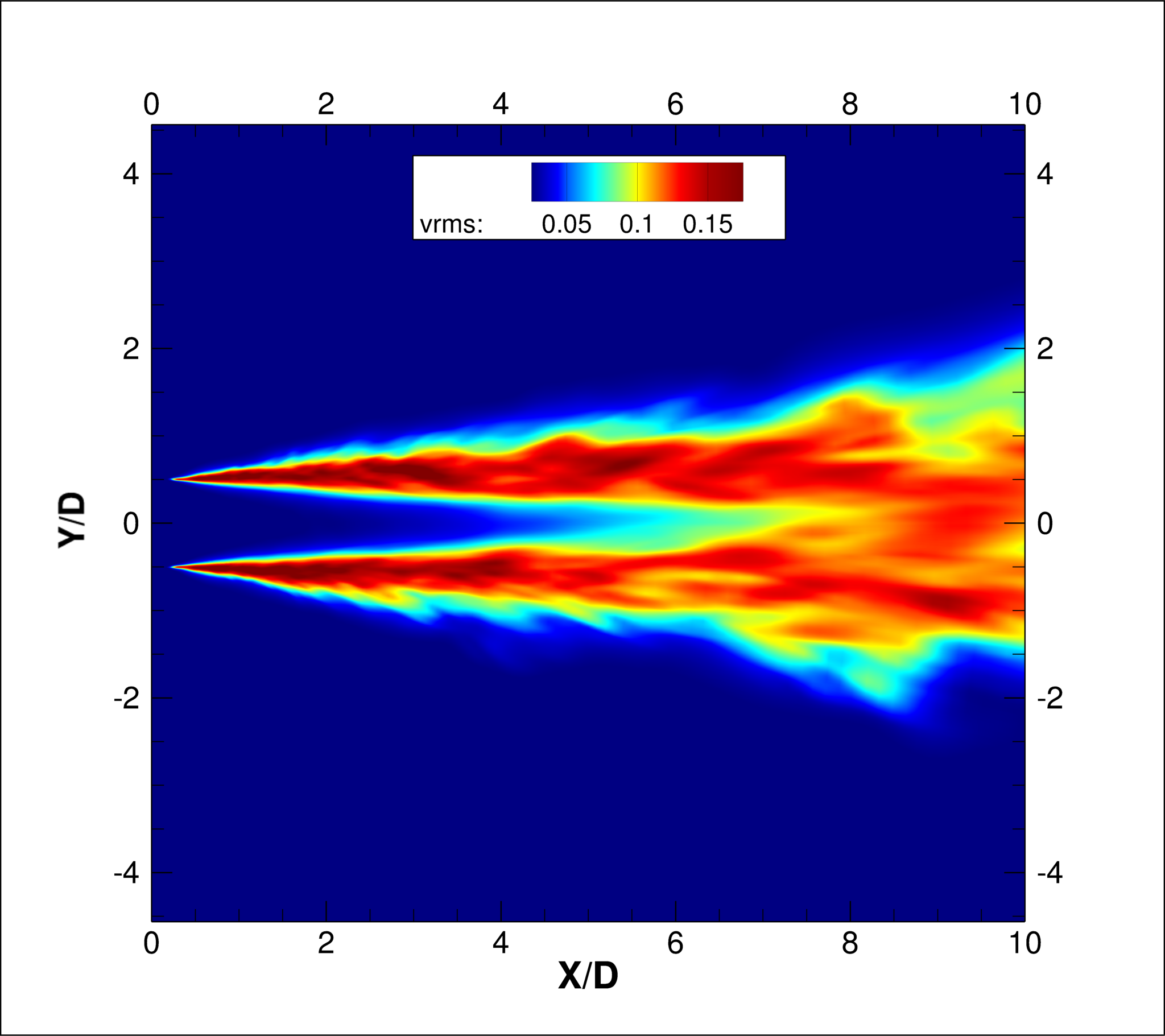}\label{subfig:vrms-sgs-s4}}
  \subfigure[Lateral view of $\langle u^{*}v^{*}\rangle$ for S2 simulation.]
    {\includegraphics[trim= 5mm 5mm 5mm 5mm, clip, width=0.32\textwidth]
	{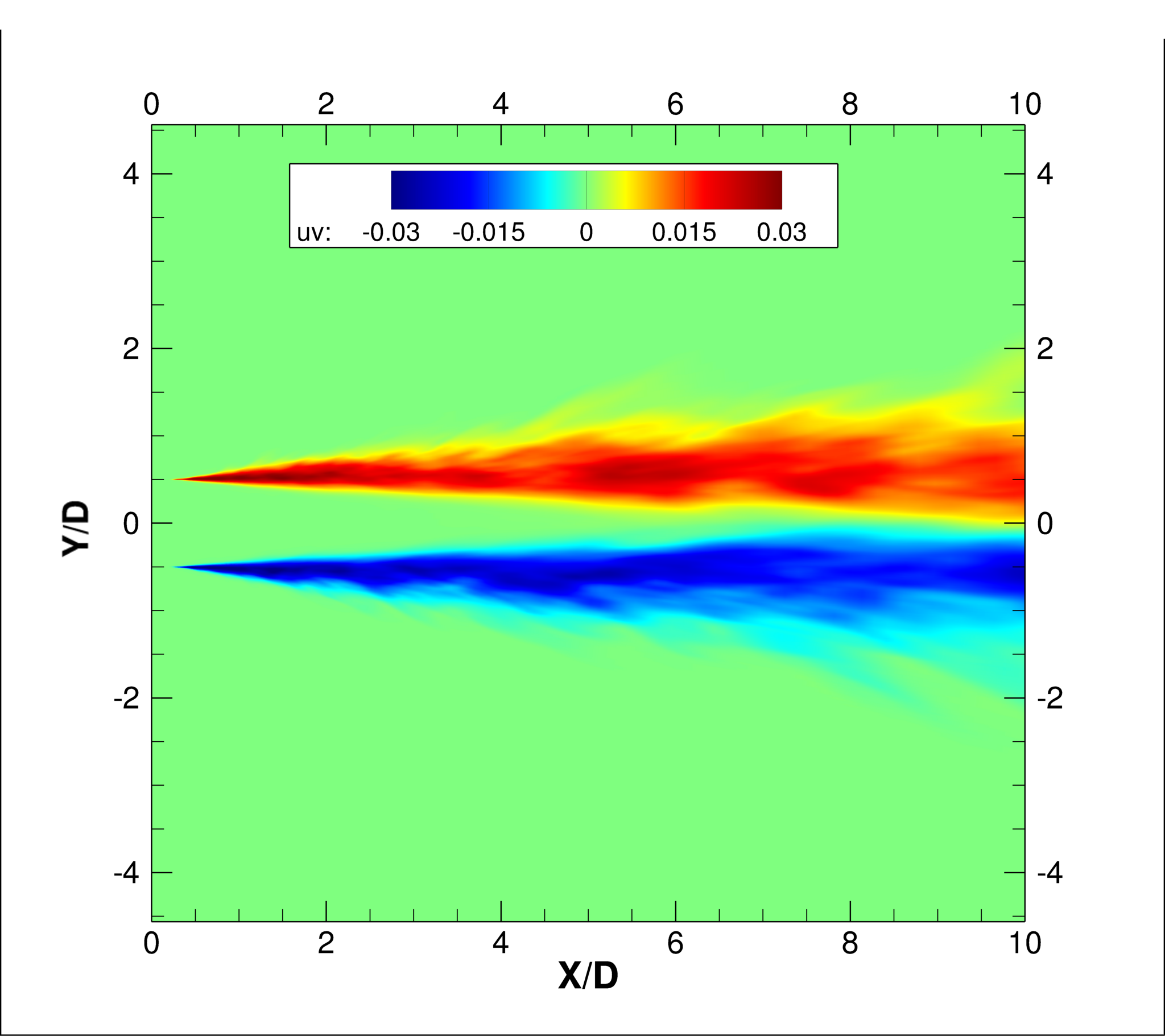}\label{subfig:uv-sgs-s2}}
  \subfigure[Lateral view of $\langle u^{*}v^{*}\rangle$ for S3 simulation.]
    {\includegraphics[trim= 5mm 5mm 5mm 5mm, clip, width=0.32\textwidth]
	{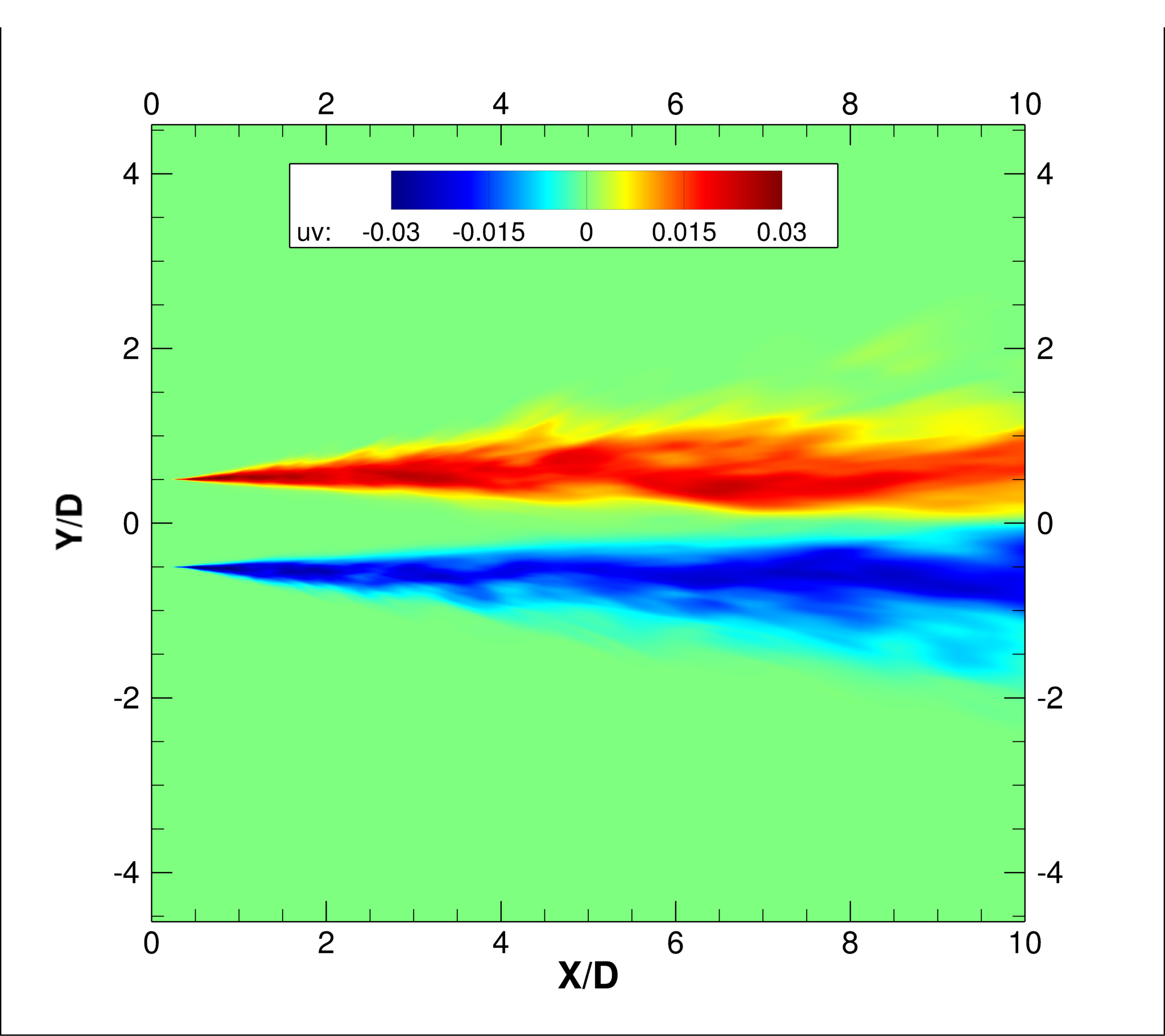}\label{subfig:uv-sgs-s3}}
  \subfigure[Lateral view of $\langle u^{*}v^{*}\rangle$ for S4 simulation.]
    {\includegraphics[trim= 5mm 5mm 5mm 5mm, clip, width=0.32\textwidth]
	{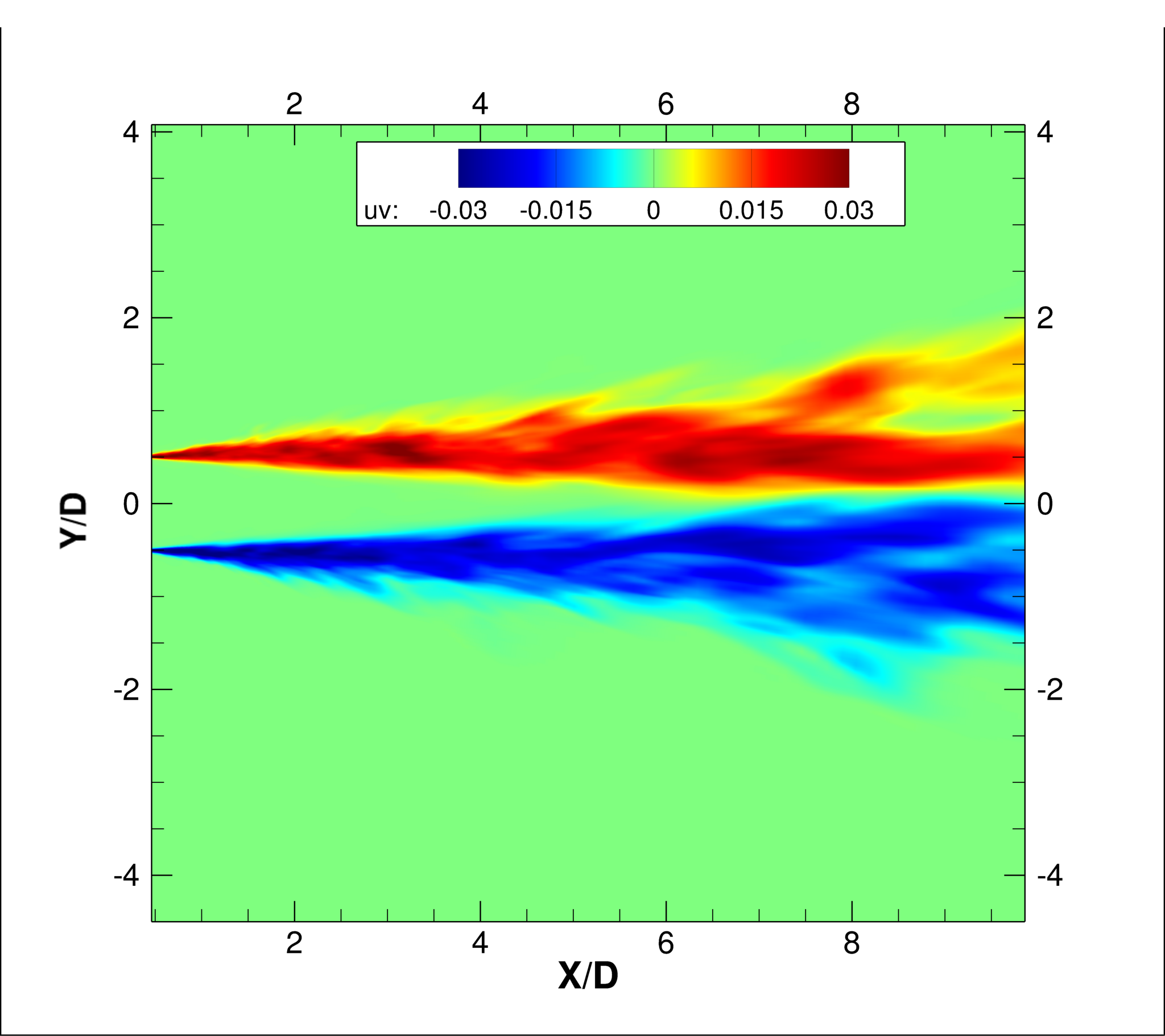}\label{subfig:uv-sgs-s4}}
	\caption{Lateral view of RMS of time fluctuation
	of axial component of velocity, $u_{RMS}^{*}$, RMS 
	of time fluctuation of radial component of velocity, 
	$v_{RMS}^{*}$ and $\langle u^{*}v^{*}\rangle$ Reynolds 
	shear stress tensor component, for S2, S3 and S4 simulations.} 
	\label{fig:lat-u-v-uv-sgs}
\end{figure}
{  The profiles of $u^{*}_{RMS}$ at $X=2.5D$ obtained by S2, S3 
and S4 simulations are in good agreement with the numerical reference until
$X=5.0D$, as one can observe in Fig.\ \ref{fig:prof-u-rms-sgs-new}. For
$X>10D$ the results achieved using the more refined mesh overestimate
the magnitude of $u^{*}_{RMS}$ in the region where $-0.5D<Y<0.5D$. One should
notice that} all simulations, including the LES reference, present difficulties 
to predict the peaks of $u^{*}_{RMS}$.
%
%
\begin{figure}[htb!]
  \centering
  \subfigure[$u^{*}_{RMS}$ - X=2.5D ; $-1.5D\leq Y\leq 1.5D$]
    {\includegraphics[trim= 5mm 5mm 5mm 5mm, clip, width=0.42\textwidth]
	{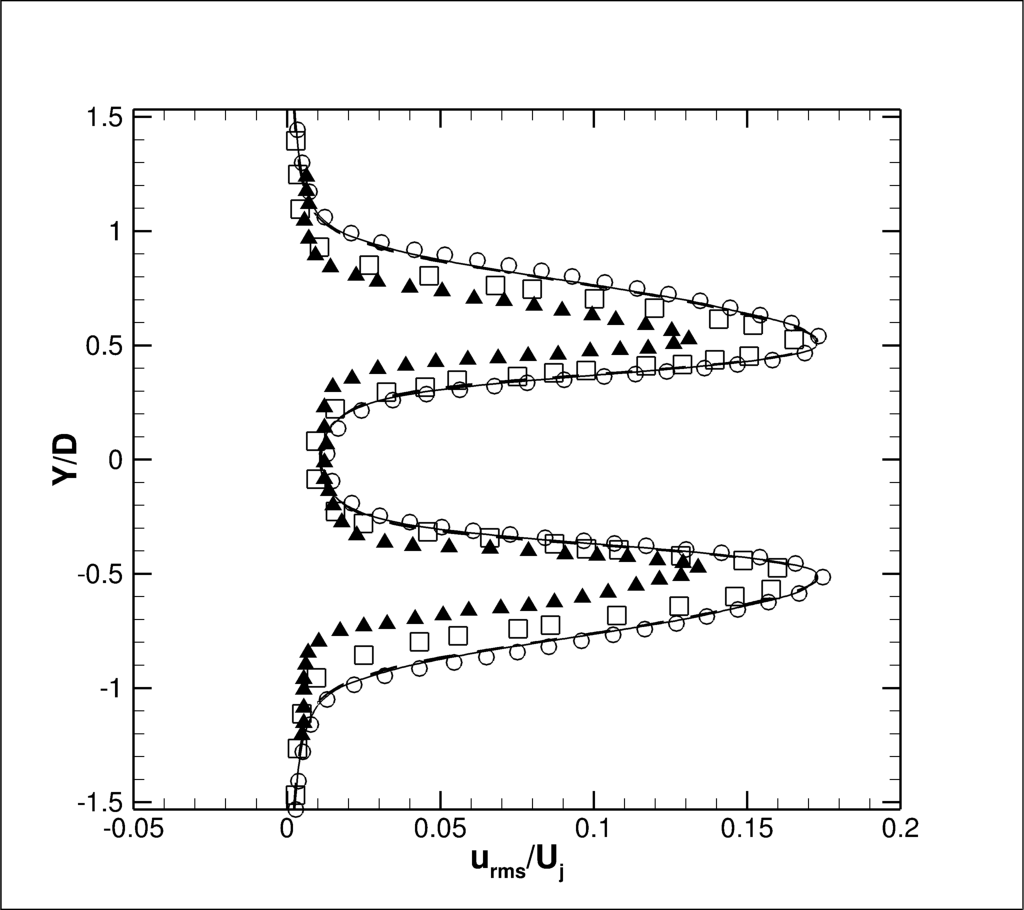}\label{fig:urms-2-5-sgs-new}}
  \subfigure[$u^{*}_{RMS}$ - X=5.0D ; $-1.5D\leq Y\leq 1.5D$]
    {\includegraphics[trim= 5mm 5mm 5mm 5mm, clip, width=0.42\textwidth]
	{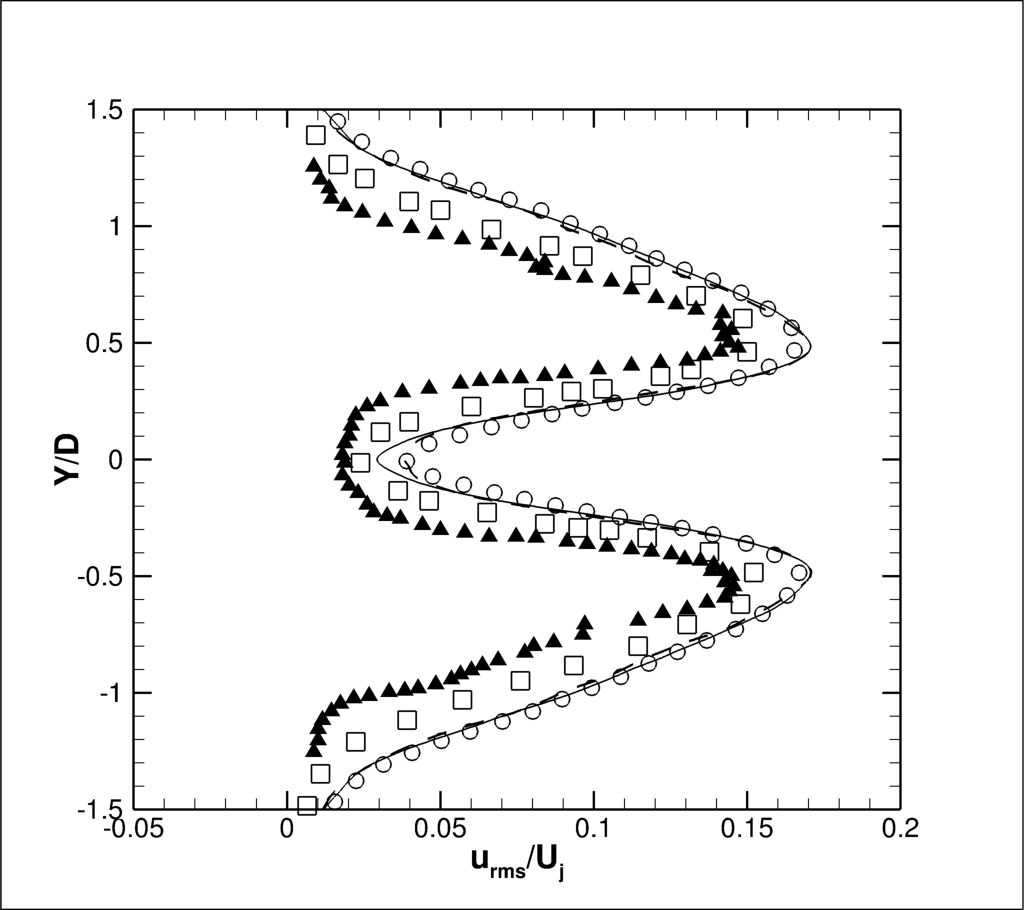}\label{fig:urms-5-0-sgs-new}}
  \subfigure[$u^{*}_{RMS}$ - X=10D ; $-1.5D\leq Y\leq 1.5D$]
    {\includegraphics[trim= 5mm 5mm 5mm 5mm, clip, width=0.42\textwidth]
	{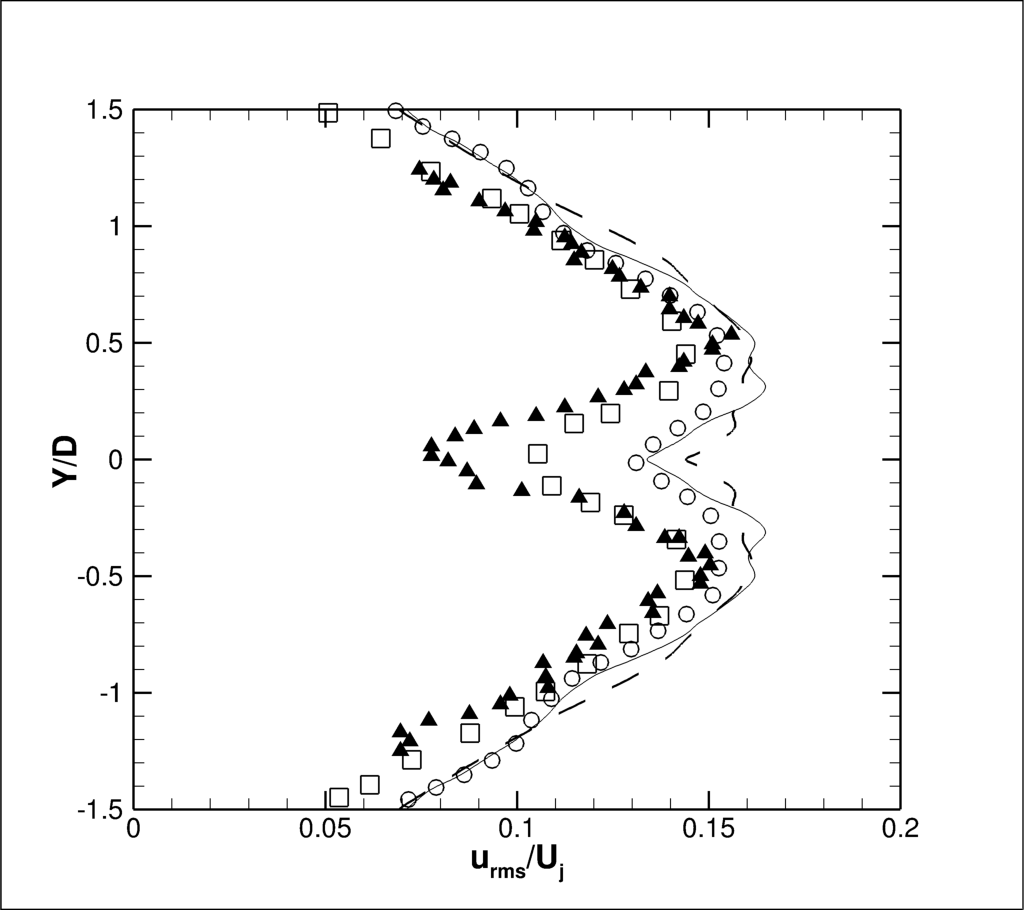}\label{fig:urms-10-0-sgs-new}}
  \subfigure[$u^{*}_{RMS}$ - X=15D ; $-1.5D\leq Y\leq 1.5D$]
    {\includegraphics[trim= 5mm 5mm 5mm 5mm, clip, width=0.42\textwidth]
	{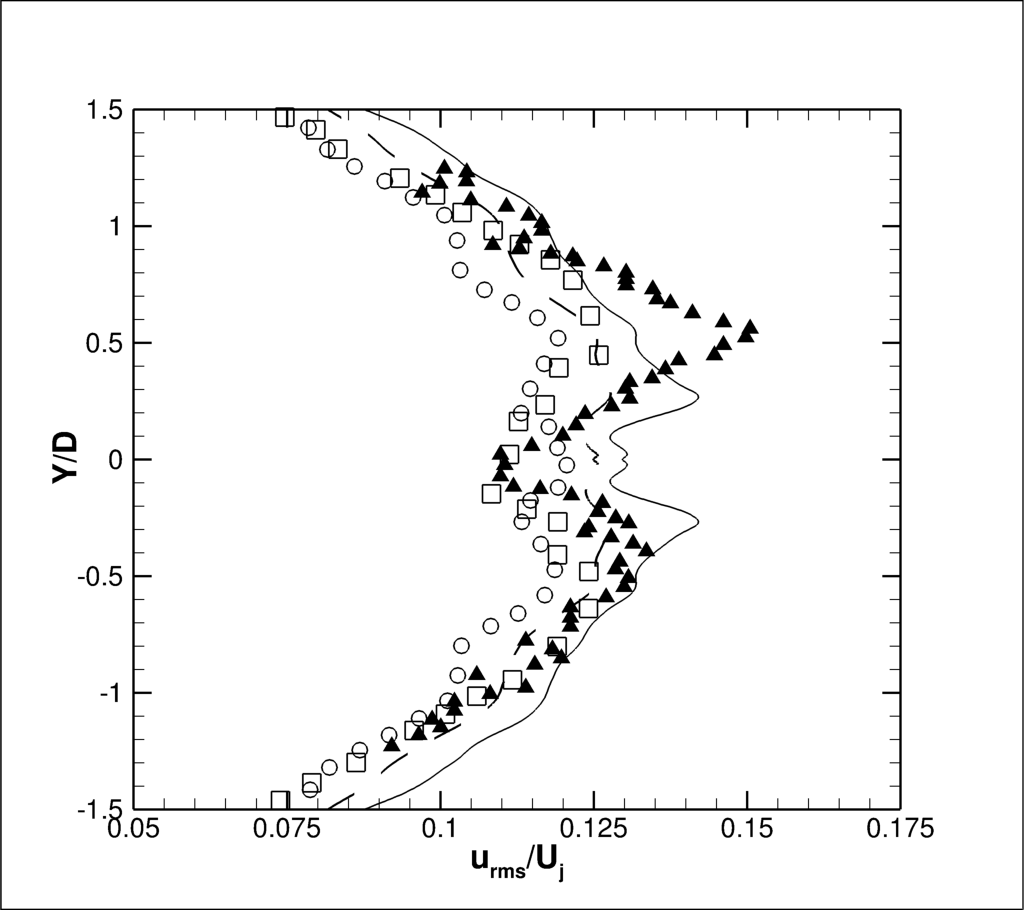}\label{fig:urms-15-0-sgs-new}}
	\caption{{ Profiles of RMS of time fluctuation of axial 
	component of velocity, $u_{RMS}^{*}$, for S2, S3 and S4,
	at different positions within the computational domain.
	(\textbf{--}), S2; 
	(\textbf{-}\textbf{-}), S3;
	($\bigcirc$), S4; ($\square$), numerical data; 
	($\blacktriangle$), experimental data.}}
	\label{fig:prof-u-rms-sgs-new}
\end{figure}
%

{  Figure \ref{fig:prof-u-rms-sgs-centerline} presents the
distribution of $u_{RMS}^{*}$ along the centerline of the domain.} 
All three simulations performed in the current work present overestimated 
distributions of $u_{RMS}^{*}$ along the centerline. However, for 
$10D<X<15D$, the Vreman model correctly reproduces the magnitude of 
$u_{RMS}^{*}$. 
\begin{figure}[htb!]
  \centering
    {\includegraphics[trim= 5mm 5mm 5mm 5mm, clip, width=0.6\textwidth]
	{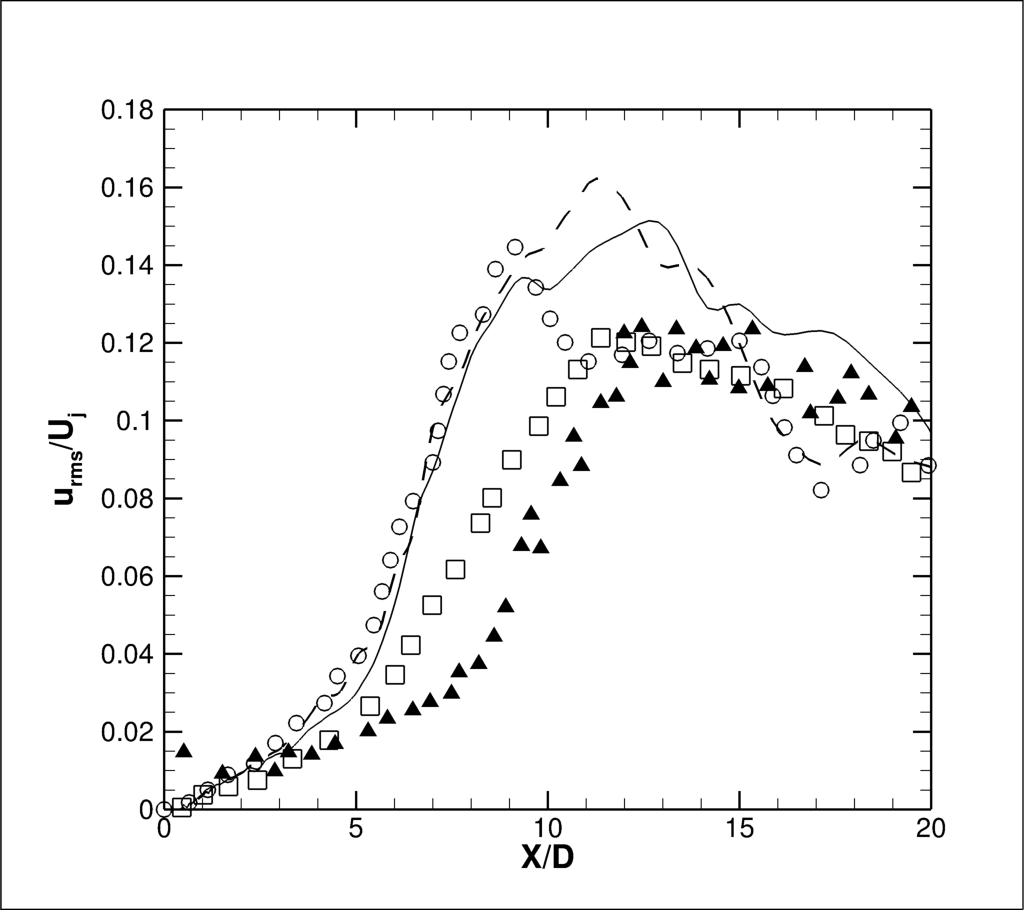}}
	\caption{{ Evolution of RMS values of time fluctuation of axial 
	component of velocity, $u_{RMS}^{*}$, along the centerline (Y=0); 
	(\textbf{--}), S2; 
	(\textbf{-}\textbf{-}), S3;
	($\bigcirc$), S4; ($\square$), numerical data; 
	($\blacktriangle$), experimental data.}}
	\label{fig:prof-u-rms-sgs-centerline}
\end{figure}

{  The same strategy used in Fig.\ \ref{fig:prof-u-av-sgs-lipline}
is used to evaluate the magnitude of $u_{RMS}^{*}$ at four different
points along the centerline. Figure \ref{fig:prof-urms-sgs-lipline} presents
the points are extracted from Figs.\ \ref{fig:u-2-5-sgs-new}, 
\ref{fig:u-5-0-sgs-new}, \ref{fig:u-10-0-sgs-new} and 
\ref{fig:u-15-0-sgs-new}. A spline is used to create the curve using the 
four points extracted from the $u_{RMS}^{*}$ at $X/D=2.5$, $X/D=5.0$, $X/D=10$
and $X/D=15$. The black line stands for the numerical data, the red line for 
the experimental, the green line for the S2 simulation, the blue line for
the S3 simulation and orange for the S4 simulation. All numerical simulations,
including the numerical reference, overpredict the fluctuations of 
$u_{RMS}^{*}$ at the lipline for $X=2.5D$ and $X=5.0D$. The Vreman model
is the SGS model which presents better results of the peak of $u_{RMS}^{*}$
at the lipline for $X=10.0D$ when compared with the experimental reference.
At $X=15.0D$ the dynamic Smagorinsky model presents the best prediction of
the fluctuation of the axial velocity component at the the lipline. However, 
the other turbulent models present results which differ about $10\%$ from
experimental data at the same point. The three SGS models provide similiar
behavior of $u_{RMS}^{*}$ at the lipline region.}
%
%
\begin{figure}[htb!]
  \centering
    {\includegraphics[trim= 5mm 5mm 5mm 5mm, clip, width=0.6\textwidth]
	{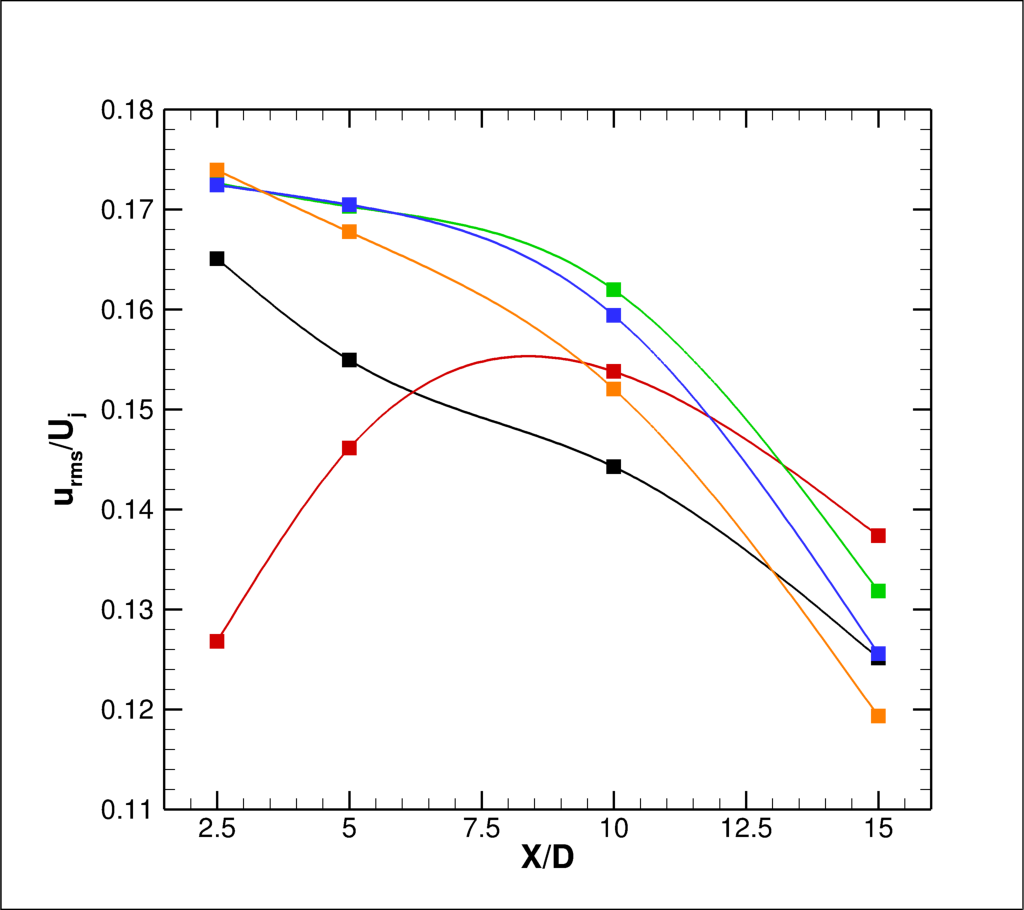}}
	\caption{{ Extractions of RMS values of time fluctuation 
	of axial component of velocity, $u_{RMS}^{*}$, along the lipline 
	(Y/D=0.5); $0\leq X \leq 20D$.
	({\color{black}$\blacksquare$}) numerical data; 
	({ $\blacksquare$}) experimental data;
	({\color{green}$\blacksquare$}) S2 simulation.
	({\color{blue}$\blacksquare$}) S3 simulation.
	({\color{orange}$\blacksquare$}) S4 simulation.}}
	\label{fig:prof-urms-sgs-lipline}
\end{figure}

\subsubsection{Root Mean Square Distribution of Time Fluctuations 
of Radial Velocity Component}

Effects of SGS modeling on the time fluctuation of the radial 
component of velocity are also compared with the reference data. 
Figures \ref{subfig:vrms-sgs-s2}, \ref{subfig:vrms-sgs-s3} and
\ref{subfig:vrms-sgs-s4} illustrate a lateral view of the 
distribution of $v_{RMS}^{*}$ computed by S2, S3 and S4 simulations, 
respectively. The choice of SGS model does not significantly affect 
the distribution of $v_{RMS}^{*}$. All distributions calculated by S2, 
S3 and S4 simulations have shown similar behavior.

Four profiles of $v_{RMS}^{*}$ at $X=2.5D$, 
$X=5.0D$, $X=10.0D$ and $X=15.0D$ are presented in Fig.\ 
{  \ref{fig:prof-v-rms-sgs-new}. 
One can observe that all the profiles calculated on S2, S3 and S4 
simulations are close to the reference at $X=2.5D$. The results correlates 
better with the experimental data than the numerical reference does at 
$X=2.5D$. The profile of $v_{RMS}^{*}$ calculated by S2, S3 and S4
at $X=5.0D$ are close to the experimental results. However, The numerical 
reference and the simulations performed in the present work, using the more 
refined mesh, present difficulties to predict the peaks of $v_{RMS}^{*}$ at 
the lipline region. For $X\geq10.0D$ the S2, S3 and S4 computations fails to 
represent the profiles of $v_{RMS}^{*}$. The results are fairly underpredicted 
when comprared with the reference results due to the mesh coarsening.}
%
%
%
\begin{figure}[htb!]
  \centering
  \subfigure[$v^{*}_{RMS}$ - X=2.5D ; $-1.5D\leq Y\leq 1.5D$]
    {\includegraphics[trim= 5mm 5mm 5mm 5mm, clip, width=0.45\textwidth]
	{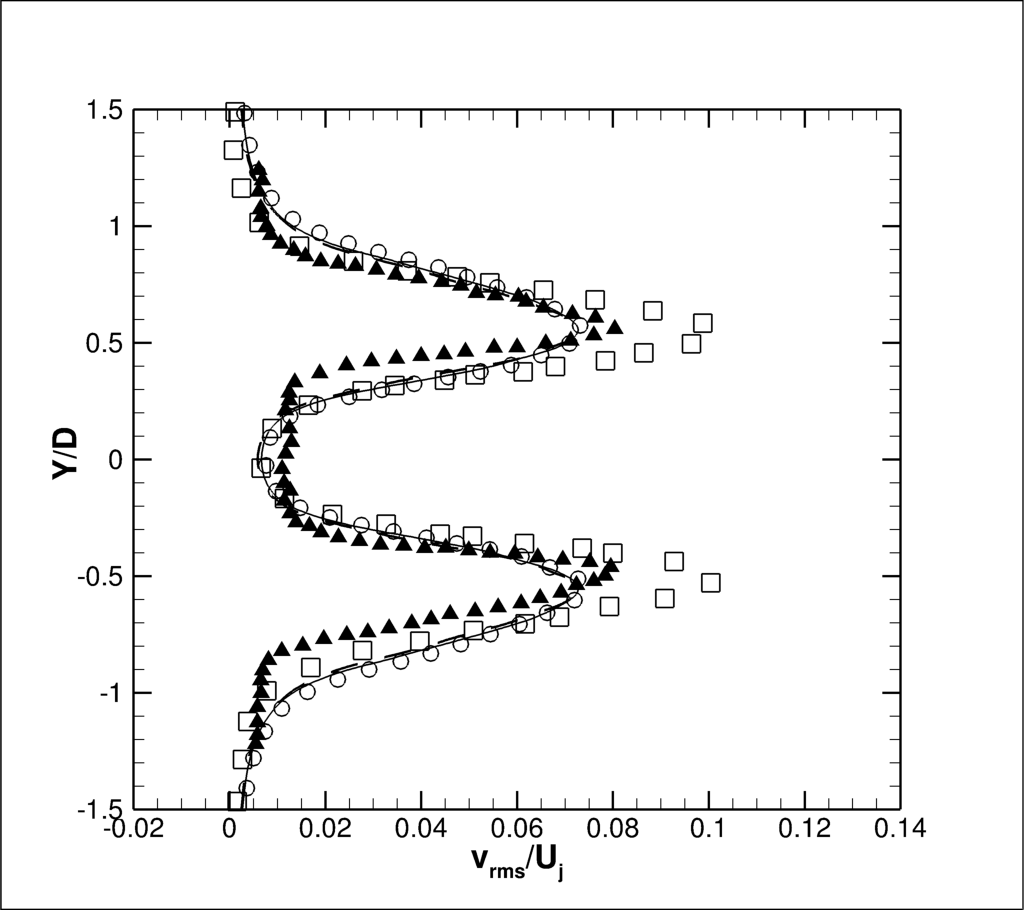}}
  \subfigure[$v^{*}_{RMS}$ - X=5.0D ; $-1.5D\leq Y\leq 1.5D$]
    {\includegraphics[trim= 5mm 5mm 5mm 5mm, clip, width=0.45\textwidth]
	{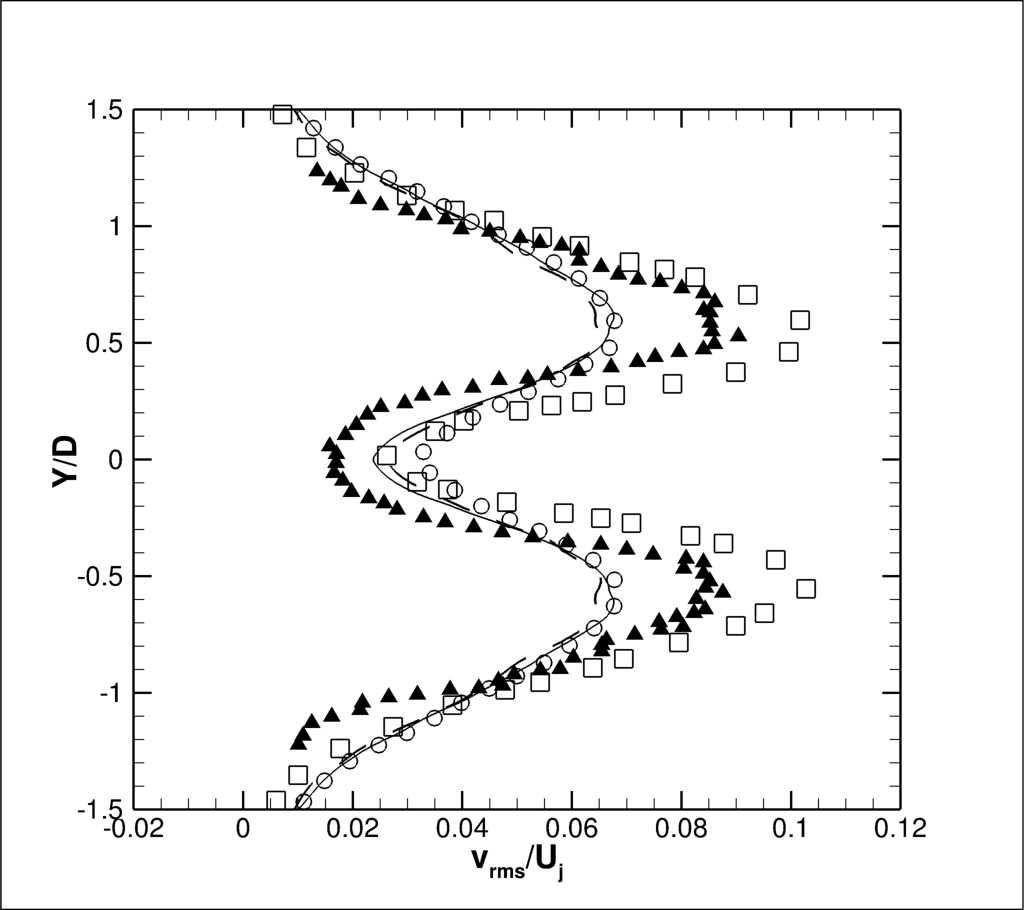}}
  \subfigure[$v^{*}_{RMS}$ - X=10D ; $-1.5D\leq Y\leq 1.5D$]
    {\includegraphics[trim= 5mm 5mm 5mm 5mm, clip, width=0.45\textwidth]
	{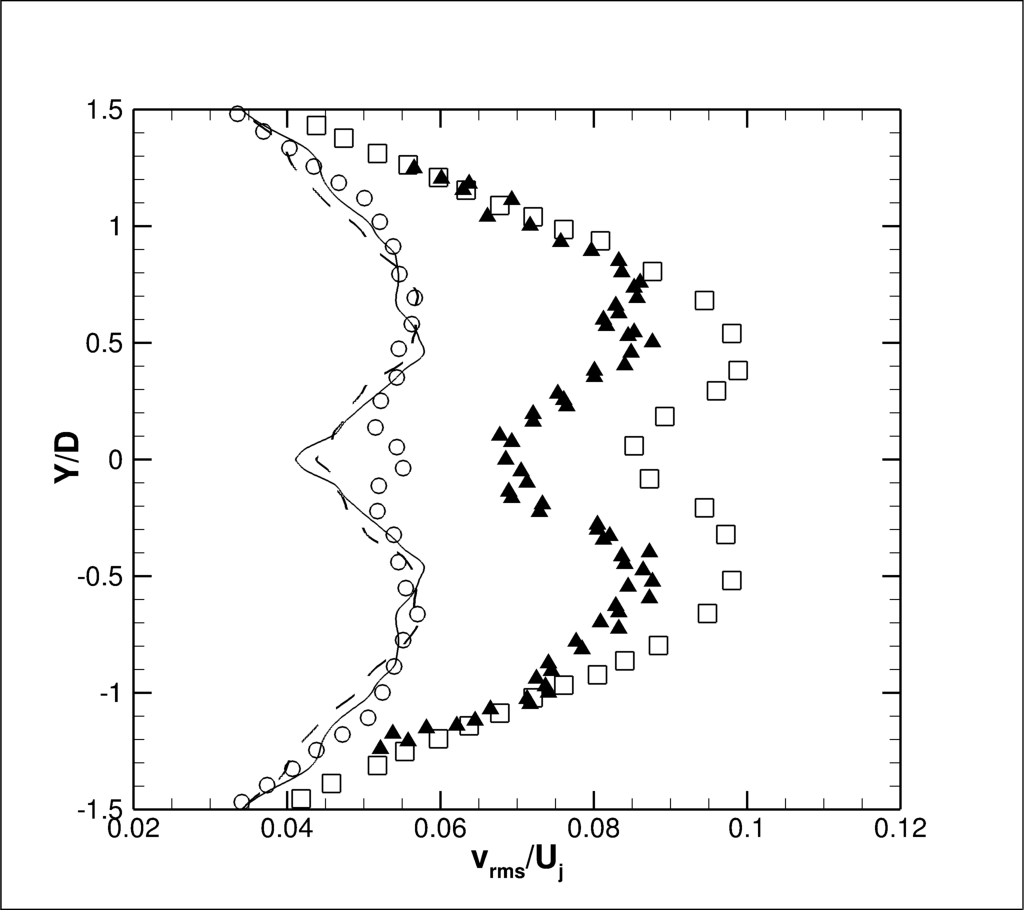}}
  \subfigure[$v^{*}_{RMS}$ - X=15D ; $-1.5D\leq Y\leq 1.5D$]
    {\includegraphics[trim= 5mm 5mm 5mm 5mm, clip, width=0.45\textwidth]
	{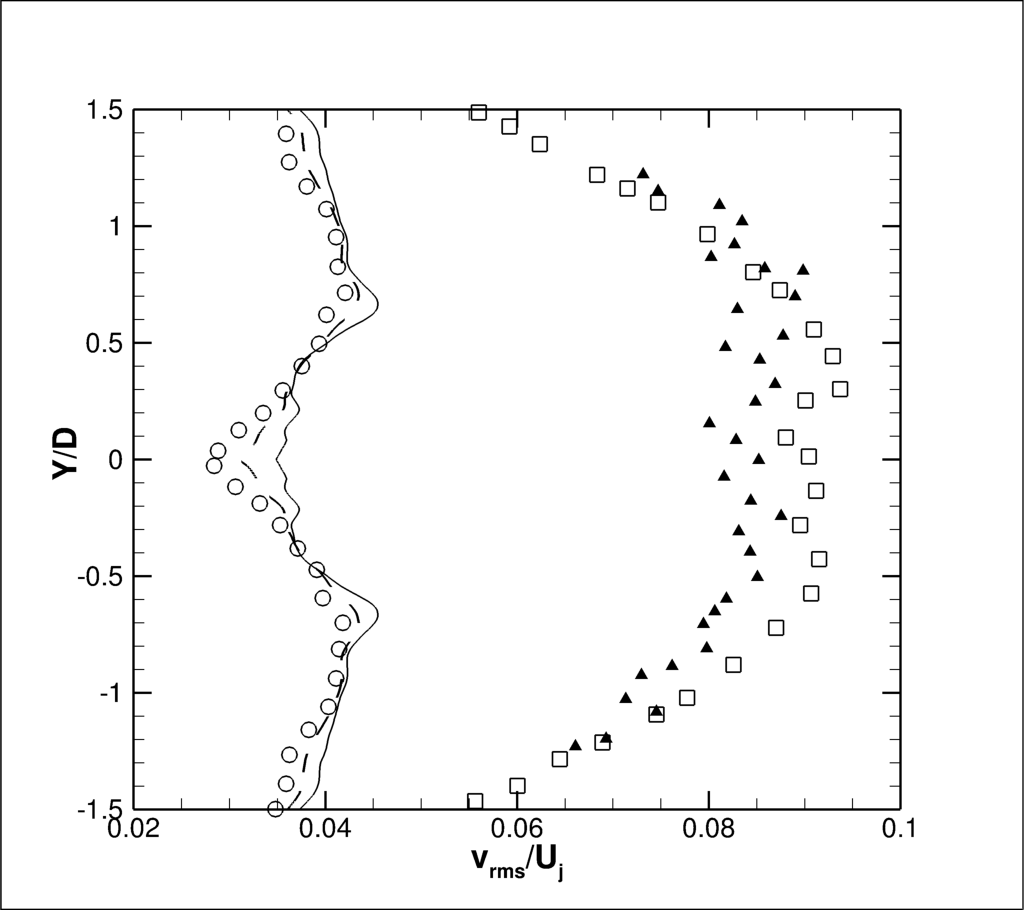}}
	\caption{{  Profiles of RMS of time fluctuation of radial 
	component of velocity, $v_{RMS}^{*}$, for S1 and S2 simulations,
	at different positions within the computational domain.
	(\textbf{--}), simulation S2; 
	(\textbf{-}\textbf{-}), simulation S3;
	($\bigcirc$), simulation S4; ($\square$), numerical data; 
	($\blacktriangle$), experimental data.}}
	\label{fig:prof-v-rms-sgs-new}
\end{figure}

\subsubsection{Component of Reynolds Stress Tensor}

Figures \ref{subfig:uv-sgs-s2}, \ref{subfig:uv-sgs-s3} and 
\ref{subfig:uv-sgs-s4} present lateral view and profiles of 
$\langle u^{*}v^{*}\rangle$ component of the Reynolds stress tensor 
computed using three different SGS models, respectively. One can 
observe that the simulation performed using different SGS models have 
produced very similar distributions of $\langle u^{*}v^{*} \rangle$. 
%
All numerical simulations performed in the present work 
have difficulties to correctly predict the profiles of 
$\langle u^{*}v^{*} \rangle$  as presented in Fig.\ 
{  \ref{fig:prof-uv-av-sgs-new}.
At $X=2.5D$, S2, S3 and S4 calculations present a profile of 
$\langle u^{*}v^{*} \rangle$ which is similar to the reference data. 
However, the simulations performed in the current work fail to correctly 
represent the peaks near the lipline regions. For $X\geq5.0D$ the 
$\langle u^{*}v^{*} \rangle$ profiles calculated in the present work
fails to reproduce the shape and peaks of $\langle u^{*}v^{*} \rangle$.
The cause of the issue could be related to an eventual lack of grid
points in the radial direction. In spite of that, more studies on the
subject are necessary in order to understand such behavior.}
%
%
\begin{figure}[htb!]
  \centering
  \subfigure[$\langle u^{*}v^{*}\rangle$ - X=2.5D ; $-1.5D\leq Y\leq 1.5D$]
    {\includegraphics[trim= 5mm 5mm 5mm 5mm, clip, width=0.45\textwidth]
	{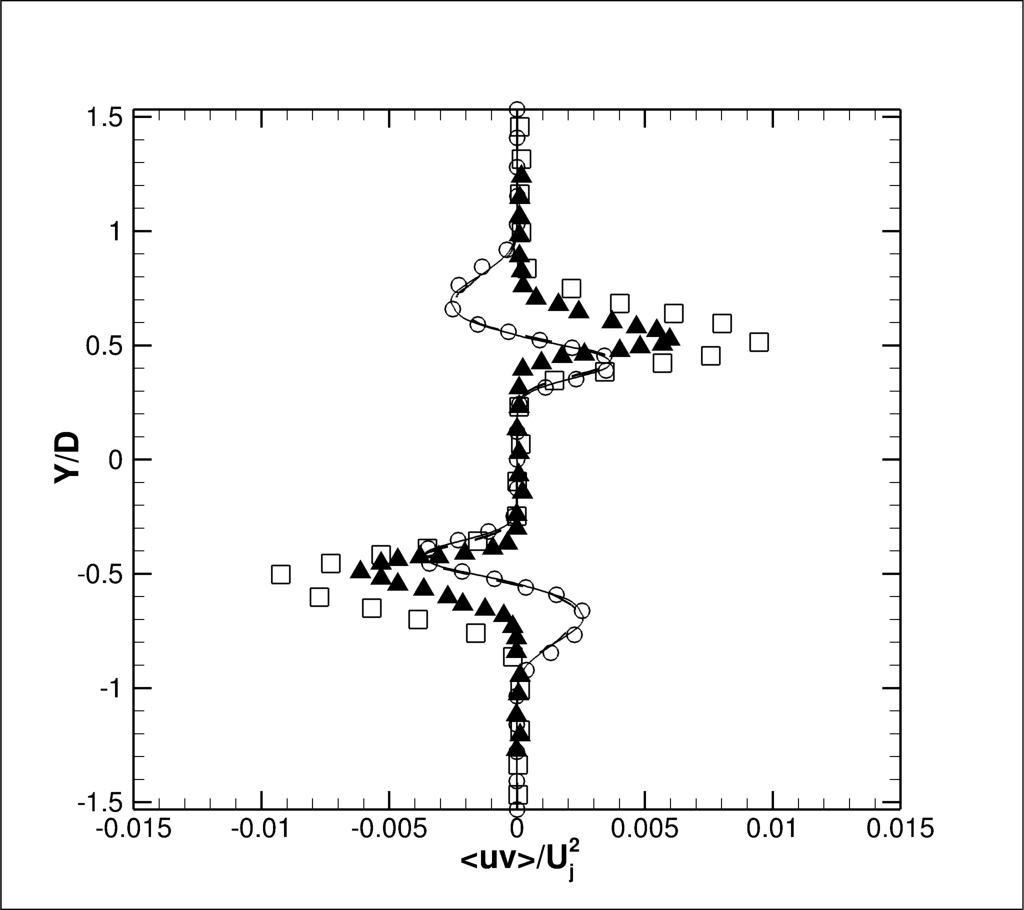}}
  \subfigure[$\langle u^{*}v^{*}\rangle$ - X=5.0D ; $-1.5D\leq Y\leq 1.5D$]
    {\includegraphics[trim= 5mm 5mm 5mm 5mm, clip, width=0.45\textwidth]
	{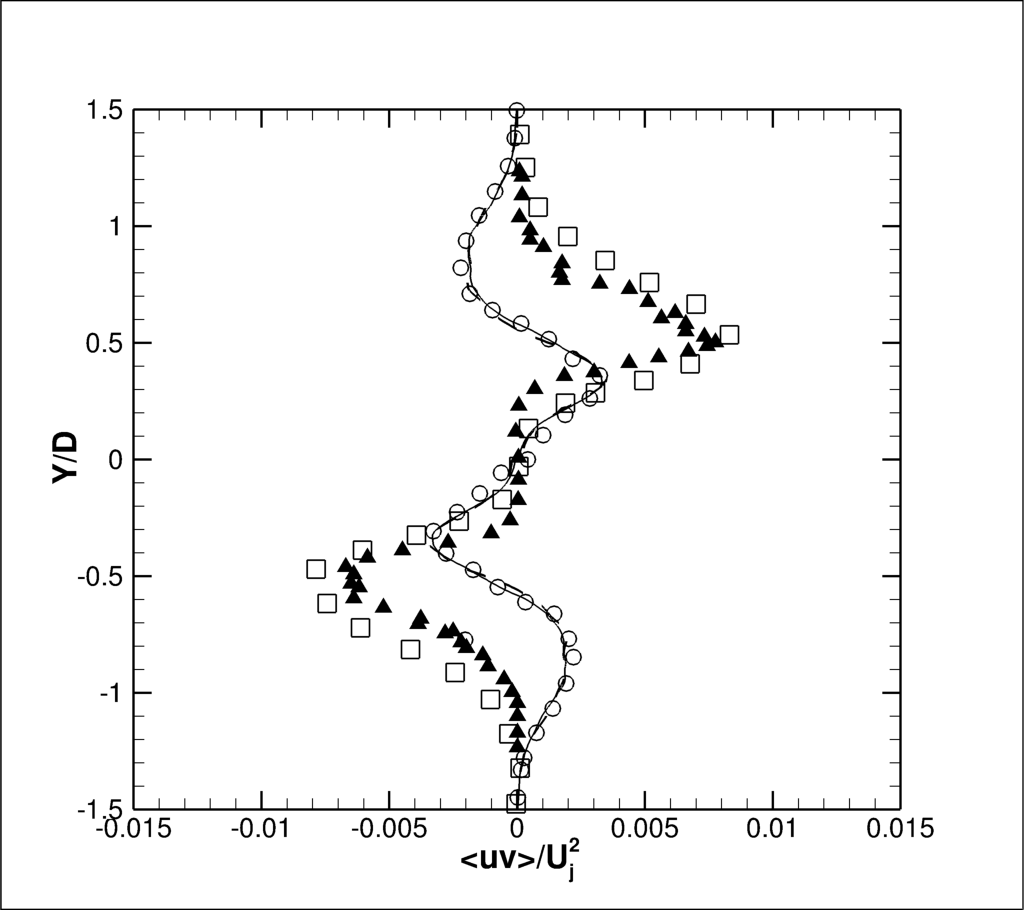}}
  \subfigure[$\langle u^{*}v^{*}\rangle$ - X=10D ; $-1.5D\leq Y\leq 1.5D$]
    {\includegraphics[trim= 5mm 5mm 5mm 5mm, clip, width=0.45\textwidth]
	{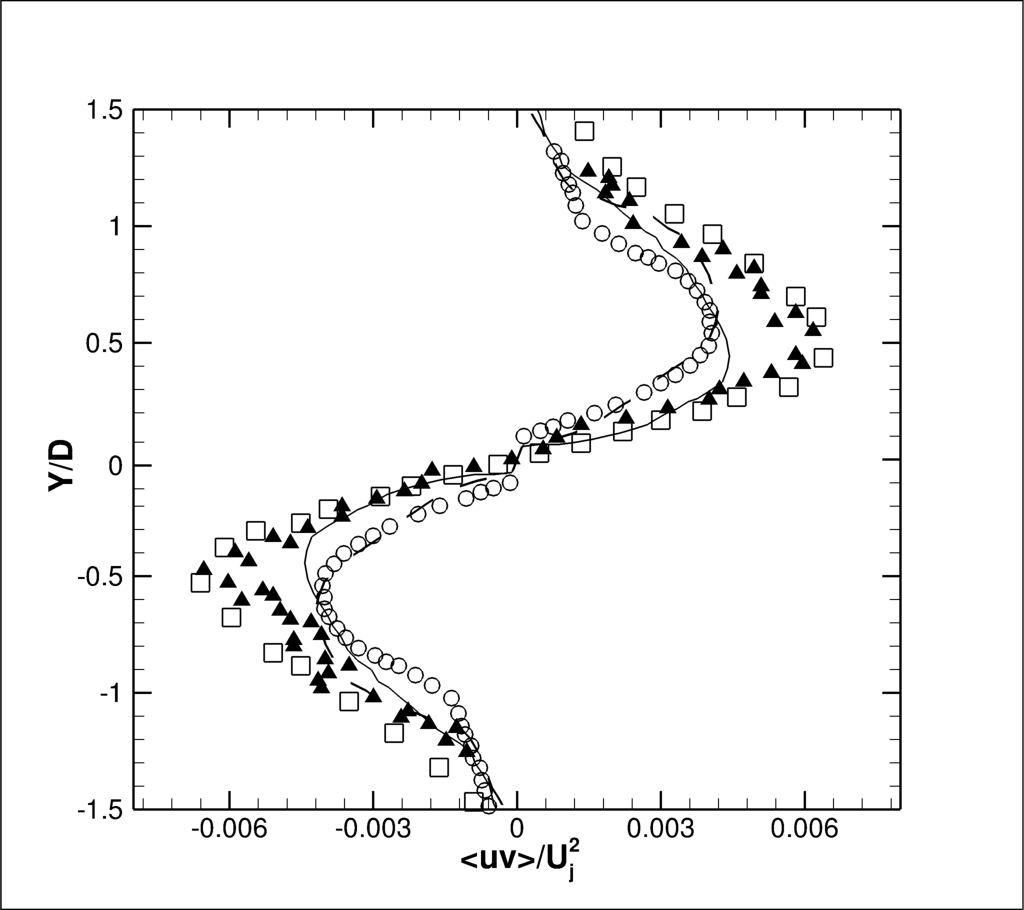}}
  \subfigure[$\langle u^{*}v^{*}\rangle$ - X=15D ; $-1.5D\leq Y\leq 1.5D$]
    {\includegraphics[trim= 5mm 5mm 5mm 5mm, clip, width=0.45\textwidth]
	{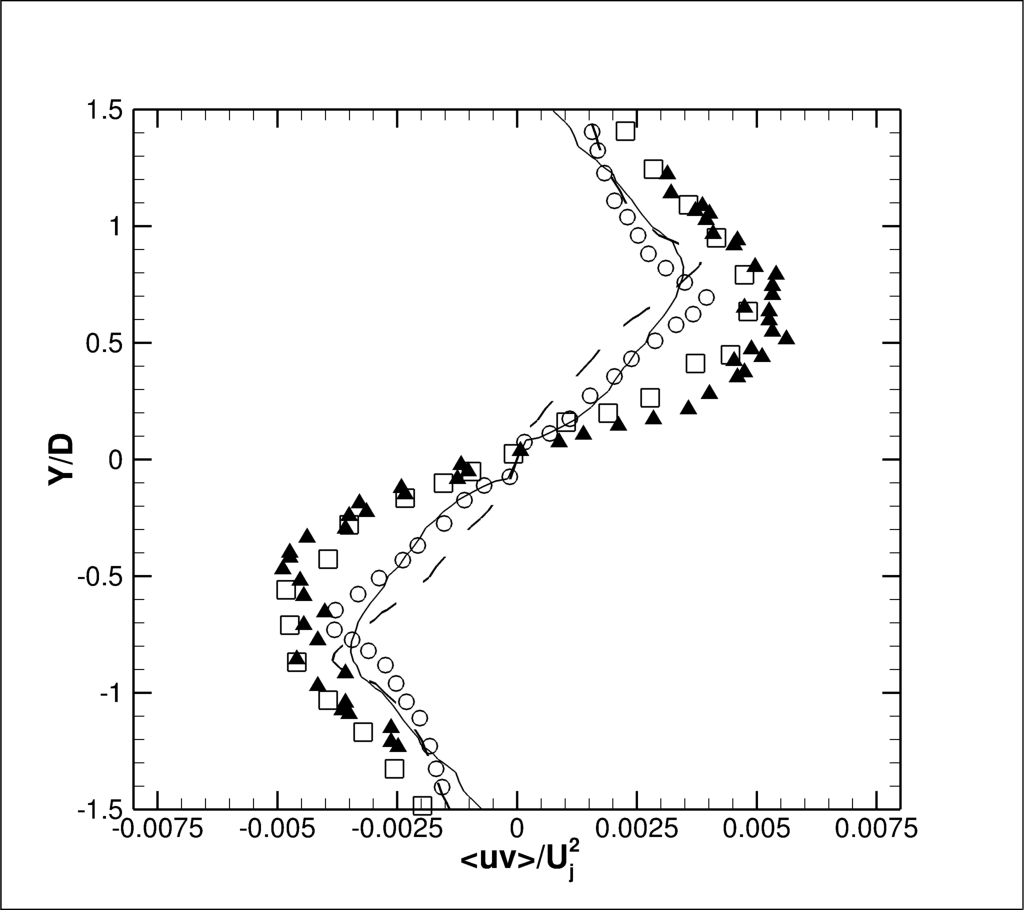}}
	\caption{{  Profiles of the $\langle u^{*}v^{*} \rangle$ 
	Reynolds shear stress tensor component, for S2, S3 and S4 simulations, at 
    different positions within the computational domain.
	(\textbf{--}), S2 simulation; 
	(\textbf{-}\textbf{-}), S3 simulation;
	($\bigcirc$), S4 simulation; ($\square$), numerical data; 
	($\blacktriangle$), experimental data.}}
	\label{fig:prof-uv-av-sgs-new}
\end{figure}

\subsubsection{Time Averaged Eddy Viscosity}

The distribution of the eddy viscosity, $\mu_{t}$, is discussed 
in the current subsection. Figure \ref{fig:lat-mut-sgs} presents 
distributions of time averaged eddy viscosity calculated using 
different SGS models. All subgrid scale closures used in the present 
work, the static Smagorinsky \cite{Smagorinsky63,Lilly65,Lilly67}, 
the dynamic Smagorinsky \cite{Germano91,moin91} and the Vreman 
\cite{vreman2004} models, are dependent of the local mesh size 
by design. This characteristic is exposed on the lateral view 
of the flow presented in Fig.\ \ref{fig:lat-mut-sgs}. The SGS 
models are only acting in the region where mesh presents a low 
resolution. Near the entrance domain, where the computational 
grid is very refined, the eddy viscosity can be neglected. 
\begin{figure}[htb!]
  \centering
  \subfigure[$\langle\mu_{t}\rangle$ - S2 simulation]
    {\includegraphics[trim= 5mm 5mm 5mm 5mm, clip, width=0.32\textwidth]
	{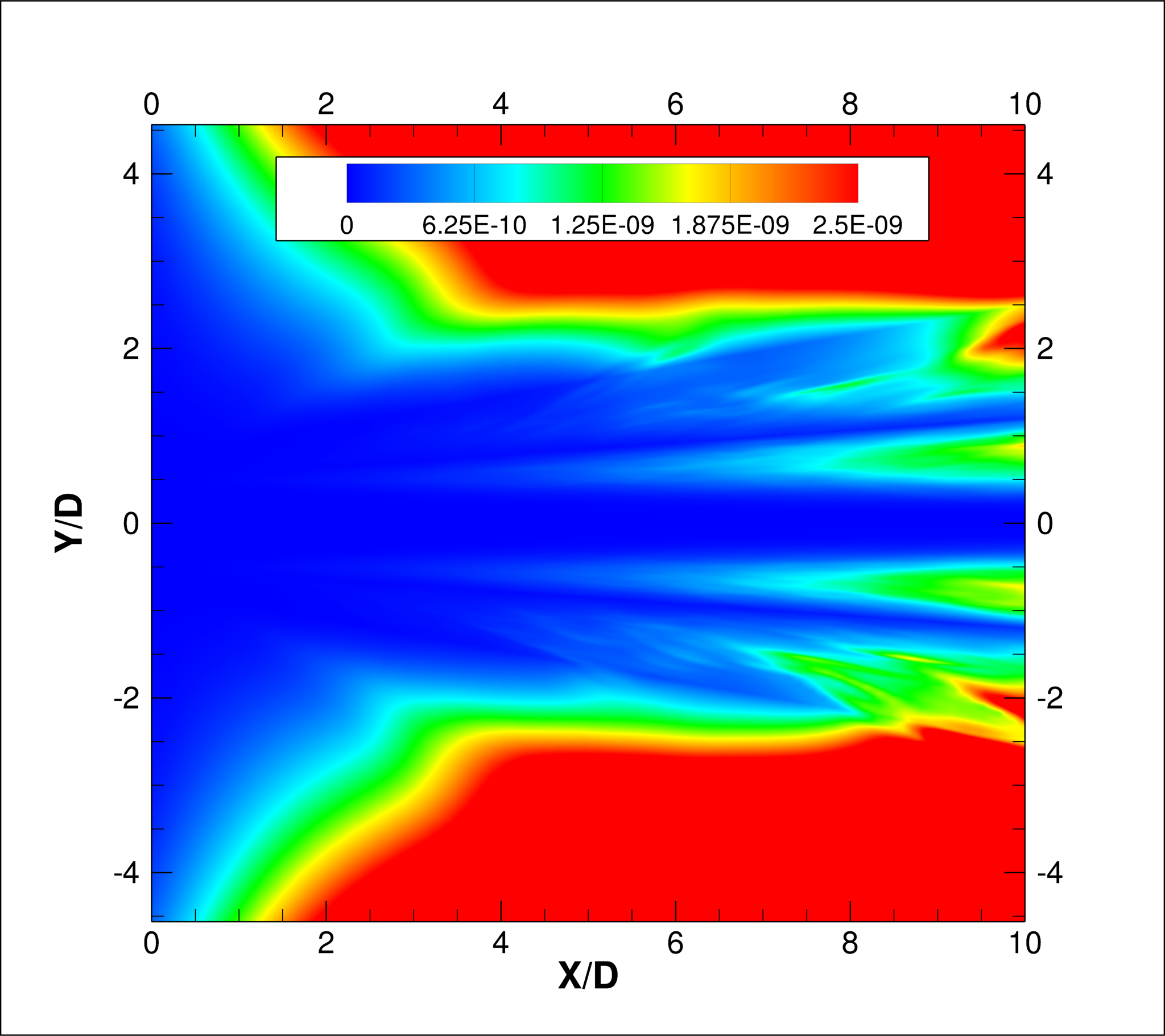}}
  \subfigure[$\langle\mu_{t}\rangle$ - S3 simulation]
    {\includegraphics[trim= 5mm 5mm 5mm 5mm, clip, width=0.32\textwidth]
	{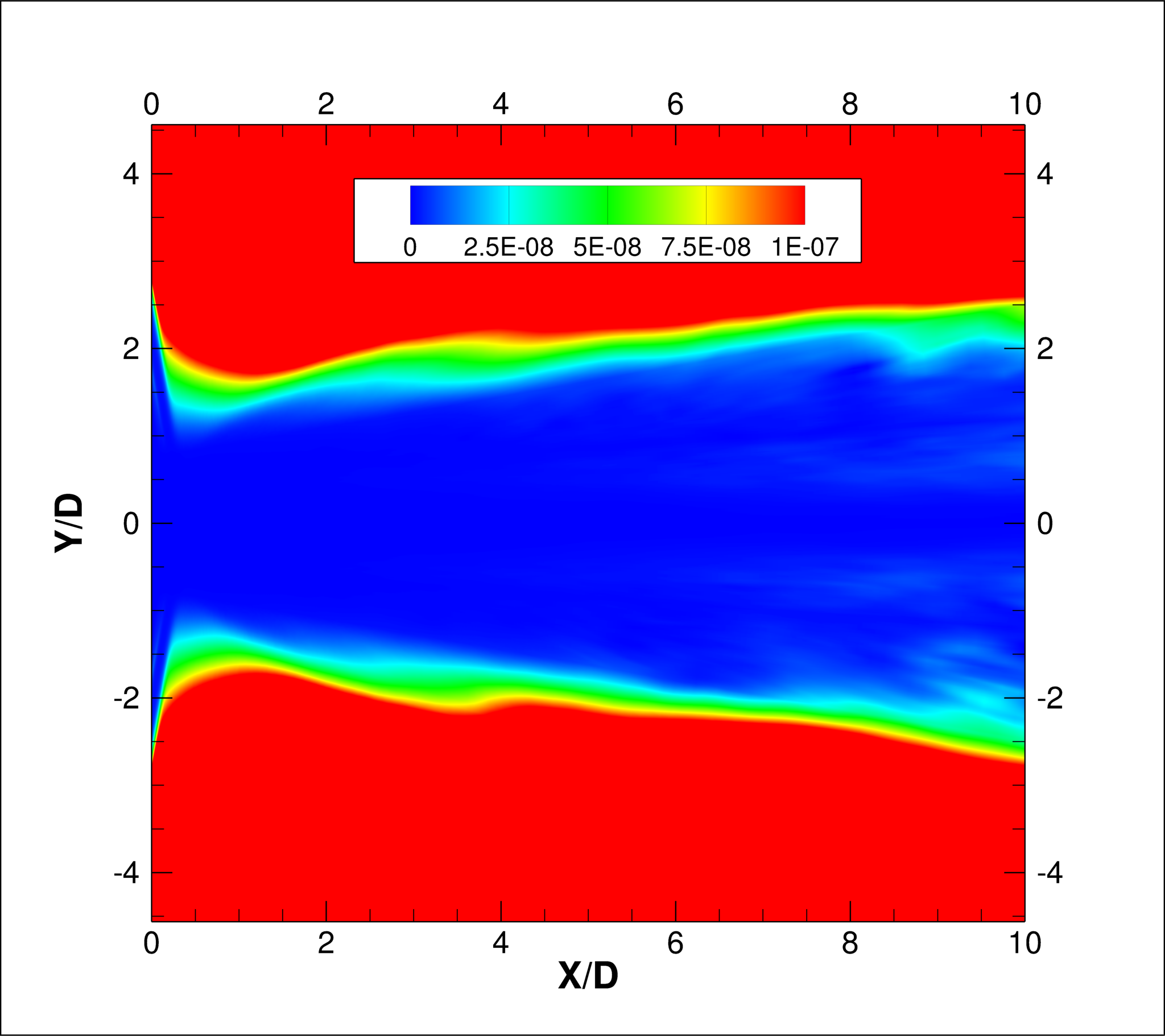}}
  \subfigure[$\langle\mu_{t}\rangle$ - S4 simulation]
    {\includegraphics[trim= 5mm 5mm 5mm 5mm, clip, width=0.32\textwidth]
	{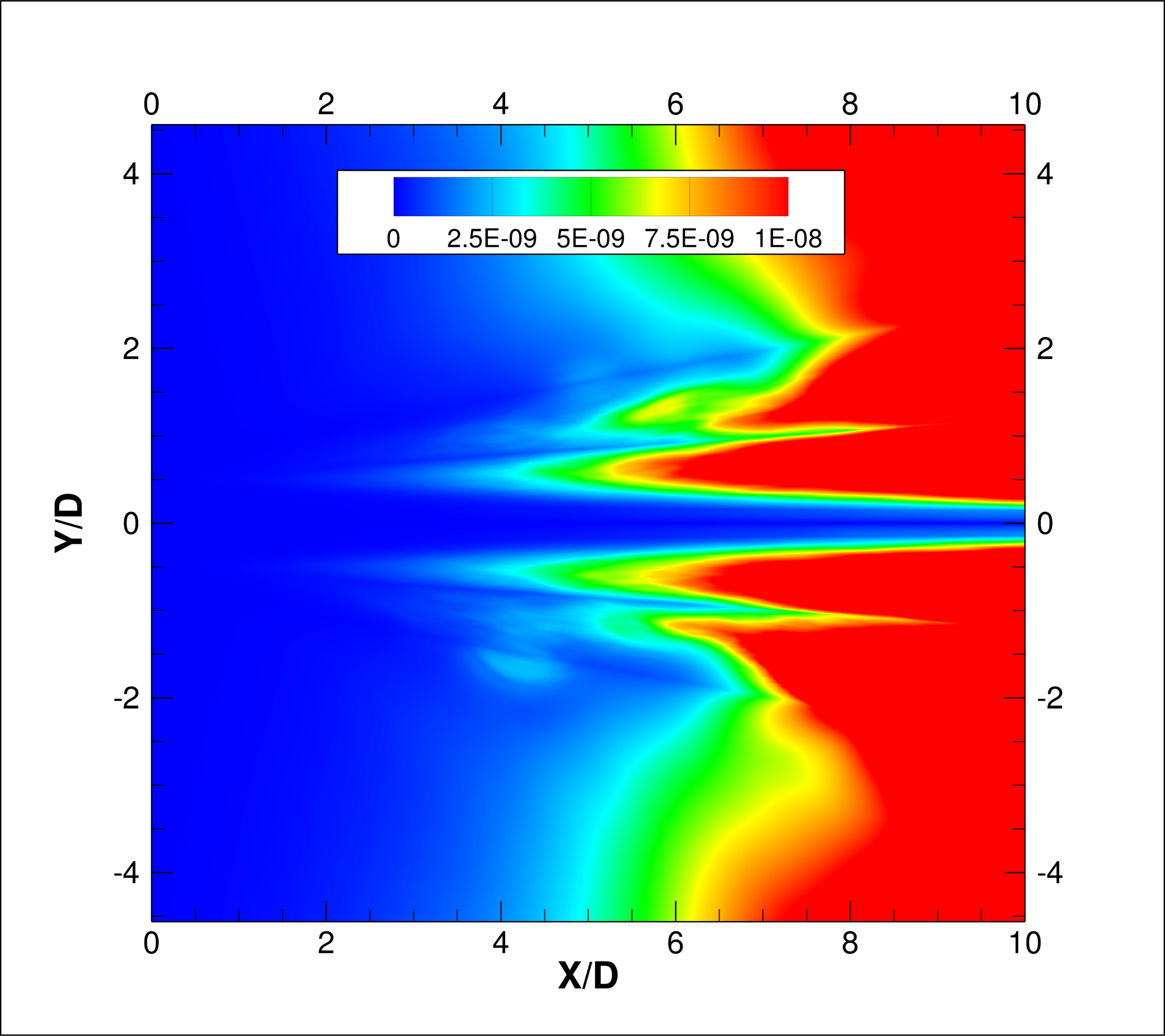}}
	\caption{Lateral view of time averaged eddy viscosity, 
	$\langle\mu_{t}\rangle$, for S2, S3 and S4 simulations.} 
	\label{fig:lat-mut-sgs}
\end{figure}

The remark goes in the same direction of the work of Li and Wang
\cite{Li15}, which indicates that SGS closures introduce numerical 
dissipation that can be used as a stabilizing mechanism. However, 
this numerical dissipation does not necessarily add more physics 
of the turbulent flow to the LES solution. Therefore, in the present 
work, the numerical truncation, which generates the dissipative 
characteristic of JAZzY solutions, has shown to overcome the 
effects of the SGS modeling. The mesh needs to be very fine in order
to achieve good results with second order spatial discretizations.
The grid refinement generates very small grid spacing. Consequently, 
the SGS models, which are strongly dependent on the filter width, cannot 
have a very decisive effect on the solution. LES of compressible flow 
configurations without the use of SGS closures would be welcome in order 
to complete such discussion.
%

%


\section{Concluding Remarks}

The current work is concerned with a study of the effects of different 
subgrid scales models on perfectly expanded supersonic jet 
flow configurations using centered second-order spatial 
discretizations. A formulation based on the System I set 
of equations is used in the present work. The time integration 
is performed using a five-stage, second-order, explicit Runge-Kutta scheme.
Four large eddy simulations of compressible jet flows are performed 
in the present research using two different mesh configurations 
and three different subgrid scale models. Their effects
on the large eddy simulation solution are compared and discussed.

{  Large eddy simulations of high Reynolds 
supersonic jet flows using a second-order spatial 
discretization require high mesh resolution. A special 
care is necessary at the shear-layer region of the the 
jet flow.} The mesh refinement study performed in the 
current study has indicated that, in the region where 
the grid presents high resolution, the simulations are 
in good agreement with experimental and numerical 
references. For the mesh with 14 million points, the 
simulation has produced good results for $X<2.5D$ 
For the other mesh, with 50 
million points, the simulations provided good agreement 
with the literature for $X<5.0D$.
The eddy viscosity, calculated by the static Smagorinsky 
model, presents very low levels in the region where 
the current results have good correlation with the 
data from the literature.

The refined grid used on the mesh refinement study, Mesh B,
is selected for the comparison of SGS model effects on the 
results of the large eddy simulations. Three compressible jet
flow simulations are performed using the classic Smagorinsky
model \cite{Lilly67,Lilly65,Smagorinsky63}, the dynamic 
Smagorinsky model \cite{Germano91,moin91} and the Vreman model
\cite{vreman2004}. All three simulations present similar 
behavior. Results yield good agreement with the references 
for $X<5.0D$. In the region where the grid is very fine and 
the results correlate well with the literature, the eddy 
viscosity coefficient, provided by the SGS model, has very 
low values. The reason for this behavior is related to the 
fact that the SGS closures used in the current work are 
strongly dependent on the filter width, which is proportional 
to the local mesh size. 

The numerical results indicate that it is possible to achieve
good results using second-order spatial discretizations for LES 
calculations. However, the mesh ought to be well resolved in order 
to overcome the truncation errors from the low order numerical scheme. 
{  Moreover, the numerical discretization can add significant 
artificial dissipation to the solution in the region where the mesh 
refinement is not well resolved. One can see this effect as a reduction 
of the local Reynolds number in the region where the mesh is not well 
refined which can be interpreted as a local overestimation of viscous 
effects in the flow. 

It also of most importance the highlight here that very fine meshes 
yield very small filter widths. Consequently, the effects of the 
eddy viscosity coefficient calculated by the SGS models on the
solution become unimportant for the numerical approach used in the
current work. The work of Li and Wang \cite{Li15} have presented 
similar conclusions for simplified problems. Li and Wang further 
emphasize that SGS closures introduce numerical dissipation that 
can be used as a numerical stabilizing mechanism. However, this 
numerical dissipation does not necessarily add more physics of 
the turbulent flow behavior to the LES solution.}
%


\section*{Acknowledgments}

The authors gratefully acknowledge the partial support for this research 
provided by Conselho Nacional de Desenvolvimento Cient\'ifico e Tecnol\'ogico, 
CNPq, under the Research Grants No.\ 309985/2013-7, No.\ 400844/2014-1 and 
No.\ 443839/2014-0\@. The authors are also indebted to the partial financial 
support received from Funda\c{c}\~{a}o de Amparo \`{a} Pesquisa do Estado de 
S\~{a}o Paulo, FAPESP, under the Research Grants No.\ 2008/57866-1, 
No.\ 2013/07375-0 and No.\ 2013/21535-0. 


\bibliography{references}

\begin{thebibliography}{32}
\providecommand{\natexlab}[1]{#1}
\providecommand{\url}[1]{\texttt{#1}}
\expandafter\ifx\csname urlstyle\endcsname\relax
  \providecommand{\doi}[1]{doi: #1}\else
  \providecommand{\doi}{doi: \begingroup \urlstyle{rm}\Url}\fi

\bibitem[{ANSYS}(2016)]{ICEM}
{ANSYS}.
\newblock http://www.ansys.com/, 2016.

\bibitem[Bigarella(2002)]{BIGA02}
E.~D.~V. Bigarella.
\newblock Three-dimensional turbulent flow over aerospace configurations.
\newblock {M.Sc.} {T}hesis, Instituto Tecnol\'{o}gico de Aeron\'{a}utica, S\~ao
  Jos\'e dos Campos, SP, Brasil, 2002.

\bibitem[Bigarella(2007)]{BIGA07}
E.~D.~V. Bigarella.
\newblock \emph{Advanced Turbulence Modeling for Complex Aerospace
  Applications}.
\newblock PhD thesis, Instituto Tecnol\'{o}gico de Aeron\'{a}utica, S\~ao
  Jos\'e dos Campos, SP, Brasil, 2007.

\bibitem[Bridges and Wernet(2008)]{bridges2008turbulence}
J.~Bridges and M.~P. Wernet.
\newblock Turbulence associated with broadband shock noise in hot jets.
\newblock In \emph{{AIAA Paper No.\ 2008-2834}, 14th AIAA/CEAS Aeroacoustics
  Conference}, Vancouver, Canada, May 2008.

\bibitem[Choi and Moin(2012)]{moin2012}
H.~Choi and P.~Moin.
\newblock Grid-point requirements for large eddy simulation: {C}hapman's
  estimates revisited.
\newblock \emph{Physics of Fluids}, 24\penalty0 (1):\penalty0 011702, Jan.
  2012.

\bibitem[Clark et~al.(1979)Clark, Ferziger, and Reynolds]{Clark79}
R.~A. Clark, J.~Z. Ferziger, and W.~C. Reynolds.
\newblock Evaluation of subgrid-scale models using an accurately simulated
  turbulent flow.
\newblock \emph{Journal of Fluid Mechanics}, 91:\penalty0 1--16, 1979.
\newblock \doi{0022-1 120/79/4207-6000}.

\bibitem[Deardorff(1970)]{Deardorff70}
J.~W. Deardorff.
\newblock A numerical study of three-dimensional turbulent channel flow at
  large reynolds numbers.
\newblock \emph{Journal of Fluid Mechanics}, 41, part 2:\penalty0 453--480,
  1970.

\bibitem[Garnier et~al.(2009)Garnier, Adams, and Sagaut]{Garnier09}
E.~Garnier, N.~Adams, and P.~Sagaut.
\newblock \emph{Large Eddy Simulation for Compressible Flows}.
\newblock Springer, 2009.
\newblock \doi{10.1007/978-90-481-2819-8}.

\bibitem[Germano(1990)]{germano90}
M.~Germano.
\newblock Averaging invariance of the turbulent equations and similar subgrid
  scale modeling.
\newblock In \emph{Center for Turbulence Research Manuscript 116}. Stanford
  University and NASA - Ames Research Center, 1990.

\bibitem[Germano et~al.(1991)Germano, Piomelli, Moin, and Cabot]{Germano91}
M.~Germano, U.~Piomelli, P.~Moin, and W.~H. Cabot.
\newblock A dynamic subgridscale eddy viscosity model.
\newblock \emph{Physics of Fluids A: Fluid Dynamics}, 3\penalty0 (7), July
  1991.
\newblock \doi{10.1063/1.857955}.

\bibitem[Jameson and Mavriplis(1986)]{jameson_mavriplis_86}
A.~Jameson and D.~Mavriplis.
\newblock Finite volume solution of the two-dimensional euler equations on a
  regular triangular mesh.
\newblock \emph{AIAA Journal}, 24\penalty0 (4):\penalty0 611--618, Apr. 1986.

\bibitem[Jameson et~al.(1981)Jameson, Schmidt, and Turkel]{Jameson81}
A.~Jameson, W.~Schmidt, and E.~Turkel.
\newblock Numerical solutions of the euler equations by finite volume methods
  using runge-kutta time-stepping schemes.
\newblock In \emph{{\em AIAA Paper 81--1259}, Proceedings of the {AIAA} 14th
  Fluid and Plasma Dynamic Conference}, Palo Alto, Californa, USA, June 1981.

\bibitem[Junqueira-Junior(2016)]{Junior16}
C.~Junqueira-Junior.
\newblock \emph{Development of a Parallel Solver for Large Eddy Simulation of
  Supersonic Jet Flow}.
\newblock PhD thesis, Instituto Tecnol\'{o}gico de Aeron\'{a}utica, S\~{a}o
  Jos\'{e} dos Campos, SP, Brazil, 2016.

\bibitem[Junqueira-Junior et~al.(2015)Junqueira-Junior, Yamouni, Azevedo, and
  Wolf]{JuniorAIAA2015}
C.~Junqueira-Junior, S.~Yamouni, J.~L.~F. Azevedo, and W.~R. Wolf.
\newblock Large eddy simulations of supersonic jet flows for aeroacoustic
  applications.
\newblock In \emph{{\em AIAA Paper No.\ 2015-3306}, Proceedings of the 33rd
  {AIAA} Applied Aerodynamics Conference}, Dallas, TX, June 2015.

\bibitem[Junqueira-Junior et~al.(2016)Junqueira-Junior, Yamouni, Azevedo, and
  Wolf]{jr16-aiaa}
C.~Junqueira-Junior, S.~Yamouni, J.~L.~F. Azevedo, and W.~R. Wolf.
\newblock {Influence of Different Subgrid Scale Models in LES of Supersonic Jet
  Flows}.
\newblock In \emph{{\em AIAA Paper No.\ 2016-4093}, 46th AIAA Fuid Dynamics
  Conference, AIAA Aviation Forum}, Washington, D.C., Jun. 2016.

\bibitem[Leonard(1974)]{Leonard74}
A.~Leonard.
\newblock {Energy Cascade in Large Eddy Simulations of Turbulent Fluid Flows}.
\newblock \emph{Adv. Geophys.}, A18:\penalty0 237--48, 1974.

\bibitem[Li and Wang(2015)]{Li15}
Y.~Li and Z.~J. Wang.
\newblock A priori and a posteriori evaluation of subgrid stress models with
  the {B}urger's equation.
\newblock In \emph{{\em AIAA Paper No.\ 2015-1283}, Proceedings of 53rd AIAA
  Aerospace Sciences Meeting}, Kissimmee, FL, Jan. 2015.

\bibitem[Lilly(1965)]{Lilly65}
D.~K. Lilly.
\newblock On the computational stability of numerical solutions of time-
  dependent non-linear geophysical fluid dynamics problems.
\newblock \emph{Monthly Weather Review}, 93\penalty0 (1):\penalty0 11--25,
  January 1965.
\newblock \doi{10.1175/1520-0493(1965)093<0011:OTCSON>2.3.CO;2}.

\bibitem[Lilly(1967)]{Lilly67}
D.~K. Lilly.
\newblock The representation of small-scale turbulence in numerical simulation
  experiments.
\newblock In \emph{{\em IBM Form No. 320-1951}, Proceedings of the IBM
  Scientific Computing Symposium on Environmental Sciences}, pages 195--210,
  Yorktown Heights, N.Y., 1967.

\bibitem[Long et~al.(1991)Long, Khan, and Sharp]{Long91}
L.~N. Long, M.~Khan, and H.~T. Sharp.
\newblock {A Massively Parallel Three-Dimensional Euler/Navier-Stokes Method}.
\newblock \emph{AIAA Journal}, 29\penalty0 (5):\penalty0 657--666, 1991.

\bibitem[Mendez et~al.(2010)Mendez, Shoeybi, Sharma, Ham, Lele, and
  Moin]{Mendez10}
S.~Mendez, M.~Shoeybi, A.~Sharma, F.~E. Ham, S.~K. Lele, and P.~Moin.
\newblock Large-eddy simulations of perfectly-expanded supersonic jets: Quality
  assessment and validation.
\newblock In \emph{{\em AIAA Paper No.\ 2010--0271}, 48th AIAA Aerospace
  Sciences Meeting Including the New Horizons Forum and Aerospace Exposition,
  Aerospace Sciences Meetings}, Jan. 2010.
\newblock \doi{10.2514/6.2010-271}.

\bibitem[Mendez et~al.(2012)Mendez, Shoeybi, Sharma, Ham, and Moin]{Mendez12}
S.~Mendez, M.~Shoeybi, A.~Sharma, F.~E. Ham, and S.~K. L.~P. Moin.
\newblock Large-eddy simulations of perfectly-expanded supersonic jets using an
  unstructured solver.
\newblock \emph{AIAA Journal}, 50\penalty0 (5):\penalty0 1103--1118, May 2012.

\bibitem[Moin et~al.(1991)Moin, Squires, Cabot, and Lee]{moin91}
P.~Moin, K.~Squires, W.~Cabot, and S.~Lee.
\newblock A dynamic subgrid-scale model for compressible turbulence and scalar
  transport.
\newblock \emph{Physics of Fluids A: Fluid Dynamics (1989-1993)}, 3\penalty0
  (11):\penalty0 2746--2757, 1991.
\newblock \doi{10.1063/1.858164}.

\bibitem[Sagaut(2002)]{Sagaut05}
P.~Sagaut.
\newblock \emph{Large Eddy Simulation for Incompressible Flows}.
\newblock Springer, 2002.

\bibitem[Smagorinsky(1963)]{Smagorinsky63}
J.~Smagorinsky.
\newblock General circulation experiments with the primitive equations: I. the
  basic experiment.
\newblock \emph{Monthly Weather Review}, 91\penalty0 (3):\penalty0 99--164,
  March 1963.
\newblock \doi{10.1175/1520-0493(1963)091<0099:GCEWTP>2.3.CO;2}.

\bibitem[Turkel and Vatsa(1994)]{Turkel_Vatsa_1994}
E.~Turkel and V.~N. Vatsa.
\newblock {Effect of Artificial Viscosity on Three-Dimensional Flow Solutions}.
\newblock \emph{AIAA Journal}, 32\penalty0 (1):\penalty0 39--45, 1994.
\newblock URL \url{http://doi.aiaa.org/10.2514/3.11948}.

\bibitem[Vreman(1995)]{Vreman1995}
A.~W. Vreman.
\newblock \emph{Direct and Large-Eddy Simulation of the Compressible Turbulent
  Mixing Layer}.
\newblock PhD thesis, Universiteit Twente, The Netherlands, 1995.

\bibitem[Vreman(2004)]{vreman2004}
A.~W. Vreman.
\newblock An eddy-viscosity subgrid-scale model for turbulent shear flow:
  Algebraic theory and applications.
\newblock \emph{Physics of Fluids}, 16\penalty0 (10), October 2004.

\bibitem[Vreman et~al.(1996)Vreman, Geurts, and Kuerten]{vreman1996}
B.~Vreman, B.~Geurts, and H.~Kuerten.
\newblock Large-eddy simulation of the turbulent mixing layer using the {C}lark
  model.
\newblock \emph{Theoretical Computational Fluid Dynamics}, 8\penalty0
  (4):\penalty0 309--324, 1996.

\bibitem[Wang(2015)]{WangAIAA2015}
Z.~J. Wang.
\newblock Large eddy simulations of turbulent flows using discontinuous high
  order methods.
\newblock In \emph{22nd AIAA Computational Fluid Dynamics Conference, {\em
  Invited talk}}, Dallas, TX, June 2015.

\bibitem[Wolf et~al.(2012)Wolf, Azevedo, and Lele]{Wolf2012}
W.~R. Wolf, J.~L.~F. Azevedo, and S.~K. Lele.
\newblock Convective effects and the role of quadrupole sources for aerofoil
  aeroacoustics.
\newblock \emph{Journal of Fluid Mechanics}, 708:\penalty0 502--538, 2012.

\bibitem[Yoshizawa(1986)]{Yoshizawa86}
A.~Yoshizawa.
\newblock Statistical theory for compressible turbulent shear flows, with the
  application to subgrid modeling.
\newblock \emph{Physics of Fluids}, 29\penalty0 (7), July 1986.
\newblock \doi{10.1063/1.865552}.

\end{thebibliography}

\end{document}